\documentclass[twocolumn]{aastex631}

\usepackage{soul}
\usepackage{comment}
\usepackage[]{threeparttable}

\newcommand\aastex{AAS\TeX}

\accepted{for publication in The Astrophysical Journal on August 14, 2023.}  % added for arxiv

\shorttitle{\aastex\ CGM of LRGs}
\shortauthors{Smailagi{\' c} et al.} 

\graphicspath{{./}{figures/}}

\begin{document}

\title{CIRCUMGALACTIC MEDIUM AT HIGH HALO MASSES - SIGNATURES OF COLD GAS DEPLETION IN LUMINOUS RED GALAXIES}

\author{Marijana Smailagi{\' c}}
\affil{University of California, 1156 High Street, Santa Cruz, CA 95064, USA \\
}
\email{msmailag@ucsc.edu}

\author{Jason Xavier Prochaska}
\affiliation{University of California, 1156 High Street, Santa Cruz, CA 95064, USA \\
}
\affiliation{
Kavli Institute for the Physics and Mathematics of the Universe (Kavli IPMU), 5-1-5 Kashiwanoha, Kashiwa, 277-8583, Japan \\
}

\author{Joseph Burchett}
\affiliation{University of California, 1156 High Street, Santa Cruz, CA 95064, USA \\
}
\affiliation{New Mexico State University, Las Cruces, NM 88003, USA \\
}

\author{Guangtun Zhu}
\affiliation{Johns Hopkins University, 3400 N. Charles Street, Baltimore, MD 21218, USA \\
}

\begin{abstract}

We study ultraviolet \ion{H}{1} and metal line transitions in the circumgalactic medium (CGM) of 15 massive, quenched luminous red galaxies (LRGs) at redshift $z\sim 0.5$ and with impact parameters up to 400 kpc. We selected 8 of LRG-CGM systems to study general properties of the CGM around LRGs, while the other 7 are already known to contain cool CGM gas from \ion{Mg}{2} optical studies (MgII-LRGs). In the general LRGs population, we detect \ion{H}{1} in 4 of 8 LRGs, in all cases with $N_{HI} < 10^{16.7} {\rm cm^{-2}}$. In contrast, all MgII-LRGs show \ion{H}{1}; for four LRGs the \ion{H}{1} column density is $N_{HI} \gtrsim 10^{18} {\rm cm^{-2}}$. The CGM of LRGs also shows low and intermediate ionized lines (such as \ion{C}{3}, \ion{C}{2}, \ion{Si}{3}, \ion{Si}{2}) and highly ionized lines of \ion{O}{6} (we detect \ion{O}{6} around 5 of 7 MgII-LRGs and 1 of 8 in the random sample). Next, we combine our sample with literature LRGs and $\lesssim L^{*}$ galaxies and we find that while for $\lesssim L^{*}$ galaxies CGM \ion{H}{1} Ly$\alpha$ absorption is stronger as galaxies are more massive, the cool CGM traced by \ion{H}{1} Ly$\alpha$ is suppressed above stellar masses of $M* \sim 10^{11.5} M_{\sun}$. While most LRG CGM systems show weak or non-detectable \ion{O}{6} (equivalent width less than 0.2 \AA), a few LRG CGM systems show strong \ion{O}{6} 1031, which in most cases likely originates from groups containing both a LRG and a blue star-forming neighboring galaxy. 

\end{abstract}


\section{Introduction}\label{sec:intro}

% CGM - observations and theory 
It is now well known that many galaxies are surrounded by a significant amount of low-density gas between the disk and virial radius, known as the circumgalactic medium
\citep[CGM; e.g.,][see also reviews \citealt{Chen17} and \citealt{Tumlinson17}]{Bergeron86,Bouche04,Tumlinson11,Rudie12,Thom12,Lehner13,X13,Tumlinson13,Werk13,Rubin14,Zhu14,Zheng17,Burchett18,Cai19,Wotta19,Lan20,Nielsen20}.
\citet{Werk14} found that for local $\sim L^{*}$ galaxies, the mass of the cool CGM gas is comparable to the stellar mass of the galaxy disk.  
This CGM gas is closely related to a galaxy's evolution 
\citep[see e.g.,][see also review \citealt{Voort17rev}]{Oppenheimer06,Keres09,Hummels13,Ford14,Thompson16,Armillotta17,Stern20,Lochhaas21}. 
As galaxies evolve, feedback processes, such as outflows from stars and active galactic nuclei, expel gas into the CGM. At the same time, galaxies accrete new gas from the intergalactic medium and CGM, which might also include recycled previously expelled gas. In some cases, this new gas cools and forms stars, and this process repeats. 
In addition, as halos become more massive, the CGM gas heats. In some cases, some of this CGM gas cools due to thermal instabilities \citep[see e.g.,][]{Oppenheimer06,Ford14,Thompson16,Tumlinson17}.  
Below we summarize the main motivations for our work:

%% motivation to study gas around massive galaxies

% CGM EW(HI) vs mass 
{\it Cool CGM gas around the most massive galaxies.}
Recently, on a lower mass scale, \citet{Bordoloi18} found that more massive galaxies show stronger \ion{H}{1} absorption from their CGM. More precisely, the authors showed that for $\sim L^{*}$ and $\lesssim L^{*}$ galaxies \ion{H}{1} 1215 equivalent widths corrected for the impact parameter increase as stellar mass increases. 
Similar results were also found in other work (for example, \citealt{Chen98} found a similar correlation by using B-band luminosity as a proxy for galaxy mass).
However, in clusters of galaxies \ion{H}{1} is more rarely detected; while for $\sim L^{*}$ galaxies \ion{H}{1} 1215 equivalent width averages $\sim 1$ \AA\ , all 13 CGM systems in clusters of galaxies studied by \citet{Burchett18} show \ion{H}{1} 1215 equivalent widths below $0.4$ \AA\ (see also \citealt{Lopez08} and \citealt{Yoon12}).  
These results indicate that \ion{H}{1} absorption increases with $M_{*}$, but is suppressed for the most massive galaxies.
In addition, current theory predicts that in massive halos at low redshift, newly accreted cool gas is shock-heated to the virial temperature, and these most massive halos contain very small amounts of cool gas \citep{Dekel06,Nelson15,FG17,Voort17}.  
These findings motivated us to study the CGM of massive galaxies and to search for a critical mass above which cool CGM gas is suppressed. 

{\it Cool CGM gas around quenched galaxies.}
Previous studies have found that both star-forming and passive galaxies show relatively similarly strong cool CGM absorption lines \citep[][but see also \citealt{Lan14}]{Thom12}. However, if galaxies contain cool gas in their CGM, the cool gas may form stars. We might ask if cool gas could survive for more than 1 Gyr without being heated or forming stars and, hence, if galaxies that are quenched more than 1 Gyr ago also show similarly strong CGM absorption lines. 
In massive halos, cool gas clouds are surrounded by hot gas and after some time may be evaporated due to thermal conduction, but other factors might also have an important role \citep[see e.g.,][]{Maller04, Gauthier11, Thompson16, GronkeOh18}.

{\it The \ion{O}{6} in the CGM around massive, quenched galaxies.}
\citet{Tumlinson13} found that \ion{O}{6} is correlated with specific star-formation rate and is rare in the CGM of passive galaxies. Some proposed explanations include that the observed red galaxies are more massive and have higher virial temperature such that oxygen is found in a more highly ionized state in these halos \citep{Oppenheimer16}, \ion{O}{6} traces cool gas beyond the accretion shock \citep{Stern18}, \ion{O}{6} traces ancient outflows \citep{Ford14}, and other. 
However, it is not clear if \ion{O}{6} is correlated with the star formation activity or (inverse of) the halo masses. One way to test those explanations is to study the CGM of massive galaxies whose environment may contain star-forming satellite galaxies, potentially offering another origin of \ion{O}{6} in the halos of these galaxies.

{\it Connection between the CGM gas at low and at high redshift around massive galaxies.}
QSO host galaxies are massive galaxies at $z\sim 2-3$ and show the strongest \ion{H}{1} absorption \citep[e.g.,][]{X13}. Currently, it is not clear how long cool gas with temperatures below the halo virial temperature could survive in massive halos. Studying the CGM of similarly massive halos at lower redshifts might constrain the timescale for the survival of the cool gas, and the evolution of cool CGM gas from high to low redshift.

\vspace{10pt}

% LRGs and LRG-CGM in MgII
% LRGs
Luminous red galaxies (LRGs) are the most massive galaxies at $z\sim 0.5$, with average stellar mass  $\sim 10^{11.5} M_{\sun}$ \citep[e.g.,][]{Maraston13}. LRGs, by selection, have negligible star formation rates \citep{Roseboom06} and have been quenched since $z \gtrsim 1$ \citep{Banerji10}. These galaxies are located in overdense regions with average halo masses $M_{\rm halo}\sim 10^{13.5} M_{\sun}$ \citep{White11,Zhu14}. 
% LRG-CGM in MgII
Previous work studying the CGM around LRGs has focused mainly on the \ion{Mg}{2} 2796, 2803 doublet, which traces cool CGM gas, and consists of the strongest resonant lines that are accessible in optical Sloan Digital Sky Survey (SDSS) spectra at $z\sim 0.5$.
These studies have found that only $\sim 5$ percent of LRGs show \ion{Mg}{2} absorption with equivalent width $> 0.3$ \AA\ in the CGM at impact parameters up to $\sim 500$ kpc  \citep[e.g.,][]{Huang16} and that, despite the small covering fraction, these \ion{Mg}{2} absorbers are correlated with LRGs \citep[e.g.,][]{Bouche04,Gauthier09,Zhu14}.
In addition, \citet{Zhu14} found that the average \ion{Mg}{2} absorption as a function of impact parameter is well described by a combination of a `1-halo' term that describes gas within the LRG halos and a `2-halo' term that describes gas from neighboring galaxies from the larger structure. 
These studies showed that although LRGs are quenched and located in massive halos, their halos contain a significant amount of cool, metal-rich gas. 

% LRG-CGM: HST UV studies 
However, these optical studies did not provide access to \ion{H}{1} and higher ionization states (such as \ion{C}{3} and \ion{O}{6}), and therefore did not provide the opportunity to study metal-poor and more highly ionized gas. Most of these lines at $z\sim 0.5$ are located in the ultraviolet (UV) spectral range, and only space-based UV spectrographs offer the opportunity to access them. 
In addition to our own, there are two related ongoing studies of samples of LRG-CGM systems. One is introduced by \citet{Chen18}, where the authors studied \ion{H}{1} and UV metal-line transitions in the CGM at $\lesssim 160$ kpc around 16 massive red galaxies, 9 of which are LRGs (see Section \ref{sec:oth}). The other study is introduced by \citet{Berg18}, where the authors examined \ion{H}{1} and UV metal-line transitions in the CGM of 21 LRGs at impact parameters up to 500 kpc. Both of these studies found a relatively high covering fraction of \ion{H}{1} absorption (for example, $\sim 50-81$ percent of those CGM systems have detected \ion{H}{1}), some of which is optically thick (covering fraction $\sim 15 - 44$ percent).

% This work and previous similar work
In an another UV study of the LRG-CGM systems, 
% In a previous work, 
\citet{Smailagic18} presented results for two LRGs with extreme properties. These two LRGs are part of our full sample of 15 LRGs. \citet{Smailagic18} found that the CGM of these two LRGs show very strong and extended \ion{H}{1} and \ion{C}{3} absorption that exceeds those of all known local $\sim L^{*}$ galaxies. 
In our work, we present the full sample of 15 LRG-CGM systems, with a median redshift $z\sim 0.5$ and impact parameters $R_{\perp} < 400$ kpc, in which we study \ion{H}{1} and UV metal-line transitions. 
One difference between our sample and \citet{Chen18} and \citet{Berg18} samples is that approximately half (7) of the LRGs in our sample were selected such that their CGM shows \ion{Mg}{2} absorption in the SDSS optical spectra, with a goal of studying cool gas around massive quenched galaxies. 

% This work - sections
In Section \ref{sec:data}, we describe our sample selection, observations, data reduction and data analysis. 
In Section \ref{sec:lines}, we describe the quantities that we measure for each LRG-CGM system. 
Section \ref{sec:oth} lists other observations with which we compare our results. 
Section \ref{sec:res} presents our results, including the covering fraction of \ion{H}{1} 1215 and metals in the LRG-CGM systems, their kinematics, and association with LRGs (equivalent widths and column densities). 
Section \ref{sec:discb} discusses our results regarding the general LRG population, including searching for a critical mass above which cool CGM gas is suppressed.
Section \ref{sec:discm} discusses our results involving the LRGs exhibiting \ion{Mg}{2} absorption, including possible connection with the CGM of higher-redshift galaxies and the detection of the warm-hot gas component (\ion{O}{6}). 
Section \ref{sec:ng} discusses the contributions of neighboring galaxies to the CGM of LRGs. 
Appendices \ref{sec:app1}, \ref{sec:appb}, \ref{sec:appc}, \ref{sec:appd}, and \ref{sec:appe} present more detail about absorption lines detected, analysis of less reliable lines, neighboring galaxies, notes about individual absorbers, and additional components associated with LRGs. 

We adopt a flat cosmology with $H_{0}=67.7$ km s$^{-1}$ Mpc$^{-1}$ and $\Omega_{m}=0.307$ \citep{Planck16}. 

\vspace{10pt}

\section{Sample of 15 LRGs} \label{sec:data}

To study the CGM of LRGs, we selected 15 pairs of $z\sim 0.5$ LRGs and UV-bright background quasi-stellar objects (QSOs) at impact parameter ($R_{\perp}$; projected QSO-LRG distance at the LRG redshift) up to 400 kpc, corresponding to $\sim 2/3$ of the LRG virial radii. 
Here, virial radius ($R_{\rm vir}$) is defined as the radius inside which the density equals 200 times the critical density at the LRG redshift, as in \citet{Tumlinson13}. The virial mass is the mass enclosed by the virial radius, approximately $M_{\rm vir} \sim 10^{13.5} M_{\sun}$ for LRGs halos \citep{White11,Zhu14}. 
We obtained HST COS spectra of the background QSOs, which cover \ion{H}{1} Lyman transitions and metal transitions, such as \ion{C}{3}, \ion{C}{2}, \ion{Si}{3}, \ion{Si}{2}, and \ion{O}{6}. 
The properties of target LRGs and background QSOs are summarized in Table~\ref{tab:lrgtab} and Table~\ref{tab:qsotab}, respectively. 
We further compare our LRG sample with local $\sim L^{*}$ galaxies and with literature LRGs in Figure \ref{fig:samplelrg}. 
LRG-CGM system names are created from the sky coordinates of QSOs, in the format 'HHMM$\pm$DDMM'. 

\begin{table*}
\center
\caption{Properties of LRGs$^{a}$
\label{tab:lrgtab}}
\begin{threeparttable}
{
 \setlength\tabcolsep{3pt} 
\makebox[\linewidth]
{
\hspace{-1.1in}
\centering 
\begin{tabular}{ccccccccccc}
\hline
Name & $\alpha$ (J2000) & $\delta$ (J2000) & $z$ & $R_{\perp}$ & $r$ & $u - r$ & $\log_{10} M_{*}$ & $\log_{10} M_{\rm halo}$ & $R_{\rm vir}$ & \ion{Mg}{2}
\\
  &  (deg)  & (deg)  &  & (kpc) &  &  &  $(M_{\sun})$ &  $(M_{\sun})$ & (kpc) & 
\\
\hline
LRG\_1059+4039  &  164.8804  &  40.6666  &  0.4489  &  29  &  -22.53  &  2.32  &  11.29  &  13.67  &  646  &  1  \\ 
LRG\_1121+4345  &  170.4643  &  43.7523  &  0.4216  &  80  &  -22.64  &  2.32  &  11.23  &  13.09  &  417  &  1  \\ 
LRG\_1144+0714  &  176.1888  &  7.2490  &  0.4892  &  98  &  -22.73  &  2.47  &  11.45  &  13.44  &  532  &  1  \\ 
LRG\_1351+2753  &  207.8403  &  27.8910  &  0.4320  &  99  &  -23.17  &  2.32  &  11.53  &  13.61  &  618  &  1  \\ 
LRG\_1520+2534  &  230.0516  &  25.5695  &  0.5385  &  225  &  -23.51  &  2.43  &  11.72  &  13.85  &  716  &  1  \\ 
LRG\_1440-0157  &  220.0786  &  -1.9505  &  0.4801  &  343  &  -22.95  &  2.44  &  11.47  &  13.49  &  554  &  1  \\ 
LRG\_1106-0115  &  166.6615  &  -1.2621  &  0.6106  &  383  &  -23.18  &  2.41  &  11.68  &  13.75  &  641  &  1  \\ 
LRG\_1237+1106  &  189.4359  &  11.1151  &  0.4731  &  23  &  -22.97  &  2.64  &  11.43  &  13.42  &  527  &  0  \\ 
LRG\_1549+0701  &  237.4905  &  7.0167  &  0.4997  &  120  &  -22.46  &  2.51  &  11.30  &  13.19  &  439  &  0  \\ 
LRG\_1306+3421  &  196.5362  &  34.3539  &  0.4690  &  184  &  -22.96  &  2.39  &  11.56  &  13.63  &  623  &  0  \\ 
LRG\_1102+4543  &  165.7055  &  45.7169  &  0.4875  &  193  &  -23.12  &  2.26  &  11.45  &  13.45  &  535  &  0  \\ 
LRG\_1217+4931  &  184.4190  &  49.5309  &  0.5230  &  208  &  -22.77  &  2.36  &  11.41  &  13.37  &  497  &  0  \\ 
LRG\_1251+3025  &  192.7581  &  30.4171  &  0.5132  &  281  &  -23.58  &  2.47  &  11.67  &  13.79  &  690  &  0  \\ 
LRG\_0855+5615  &  133.9658  &  56.2505  &  0.4399  &  315  &  -22.62  &  2.52  &  11.34  &  13.28  &  481  &  0  \\
LRG\_0226+0014  &  36.5509  &  0.2444  &  0.4731  &  357  &  -23.37  &  2.42  &  11.53  &  13.58  &  596  &  0  \\ 
\hline
\end{tabular}
}
\begin{tablenotes}
   \item[a] {Columns contain LRG-CGM name, coordinates ($\alpha$, $\delta$), redshift ($z$), impact parameter ($R_{\perp}$), absolute $r$-magnitude, rest-frame $u-r$ color, stellar mass ($\log_{10} M_{*}$), halo mass ($\log_{10} M_{\rm halo}$), virial radius ($R_{\rm vir}$), and information about \ion{Mg}{2} detection (1 - detection, 0 - non detection).}
\end{tablenotes}
}
\end{threeparttable}
\end{table*}

\begin{table*}
\center
\caption{Properties of QSOs$^{a}$ 
\label{tab:qsotab}}
\begin{threeparttable}
{
\centering 
\begin{tabular}{cccccccc}
\hline
Name & $\alpha$ (J2000) & $\delta$ (J2000) & $z_{\rm QSO}$ & $m_{\rm NUV}$ & $m_{\rm FUV}$ & $N_{\rm orb}$ & $t_{\rm exp}$ 
\\
 &  (deg)  & (deg)  &  &  &  &   & (s)
\\
\hline
J105930+403956 & 164.8789 & 40.6657 & 1.2099 & 18.65 & 20.40 & 5 & 11689 \\
J112150+434455 & 170.4619 & 43.7487 & 0.6700 & 18.42 & 19.03 & 2 & 5417 \\
J114444+071443 & 176.1860 & 7.2455 & 0.9227 & 18.87 & 19.80 & 4 & 11308 \\
J135122+275327 & 207.8458 & 27.8910 & 1.2304 & 18.96 & 20.03 & 4 & 11359 \\
J152013+253438 & 230.0583 & 25.5773 & 1.4639 & 18.89 & 19.89 & 5 & 11375\tnote{b}  \\   
J144021-015627 & 220.0913 & -1.9408 & 0.7174 & 18.22 & 19.06 & 2 & 5226 \\
J110636-011454 & 166.6533 & -1.2485 & 0.9972 & 17.40 & 18.22 & 2 & 5057 \\
J123744+110658 & 189.4357 & 11.1161 & 0.9465 & 17.93 & 18.99 & 2 & 5243 \\
J154956+070044 & 237.4870 & 7.0125 & 0.7984 & 17.77 & 19.33 & 1 & 5007 \\
J130608+342043 & 196.5349 & 34.3453 & 0.7759 & 17.42 & 18.12 & 1 & 2092 \\
J110250+454230 & 165.7098 & 45.7086 & 0.7655 & 17.66 & 18.73 & 1 & 2156 \\
J121740+493118 & 184.4200 & 49.5217 & 0.7317 & 17.54 & 18.71 & 1 & 2180 \\
J125100+302541 & 192.7513 & 30.4283 & 0.6528 & 17.43 & 18.21 & 1 & 2092 \\
J085557+561534 & 133.9880 & 56.2597 & 0.7170 & 17.57 & 18.54 & 1 & 2276 \\
J022614+001529 & 36.5603 & 0.2583 & 0.6156 & 17.47 & 17.91 & 1 & 2052 \\
\hline
\end{tabular}
}
\begin{tablenotes}
    \item[a] {Columns contain LRG-CGM name, QSO coordinates ($\alpha$, $\delta$), redshift ($z_{\rm QSO}$), GALEX NUV and FUV apparent magnitudes ($m_{\rm NUV}$ and $m_{\rm FUV}$), number of orbits ($N_{\rm orb}$), and COS exposure time ($t_{\rm exp}$).}
    \item[b] {Two exposures were taken, with $t_{\rm exp}$ equals 11375$s$ and 2227$s$. 
Since the shorter exposure has much worse $S/N$, we use only the 11375$s$ exposure data.}
  \end{tablenotes}
\end{threeparttable}
\end{table*}

\subsection{Sample Selection} \label{subsec:sample}

%% LRGs and QSOs selection 
We now detail the precise criteria and strategy for our sample selection.
Our sample was drawn from a parent sample of LRG-QSO pairs that were selected as in \citet{Zhu14}. The parent sample of LRGs was selected from the SDSS-III/Baryon Oscillation Spectroscopic Survey (BOSS). The BOSS survey contains spectra of more than one million LRGs at average redshift $z\sim 0.5$ ($\sim 840,000$ at $0.4<z<0.6$), and includes two subsamples of galaxies: the LOWZ (LRGs at $z \sim 0.15 - 0.43$) and CMASS (“constant mass”; LRGs at $z \sim 0.43 - 0.7$) subsamples \citep{Dawson13,Alam15,Eisenstein01,Cannon06}. 
Stellar mass completeness of CMASS sample at stellar mass of $\sim 10^{11.46} M_{\sun}$ at redshifts $z\sim 0.46, 0.51, 0.56, 0.61$
is 51\%, 85\%, 82\%, 58\%, respectively \citep{Leauthaud16} . However, at the time of sample selection for observations, the BOSS survey was the largest LRGs sample known. For this reason, we chose to use the BOSS survey to select our sample of LRGs.
We selected LRGs at redshift $z > 0.4$, which placed the LRG \ion{H}{1} Lyman series transitions redward of the geocoronal \ion{H}{1} Ly$\alpha$ emission line, and enabled us to study these transitions.
QSOs were selected from SDSS DR12 \citep{Alam15} and chosen such that they have GALEX magnitudes NUV $< 19.0$ and FUV $< 20.0$, with the exception of one QSO (J105930+403956) that has $FUV = 20.4$. We decided to select J1059+4039 because it is the only QSO in SDSS with cold enriched gas (traced by \ion{Mg}{2} in the QSO SDSS spectrum) at impact parameter $R_{\perp} < 50$ kpc from a LRG and that is relatively bright in the GALEX NUV band ($NUV<19$). 
Then, we further selected all QSO - LRG pairs such that the LRG's physical impact parameter is less than 400 kpc. 

%% subsamples
From this set of $z > 0.4$ LRGs with coincident background QSOs at $R_{\perp} < 400$ kpc, we generated two subsamples for follow-up HST observations:
7 with detected \ion{Mg}{2} in the SDSS QSO spectra (to study the cool gas around LRGs) and 
8 without positive detections (to study the general properties of the gas around LRGs).  
These subsamples are: 

1) {\it{MgII-LRG sample}:}
These are pairs in which QSO SDSS spectra have detected \ion{Mg}{2} absorption with equivalent width EW(\ion{Mg}{2} 2796 \AA ) $\gtrsim 0.2$ \AA\ within 1200 km s$^{-1}$ of the LRG redshift\footnote{The justification for such a velocity window is to take into account CGM gas that is gravitationally bound to the host LRG halos. Other authors \citep[e.g.][]{Berg18} used a similarly large velocity window. We also note that scaling down our search window by $\sqrt{3}$ would not impact our results, since the maximum absolute values of our centroid velocities are 480 km/s (in this case there is also absorption at LRG redshift) and 263 km/s, as it could be later seen from Figure \ref{fig:kin}.} and do not show strong \ion{Mg}{2} at any other redshift.
%% motivation 
Thus, we are selecting QSO-LRG pairs in which there is already a confirmed presence of cold metal enriched gas at the LRG redshift. In these LRG-CGM systems, we expect to detect \ion{H}{1} lines and additional metal lines characteristic of cool CGM gas \citep[see][]{Churchill00,Rao06}.
At the time of the proposal submission, it was not known if any \ion{H}{1} or cool gas metal lines may be detected 
in a random sample of approximately 15 LRG-CGM systems. %in the LRG-CGM. 
The literature only demonstrated that approximately 5 \% (or 1 in 20) LRG-CGM systems show \ion{Mg}{2} absorption, and therefore contain cool gas. Selecting by MgII-LRGs ensures that cool LRG-CGM gas was present and we could assess its properties including the co-existence of more highly ionized gas (e.g., \ion{O}{6}).” 
We also expect that these absorption lines will be on average stronger and, hence, easier to detect and analyze than in the general population of LRG-CGM systems. 
For example, as we will see in the next sections, we detect \ion{O}{6} more commonly in the CGM of MgII-LRGs than in the CGM of Baseline LRGs (5 MgII-LRGs in comparison to only one Baseline LRG), which helped us to analyze the origin of \ion{O}{6} in the LRG-CGM. 

2) {\it{Baseline sample for general LRG-CGM analysis}:}   
These are pairs in which the QSO SDSS spectrum does not show \ion{Mg}{2} absorption at any redshift. 
Previous studies \citep[e.g.,][]{Huang16} have found that QSO-LRG pairs with \ion{Mg}{2} absorption are rare; even within 0.5 $R_{\rm vir}$, the covering fraction is only $\sim  5\%$. 
We note that \citet{Huang16} found that the \ion{Mg}{2} covering fraction at $<120$ kpc is $\sim 15$\% for passive LRGs and up to $\sim 40$\% for [\ion{O}{2}]-emitting LRGs. However, most LRGs are passive LRGs: the \citet{Huang16} sample contains 1575 [\ion{O}{2}]-emitting LRGs and 11755 passive LRGs, i.e. only 12\% of LRGs are [\ion{O}{2}]-emitting. This implies that the \ion{Mg}{2} covering fraction for a random LRG at $<120$ kpc is up to ~18\%. In addition, only 2 of our 8 Baseline LRGs are located at $\leq 120$ kpc (one of them at 120 kpc), and, as we will see later, the one with the impact parameter $<120$ kpc has \ion{H}{1} detected ($EW=0.57\AA$), as MgII-LRGs. For these reasons, we consider our Baseline LRGs are 
 consistent with the general LRGs population, especially given the small sample size. 

We restrict both subsamples to QSOs whose spectrum does not contain strong \ion{Mg}{2} absorption at any other redshift with a goal to mitigate strong \ion{Mg}{2} absorbers attenuating the FUV flux if they are also Lyman-limit systems. 

%% MgII 
We identify background QSOs that have \ion{Mg}{2} absorption in their SDSS spectra by applying the algorithm from \citet{ZhuM13} to SDSS DR12 data.  \citet{ZhuM13} used technique of nonnegative matrix factorization to find continua, identified \ion{Mg}{2} 2796, 2803 doublet absorption lines in the QSO spectra, and measured their equivalent widths.  

\vspace{0.4cm}

%% final sample 
With the goal of constructing a survey of LRG-CGM systems from these two subsamples, we selected targets as follows: 

1. The brightest QSOs in the GALEX NUV and FUV bands (typical GALEX NUV magnitudes are $\sim 17.5 - 19$). This minimized the HST orbits for obtaining the COS spectra.

2. CGM systems uniformly distributed in $\log_{10} R_{\perp}$.

3. The full sample should contain approximately half of each subsample with a goal of at least 15 sightlines. 

\begin{figure*}[t!]
\centering
\includegraphics[width=0.49\textwidth]{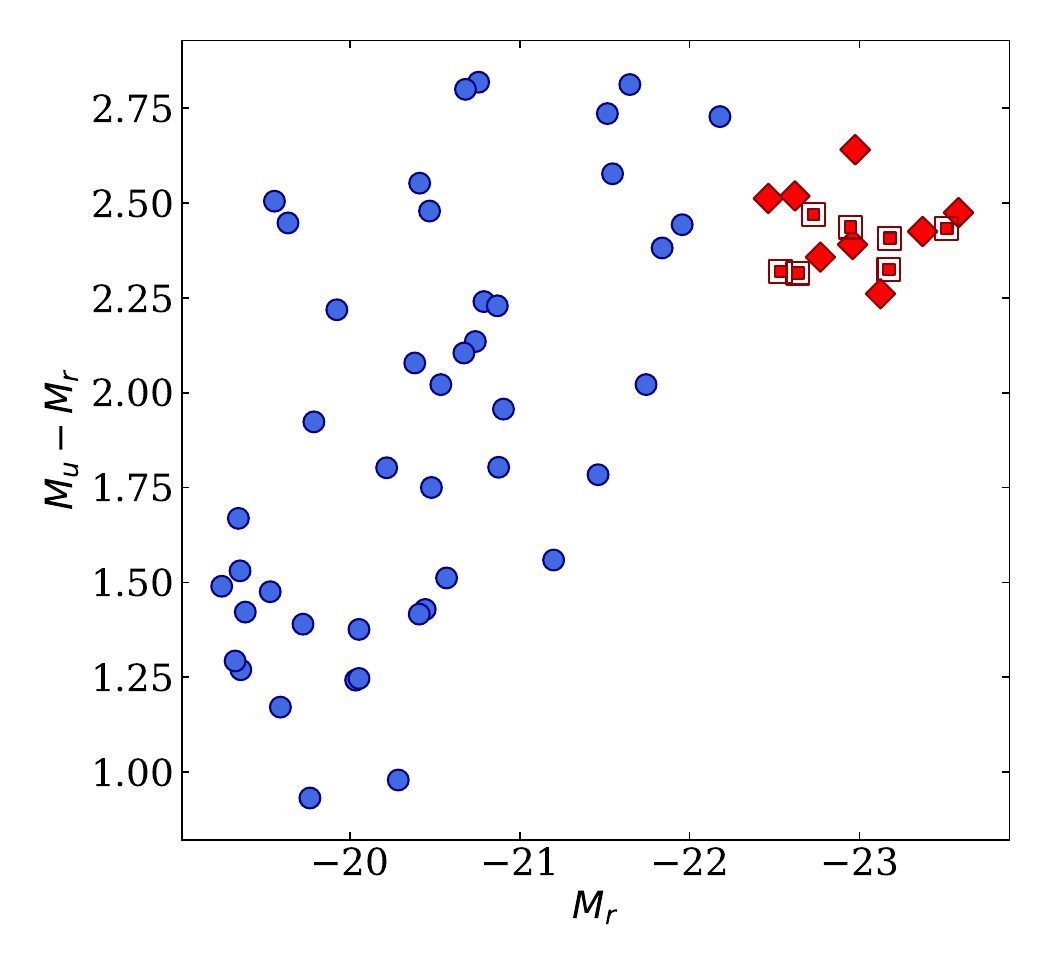}
\includegraphics[width=0.49\textwidth]{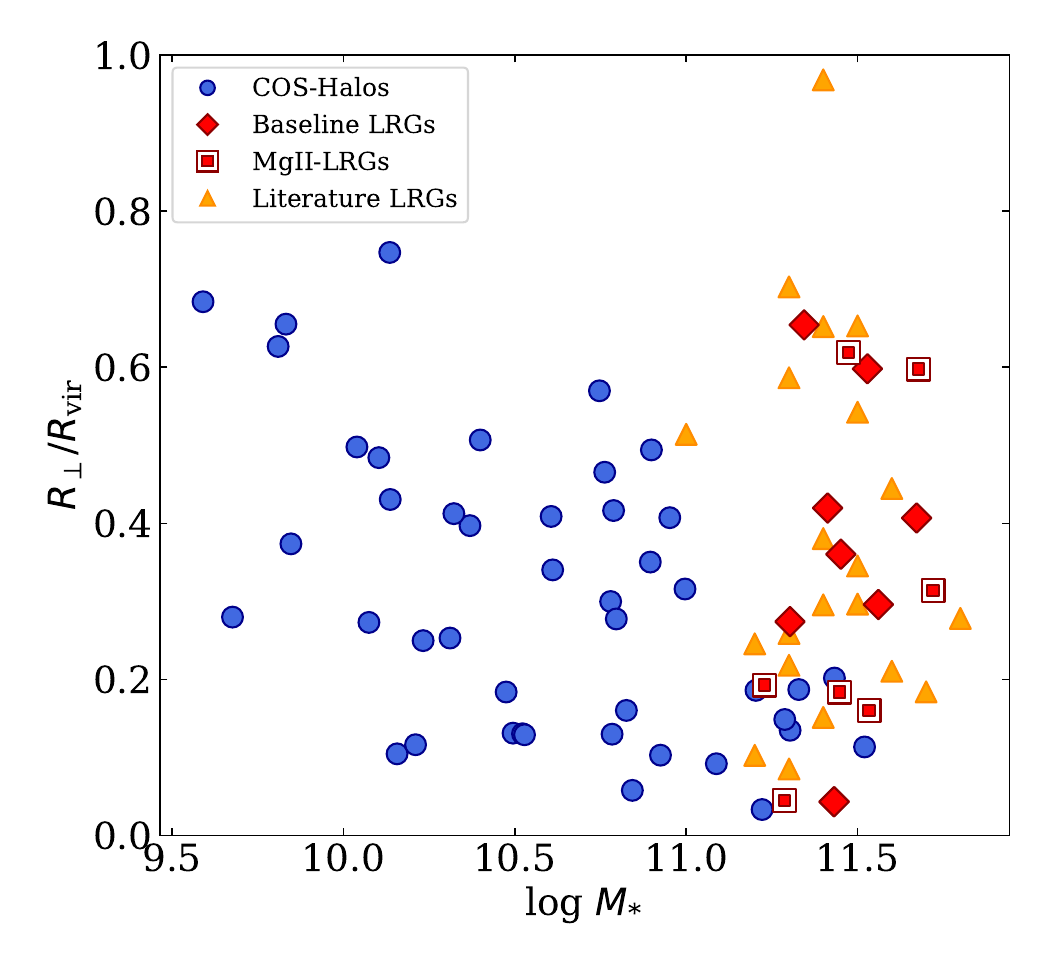}
\caption{\small 
Our LRG sample compared with the COS-Halos sample. {\it Left panel:} 
Color $u - r$ versus absolute $r$-magnitude, for our LRGs with (red squares) and without (red diamonds) \ion{Mg}{2}. 
COS-Halos are shown as blue circles. 
{\it Right panel:} 
Impact parameter scaled to the virial radius versus stellar mass for our LRGs with (red squares) and without (red diamonds) \ion{Mg}{2}. For reference, we also show the COS-Halos galaxies (blue circles) and the LRG samples from \citet[][]{Chen18} and \citet[][]{Berg18} (orange triangles). 
}
\label{fig:samplelrg} 
\end{figure*}

\begin{figure*}[t!]
\centering
\includegraphics[width=0.49\textwidth]{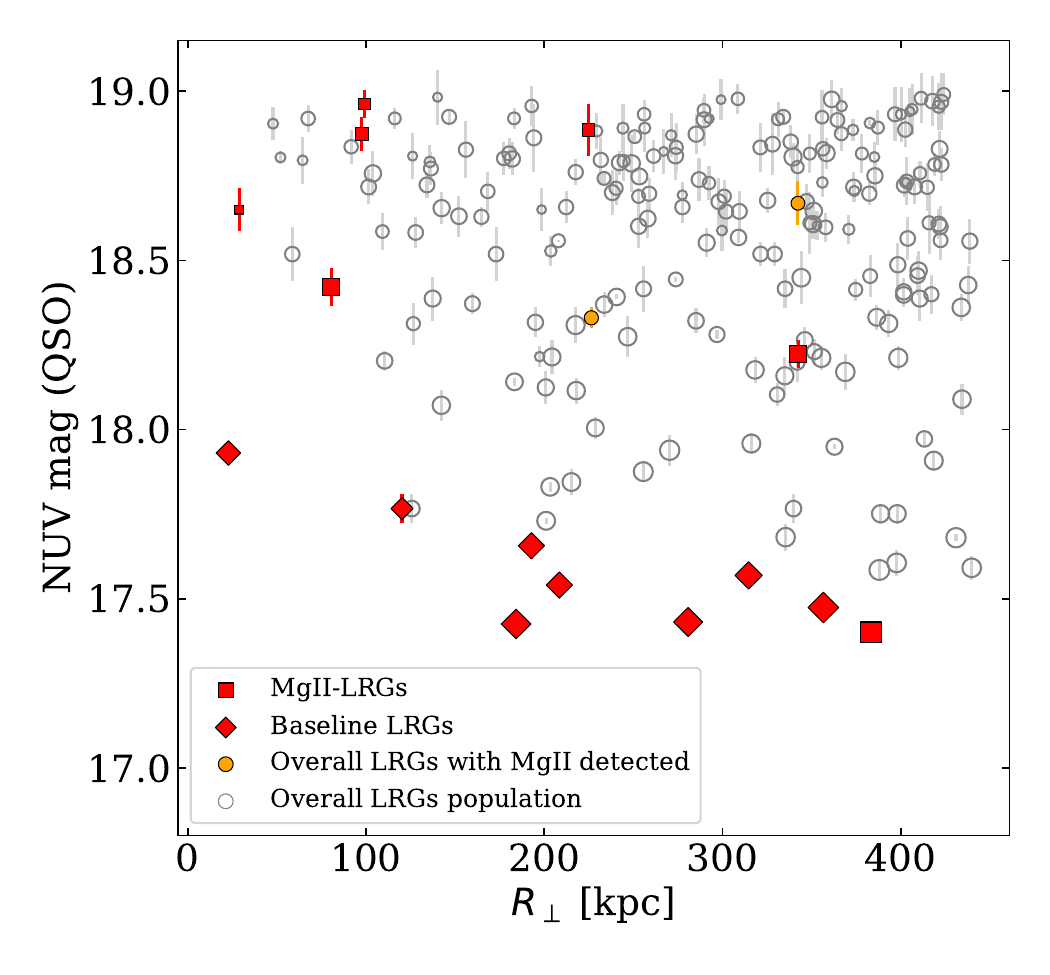}
\includegraphics[width=0.49\textwidth]{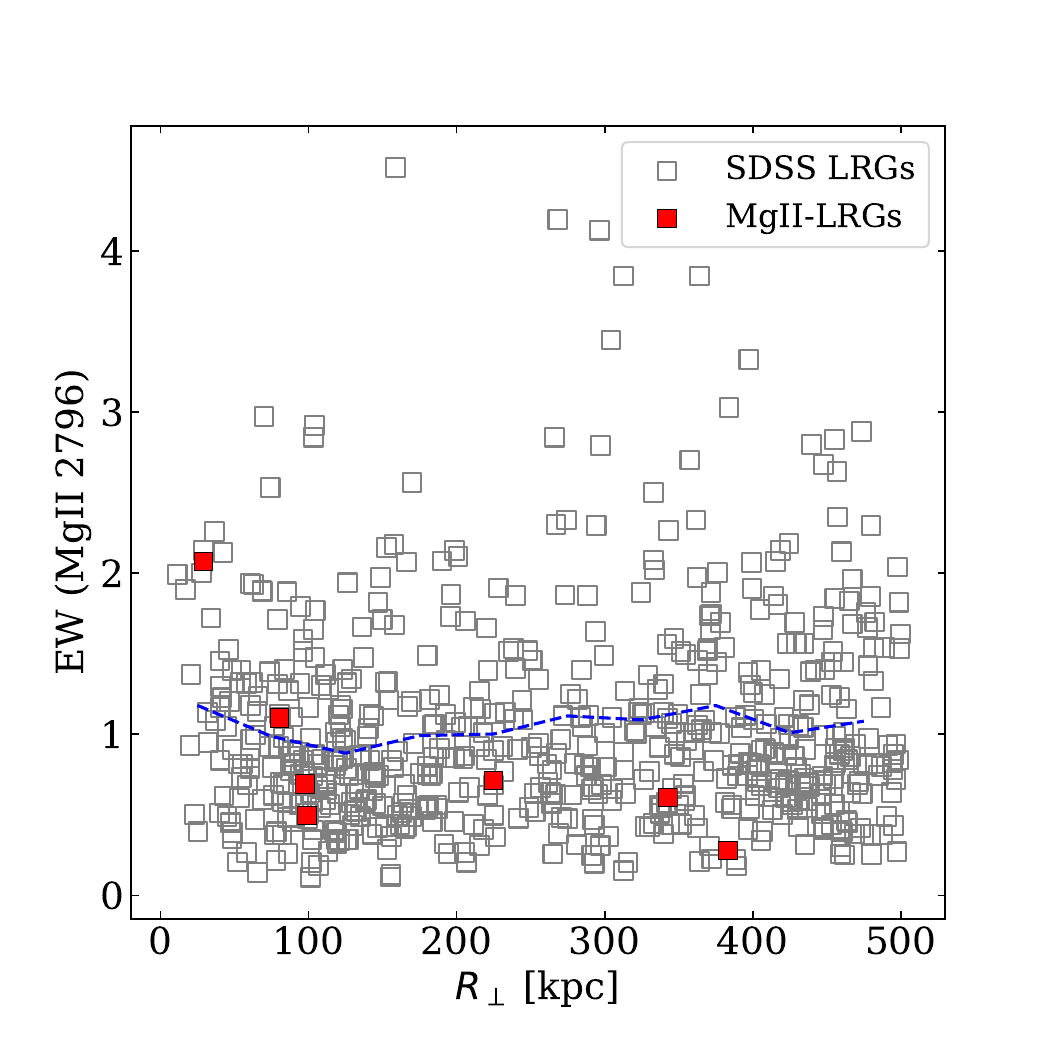}
\caption{\small 
Our LRG sample compared against subsets of the overall LRG population. 
{\it Left panel:}
QSO NUV magnitude versus impact parameter for QSO-LRG pairs with bright (in GALEX NUV and FUV bands) background QSOs. 
Marked are our QSO-LRGs with (red squares) and without \ion{Mg}{2} (red diamonds), and QSO-LRGs pairs with (orange circles) and without \ion{Mg}{2} (open gray circles) in the overall LRGs population. Larger symbols correspond to QSOs brighter in FUV. 
{\it Right panel:} 
EW(\ion{Mg}{2}) versus impact parameter, for the SDSS data. The measurements are provided by \citet{Huang16}. Marked are our LRGs with \ion{Mg}{2} (red squares). The blue dashed line shows the average EW(\ion{Mg}{2}), when it is detected. 
}
\label{fig:samplelrg3}
\end{figure*}

Figure \ref{fig:samplelrg} shows our 15 LRGs in a color-magnitude diagram and the distribution of LRG-CGM systems in impact parameters and stellar masses. For comparison, we also show the same for COS-Halos ($z \sim 0.2$, $\sim L^{*}$ galaxies).
Figure~\ref{fig:samplelrg3} compares our sample of 15 LRG-QSO pairs to the overall LRG population. The Figure also shows EW(\ion{Mg}{2}) for our LRGs, in comparison to the full SDSS sample of QSO-LRGs with \ion{Mg}{2}. One notes that QSOs of the Baseline sample are on average much brighter in the GALEX NUV band. 

\vspace{0.4cm}

% LRGs properties
Tables \ref{tab:lrgtab} and \ref{tab:qsotab} list properties of LRGs and QSOs in our sample. We adopted stellar masses from the Wisconsin group \citep{Chen12}\footnote{\url{http://www.sdss.org/dr16/spectro/galaxy_wisconsin/}}, from the configuration based on \citet{MarastonS11} stellar population synthesis models applied on BOSS DR12 data. 
%%%
Halo masses are estimated from a parametrized relation between stellar and halo masses from \citet{Rodriguez17}. The same relation was used by \citet{Berg18}. 
For one of our LRGs (LRG\_1059+4039), there is another nearby LRG (see Section \ref{sec:ng}) with somewhat higher stellar mass. We use this stellar mass as the input stellar mass when calculating the halo mass of that LRG (LRG\_1059+4039). 
The mean halo mass for our sample is $\log M_{\rm halo} / M_{\sun} = 13.51$, comparable to the mean LRGs halo mass calculated from clustering for $z\sim 0.7$ LRGs, $\log M_{\rm halo} / M_{\sun} = 13.52 $ \citep{Zhai17}.

\subsection{Observations, Data Reduction, and Analysis}\label{sec:dataredan} 

%% observations 
We obtained Hubble Space Telescope (HST) Cosmic Origins Spectrograph (COS) spectra of the 15 background QSOs, in program GO-14171 (PI: Zhu) in Cycle 23. 
We used the G140L grating with a central wavelength of 1280 \AA . 
The COS spectrograph contains two detector segments - segment A and segment B. The central wavelength (1280 \AA ) was chosen such that the atmospheric Ly$\alpha $ line falls in the gap between the two segments. In the rest frame of a source at $z \sim 0.5$, segment A spans $\sim 850 - 1400$ \AA , and segment B spans $\sim 630 - 790$ \AA . For the wavelengths of interest, the resolving power is $R \sim 2000 - 3000$ (spectral resolution $\sim 150$ km s$^{-1}$). The exposure time was estimated from the HST online exposure time calculator 
% (\url{http://etc.stsci.edu/etc}) 
to give signal - to - noise ratio $S/N \sim 10$ at Ly$\beta$ in the observed frame for our LRGs. The $S/N$ is typically higher at shorter wavelengths. 

\vspace{0.4cm}

% data reduction 

% 0) General information
We reduced the data by using CALCOS v2.21 and our own custom Python codes\footnote{\url{https://github.com/pypit/COS_REDUX}, or in the future \url{https://github.com/pypeit/COS_REDUX}}. 
Based on \citet{Worseck16} and our private communication with G{\' a}bor Worseck, our custom Python code combined with CALCOS performs a customized data reduction, which includes customized dark subtraction and co-addition of sub-exposures 
(for more details, see Appendix \ref{sec:app1}). 

\vspace{10pt}

%% analysis - continuum, line identification, and EWs

%% continuum
For each reduced QSO spectrum, we estimated the QSO continuum level using the {\it lt\_continuumfit} GUI from the {\sc linetools} package\footnote{\url{https://github.com/linetools/linetools}}.
We selected points on the continuum level by eye through which a cubic spline was then interpolated. 

%% line identification
We identified absorption lines using the {\it igmguesses} GUI from the {\sc PYIGM} package\footnote{\url{https://github.com/pyigm/pyigm}}.  
We first identified strong absorption lines associated with Milky Way (redshift $z \sim 0$) and with the background QSO (adopting its redshift from SDSS). 
Then, we searched for the Ly$\alpha$ line in a spectral window of [-1200, 1200] km s$^{-1}$ around the LRG redshift\footnote{1200 km s$^{-1}$ corresponds to the approximate escape velocity for a $\sim 10^{13.5} M_{\sun}$ halo at $\sim 30 - 400$ kpc; for the most massive halos, the escape velocity is up to
$\sim 1700$ km s$^{-1}$}. 
If we find a potential Ly$\alpha$ line, we fit its profile with a Gaussian, and predict Gaussian line profiles for the other \ion{H}{1} Lyman transitions given their relative intrinsic strengths. If we obtain consistent line profiles for these transitions, we consider the \ion{H}{1} detection reliable, and similarly searched for strong metal lines near the redshift of the \ion{H}{1} lines. We also search for absorption lines associated with absorbers at other redshifts. We find the strongest absorption lines in a spectrum, for each of these lines assume that it represents \ion{H}{1} Ly$\alpha$ or some of the other strong transitions at some redshift, and repeat the same steps as we did for the lines associated with the LRG.

%% line identification: boundaries
The line boundaries were defined by-eye and selected as the velocity range over which the optical depth is significant. We excluded parts of lines that are blended with other interloping transitions, or we mark the line as blended. 

%% line identification: aodm
For ions with only one transition available, we compare the scaled apparent column density profiles with \ion{H}{1} or other available transitions. 
For ions with at least two transitions available, where only one transition is significantly detected (its EW is higher than $3\sigma_{\rm EW}$), we compare the apparent column density profiles of these available transitions (e.g., \ion{O}{6} 1031 and \ion{O}{6} 1037 in some LRG-CGM systems). 
If the apparent column density profiles are closely aligned in velocity space, we consider this transition reliable. Otherwise, we mark the transition as an upper limit as it is likely blended with another interloping line.  
Further details are given in Appendix \ref{sec:appb}. 

%% EW measurements 
For each absorption line, we calculated its rest-frame equivalent width (EW) by applying boxcar integration across the velocity range of the line. The line detection is considered reliable if its EW is greater than three times the uncertainty in the EW, $\sigma$(EW). If the EW is below $3 \sigma$(EW), we treat the measurement as an upper limit, with value $<$ $2\sigma$(EW). If the $\rm EW > 3\sigma$(EW), but the line is blended, we also treat the line as an upper limit, but with value $<$ EW$ + \sigma$(EW). Absorption lines that overlap with weak lines as determined from visible inspection (e.g. high order Lyman series lines) only are not considered as blended.

The COS grating G140L has a spectral resolution of $\sim 150$ km s$^{-1}$.
When measuring upper limits, we typically used velocity windows of $\sim 250$ km s$^{-1}$ or more. Velocity limits are marked on 
our Figure \ref{fig:stacksmgii1} and on Figures \ref{fig:stacksmgii2} - \ref{fig:stacksno} in Appendix \ref{sec:appd}. % added for arxiv
%Figure Set \ref{fig:stacksmgii1}. % deleted for arxiv
It is known that HST COS line spread function exhibits broad wings; however, we do not expect that this has a significant impact on our results \citep[see][]{Ghavamian09cos}.

\vspace{10pt}
 
\section{\ion{H}{1} and metal line measurements}\label{sec:lines} 

\begin{figure*}[t!]
\centering
\includegraphics[width=0.55\textwidth]{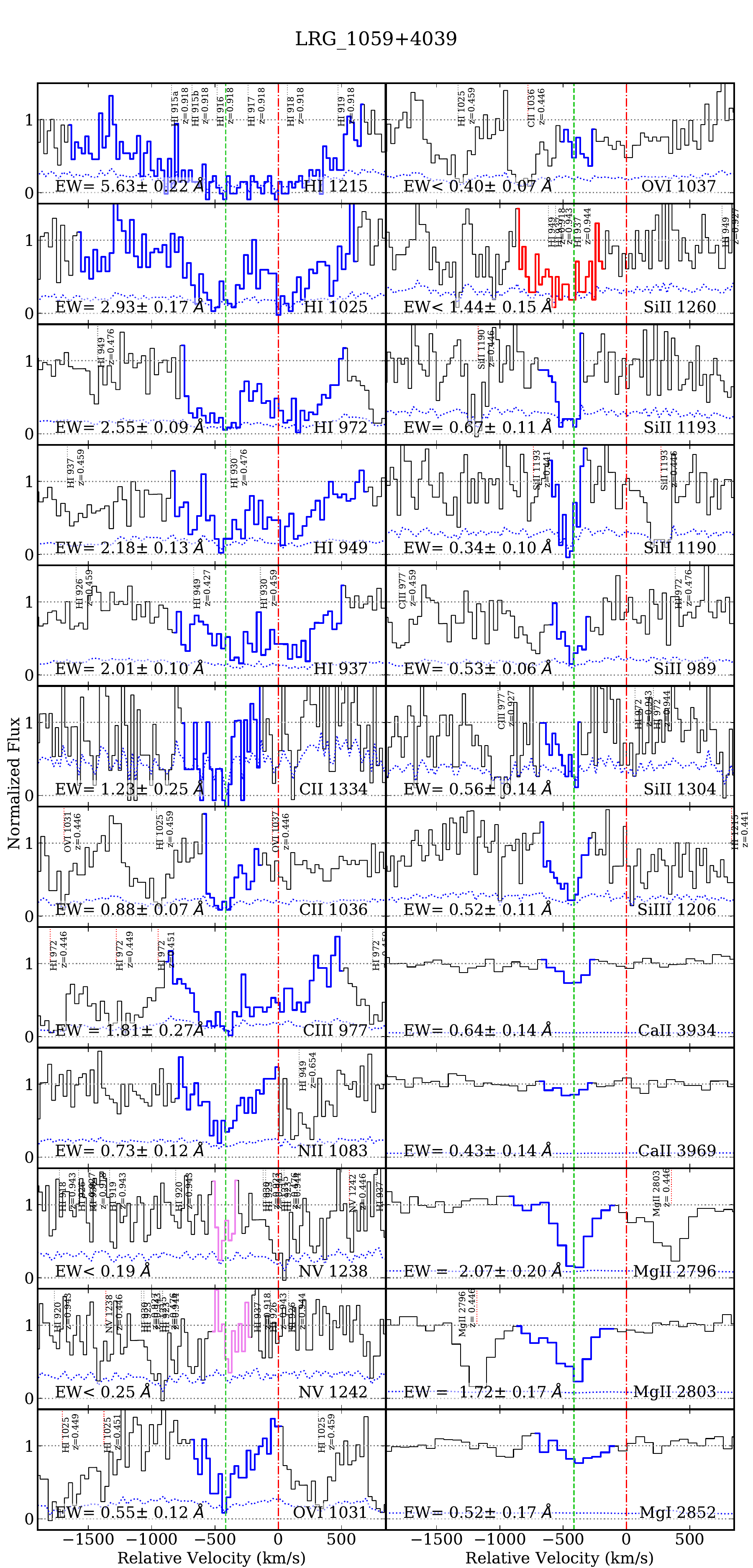}
\caption{\small 
Line profiles, shown as continuum normalized flux vs. velocity in the LRG rest frame, for one of our MgII-LRGs, LRG\_1059+4039. 
The same figures, but for other MgII-LRGs and for Baseline LRGs (15 total) are available in 
% a figure set in the online journal (ApJ; Figure Set 3) % added/modified/commented for arxiv
% and in 
Appendix \ref{sec:appd}. % added for arxiv  
Transitions are labeled in the lower right corner, and rest-frame EWs are shown in the lower left corner. The analyzed velocity ranges are marked in blue color for reliable transitions, in red color for likely blended detections, and in magenta color for $< 3\sigma$ significant transitions. 
Red dot-dashed and green dashed vertical lines denote LRG redshift and velocity offset of LRG-CGM systems (for MgII-LRGs determined from the central velocity of \ion{Mg}{2} absorption), respectively. The blue dotted line shows the flux uncertainty. Interloping transitions are marked with red and gray vertical dotted lines, if they originate from the CGM around the LRG, or from absorption systems found at other redshifts, respectively. We consider that transitions within 1200 km s$^{-1}$ of the LRG redshift are associated with the CGM around the LRG. 
Name and redshift of these interloping transitions are labeled alongside of the vertical lines. 
}
\label{fig:stacksmgii1}
\end{figure*}

%% figure with stack plots
In Figure \ref{fig:stacksmgii1} and similar figures in 
%a figure set, % commenting for arxiv 
Appendix \ref{sec:appd},
we show velocity profiles for line transitions with reliable, blended, and possible detections in the CGM around our LRGs. The transitions shown include detections from the COS and SDSS spectra. 
The Figure also shows that the velocity corresponding to the largest optical depth of \ion{Mg}{2} from the SDSS QSO spectra coincides with the velocity corresponding to the largest optical depth of \ion{H}{1} and metal line transitions from our COS spectra. The absorption line measurements (such as EWs) are listed in Table \ref{tab:lines}. 

%% results for our LRGs 
We found that all MgII-LRGs have detected \ion{H}{1}, \ion{C}{3} 977, and one or more other metal lines, such as \ion{C}{2} 1334, \ion{Si}{3} 1206, \ion{Si}{2} 1260, \ion{O}{6} 1031, \ion{N}{2} 1083. 
In the Baseline LRG spectra we detect reliable \ion{H}{1} 1215 absorption in four of eight cases, and we also detect metal lines including \ion{C}{3} 977, \ion{O}{6} 1031, and \ion{Si}{3} 1206 for three of these LRGs. 
In a few cases, besides \ion{Mg}{2}, we also detect other lines in the SDSS spectra, such as \ion{Mg}{1} and \ion{Ca}{2}.
We will see in Section \ref{sec:nhi} that four of the seven MgII-LRGs show significant Lyman-limit opacity, while none of the Baseline LRGs has significant Lyman-limit absorption.

\vspace{0.4 cm}

%% measured quantities
For each LRG, we measure the following: 

\begin{itemize}

\item Rest-frame equivalent widths (EW) for all lines (see section \ref{sec:dataredan} for details).  

\item Column density $N_{HI}$ from the Lyman-limit (LL) decrement (see section \ref{sec:nhi}).  

\item Line centroids, calculated for one of the least strong but reliably detected \ion{H}{1} or metal line transitions, and defined as velocity weighted by absorbed normalized flux:
\begin{equation}
dv = \Big[\sum (1 - F(i)) v(i)\Big] / \Big[\sum (1 - F(i)))\Big]\, ,
\end{equation}
where $F$ is continuum-normalized flux, and the sum is calculated over the selected velocity range of the line. 

\item Velocity extent $\Delta v_{90}$ for the \ion{H}{1} 1215 line. The $\Delta v_{90}$ was initially defined in \citep{PW97} as the velocity interval that contains 90\% of the total (apparent) optical depth. 
Since the resolution of our data not high enough to precisely measure the optical depth, in this work, we define $\Delta v_{90}$ as the velocity interval that contains 90\% of the optical depth that we measured for the \ion{H}{1} 1215 line. The $\Delta v_{90}$ approximately measures the velocity extent. 
Here, pixels with flux below the noise level are replaced with the noise. The apparent optical depth in a given pixel is defined as $\tau_{a} = \ln (1/F_{i})$, where $F_{i}$ is the continuum-normalized flux. 

\item Apparent column density ($N_a$) profiles, calculated as
\begin{equation}
N_a (v) = \frac{\tau_{a}(v) dv}{2.654\times 10^{-15} f \lambda}\, ,
\end{equation}
where $\tau_{a}$ is apparent optical depth, $\lambda$ is rest-frame wavelength in angstroms, $f$ is oscillator strength, and $v$ is velocity, as in \citet{SavageSem91}. We note that the resolution of our data is not high enough to calculate column densities by integrating $N_a$ because the lines are not resolved. 

\end{itemize}

\begin{table*}
\begin{rotatetable*}
\center
\caption{Rest-frame equivalent width measurements in our sample of LRG-CGM systems$^{a}$
\label{tab:lines}}
\begin{threeparttable}
{
\renewcommand\TPTminimum{\linewidth}
\makebox[\linewidth]
{
\hspace{-1.1in} 
\centering 
\begin{tabular}{ccccccccccc}
\toprule 
\hline
Name & $z$ & \ion{H}{1} 1215 & \ion{H}{1} 1025 & \ion{C}{3} 977 & \ion{C}{2} 1036 & \ion{Si}{3} 1206 & \ion{Si}{2} 1260 &\ion{O}{6} 1031 & \ion{Mg}{2} 2796 & $dv$
\\   
 & & (\AA )    & (\AA )  & (\AA ) & (\AA ) & (\AA ) & (\AA )  & (\AA ) & (\AA ) & (km s$^{-1}$)
\\
\hline
LRG\_1059+4039  & 0.4489 &  $ 5.63 \pm  0.22$ & $ 2.93 \pm  0.17$ & $ 1.81 \pm  0.27$ & $ 0.88 \pm  0.07$ & $ 0.52 \pm  0.11$ & $ < 1.6$ & $ 0.55 \pm  0.12$ & $ 2.07 \pm  0.20$ & $-480$ \\ 
LRG\_1121+4345  & 0.4216 & $ 1.72 \pm  0.12$ & $ 0.84 \pm  0.07$ & $ 0.57 \pm  0.05$ & $ 0.43 \pm  0.05$ & ... & $ 0.46 \pm  0.07$ & $ 0.39 \pm  0.05$ & $ 1.10 \pm  0.16$ & $80$ \\ 
LRG\_1144+0714  & 0.4892 & $ 2.06 \pm  0.11$ & $ 1.45 \pm  0.05$ & $ 0.96 \pm  0.04$ & $ 0.35 \pm  0.04$ & $ 0.50 \pm  0.06$ & $ 0.35 \pm  0.07$ & $ 0.22 \pm  0.04$ & $ 0.69 \pm  0.16$ & $-165$ \\ 
LRG\_1351+2753  & 0.4320 & $ 1.45 \pm  0.11$ & $ 1.01 \pm  0.06$ & $ 0.70 \pm  0.06$ & $ 0.27 \pm  0.05$ & $ 0.34 \pm  0.09$ & $ < 0.23$ & $ 0.26 \pm  0.05$ & $ 0.50 \pm  0.16$ & $-180$ \\ 
LRG\_1520+2534  & 0.5385 & $ 7.41 \pm  0.22$ & $ 2.49 \pm  0.18$ & $ 0.57 \pm  0.07$ & $ < 0.66$ & $ 0.78 \pm  0.16$ & $ < 0.29$ & $ < 0.19$ & $ 0.71 \pm  0.14$ & $-75$ \\ 
LRG\_1440-0157  & 0.4801 & $ 1.70 \pm  0.15$ & $ 1.14 \pm  0.08$ & $ 1.23 \pm  0.06$ & $ < 0.19$ & $ 0.41 \pm  0.10$ & $ < 0.28$ & $ 0.39 \pm  0.06$ & $ 0.61 \pm  0.10$ & $263$ \\ 
LRG\_1106-0115  & 0.6106 & $ 0.55 \pm  0.13$ & $ 0.49 \pm  0.05$ & $ 0.46 \pm  0.04$ & $ 0.14 \pm  0.05$ & $ 0.44 \pm  0.09$ & $ < 0.52$ & $ < 0.12$ & $ 0.28 \pm  0.06$ & $-196$ \\ 
LRG\_1237+1106  & 0.4731 & $ 0.57 \pm  0.12$ & $ < 0.11$ & $ < 0.15$ & $ < 0.13$ & $ < 0.28$ & $ < 0.35$ & $ < 0.16$ &  $ < 0.18 $ &  $22$ \\ 
LRG\_1549+0701  & 0.4997 & $ < 0.26$ & $ < 0.15$ & $ < 0.1$ & $ < 0.21$ & $ < 0.18$ & $ < 0.22$ & $ < 0.38$ &  $< 0.19$ & ... \\ 
LRG\_1306+3421  & 0.4690 & $ < 0.19$ & $ < 0.17$ & $ < 0.13$ & $ < 0.54$ & $ < 0.23$ & $ < 0.56$ & $ < 0.15$ &  $< 0.13$ & ... \\ 
LRG\_1102+4543  & 0.4875 & $ 2.52 \pm  0.19$ & $ 1.09 \pm  0.18$ & $ 0.23 \pm  0.07$ & $ < 0.28$ & $ 0.47 \pm  0.15$ & $ < 0.46$ & $ < 0.18$ &  $< 0.19$ & $-167$ \\ 
LRG\_1217+4931  & 0.5230 & $ 1.45 \pm  0.15$ & $ 0.56 \pm  0.11$ & $ 0.37 \pm  0.08$ & $ < 0.2$ & $ < 0.26$ &  ...  & $ < 0.59$ &  $< 0.14$ &  $-254$ \\ 
LRG\_1251+3025  & 0.5132 & $ < 0.22$ & $ < 0.12$ & $ < 0.11$ &  ...  & $ < 0.22$ & $ < 0.29$ & $ < 0.11$ &  $< 0.10$ & ... \\ 
LRG\_0855+5615  & 0.4399 & $ 1.32 \pm  0.09$ & $ 0.71 \pm  0.05$ & $ <0.6$ & $ < 0.31$ & $ < 0.15$ & $ < 0.19$ & $ 0.42 \pm  0.05$ &  $< 0.11$ & $-196$ \\ 
LRG\_0226+0014  & 0.4731 & $ < 0.38$ & $ < 0.1$ & $ < 0.09$ & $ < 0.11$ & $ < 0.12$ & $ < 0.18$ & $ < 0.11$ &  $< 0.07$ & ... \\ 
\hline
\end{tabular}
}
\begin{tablenotes}
    \item[a] {Table contains information about a few \ion{H}{1} and metal line transitions,  
    LRGs redshifts, and velocity offsets. 
In a few cases when 

additional line components at higher or lower velocities are detected (see section \ref{sec:kin} and appendix \ref{sec:appe}), the listed measurements

correspond only to the primary line component.
Velocity offsets are calculated later, in section \ref{sec:kin}.
}
  \end{tablenotes}
}
\end{threeparttable}
\end{rotatetable*} 
\end{table*}

\vspace{0.4 cm}

\section{Other surveys}\label{sec:oth}

Here we summarize the main surveys with which we compare our results. 
We combine some of our data with two other UV studies of the CGM around LRGs, i.e., \citet{Chen18} and \citet{Berg18}. We also compare our data with a 
study of the CGM of local $\sim L^{*}$ galaxies and with CGM studies of galaxies at lower and higher redshift, that could be evolutionary connected to LRGs.   

\begin{itemize}

\item \citet[][C-LRGs]{Chen18}: 
The authors studied the CGM around 16 massive red galaxies at $z\sim 0.5$ by using HST COS observations. Nine of these galaxies are classified as LRGs \citep[see, e.g., list of LRGs used in][i.e. \url{http://www.sdss.org/dr16/spectro/galaxy_wisconsin/}]{Chen12}, five are red COS-Halos galaxies \citep{Tumlinson13}, two are additional massive red galaxies.   

We chose to combine our sample only with the subset of nine \citet{Chen18} galaxies that are classified as LRGs. 
The galaxies that we did not include in our analysis are galaxies that were not BOSS targets (SDSSJ\,080357.74$+$ 433309.9, SDSSJ\,095000.86$+$ 483102.2, SDSSJ\,155047.70$+$ 400122.6, SDSSJ\,091027.70$+$ 101357.2) and galaxies that are not in LOWZ, CMASS and near QSO subsamples (SDSSJ\,140625.97$+$ 250923.2, SDSSJ\,092554.18$+$ 400353.4, SDSSJ\,024651.20$-$ 005914.1).
In this way, we combine our LRGs with other LRGs only. 
Previous studies showed that LRGs typically stopped forming stars more than $\sim 1-2$ Gyr ago. However, we do not know when the seven massive red galaxies that we did not include in our analysis stopped forming stars, or if some of these galaxies were forming stars within the last 1-2 Gyr. To answer such question, one could model star formation histories of these galaxies or make estimate of the 4000 \AA\, decrement and $H\delta_{A}$ index from the spectra of these galaxies \citep[e.g.][]{Bruzual83,Kauffmann03}. 
We leave a more detailed analysis for the future work. 

We further note that the \citet{Chen18} LRGs by selection have smaller maximum impact parameters than our sample ($\sim 160$ kpc compared with $\sim 400$ kpc).
The average logarithm of the stellar mass (in units of $M_{\sun}$) of the \citet{Chen18} subsample of 9 LRGs is $\sim 11.36$, similar to our sample ($\sim 11.47$).
The redshift range of the \citet{Chen18} subsample of 9 LRGs is $\sim$ 0.3 - 0.6, which is a bit lower than for our sample ($\sim$ 0.4 - 0.6).

\item  \citet[][B-LRGs]{Berg18}: The authors used archival HST data to study the CGM around LRGs at $z\sim 0.5$. Their sample contains 21 galaxies with similar impact parameters. Their average stellar mass is similar as in our sample, $\log M_{*} /M_{\sun} \sim 11.4 $. The redshift range for this sample is $\sim$ 0.3 - 0.6. 
Five of B-LRG sightlines are included in the C-LRGs. 

We downloaded the B-LRGs data, and measured the EW(\ion{H}{1} 1215) and EW(\ion{O}{6} 1031) for the LRGs that were not a part of the C-LRGs sample, using the same velocity windows as in \citet{Berg18}. When multiple data files were available, we calculated the average EW value for all of the available data, excluding low signal-to-noise data. 
However, most of these LRG-CGM systems do not have available or statistically significant \ion{H}{1} 1215 line detected.
Only one of these LRG-CGM systems has statistically significant \ion{H}{1} 1215 line detected, and in only one another case the upper limit in the EW(\ion{H}{1} 1215) is smaller than 0.3 \AA . 
In most of our analysis, we do not use B-LRGs with upper limits exceeding 0.3 \AA .
For the B-LRGs subsample that does not overlap with C-LRGs, the \ion{O}{6} 1031 line is available for 11 B-LRGs, and we measure only upper limits in all of these cases.   

Hereafter, we will simply refer to the the combined sample of the C-LRGs and B-LRGs as `literature LRGs'. 

\item  COS-Halos \citep{Tumlinson13}: The authors studied the CGM of $\sim$ L$^{*}$ galaxies at $z \sim 0.2$. The impact parameters are, by selection, $\lesssim 160$ kpc; however the virial radii are also smaller on average than for the LRGs halos ($\sim 300$ kpc, in comparison to $\sim 600$ kpc). 
COS-Halos galaxies were selected to be in relatively isolated environments, given the data available. No such selection was imposed for our LRGs. 

\item  Quasars probing quasars \citep[QPQ; ][]{X13}:  The authors studied the CGM of QSO host galaxies at redshift $z \sim 2-3$. Impact parameters are up to 1 Mpc.    
The higher-redshift QSO sample places many of the lines covered by our COS data at wavelengths observable from the ground. However, blending from the Ly$\alpha$ forest affects the blue lines significantly.
QSO host galaxies are massive galaxies at $z\sim 2-3$, with halo masses of $\sim 10^{12.5} M_{\sun}$, and previous studies predict that a significant fraction of QSO hosts evolve into LRGs \citep{White12}. 

\item  Sub-millimeter galaxies \citep{Fu16}: This study includes three sub-millimeter galaxies (SMGs) at $z \sim 2 - 2.5$ with impact parameters $\sim 100 - 200$ kpc. Besides QSO host galaxies, SMGs also have halo masses of $10^{12} - 10^{13} M_{\sun}$ \citep[see references in ][]{Fu16}, and some fraction of them may also evolve into LRGs.

\end{itemize}

\vspace{10pt}

\section{Results} \label{sec:res}

\subsection{Covering Fractions of \ion{H}{1} and Metals around LRGs}\label{sec:cf1}

%% CF
In this section, we estimate the covering fraction of \ion{H}{1} 1215 absorbers with EW $>0.3$ \AA\ for our LRGs combined with the literature LRGs. 
Since only $\sim 5$ \% of LRGs contain \ion{Mg}{2} in their CGM \citep[e.g.,][]{Huang16} , we will consider that our subsample without \ion{Mg}{2} (Baseline LRGs) is consistent with the general LRGs population. Taking this into account, all Baseline LRGs covering fractions ($C_{F}$) might be higher by $\lesssim 0.05$ (this still holds if we take into account that the \ion{Mg}{2} covering fraction is higher at $<120$ kpc). 
We choose a minimum EW of $>0.3$ \AA , based on our \ion{H}{1} 1215 detection threshold. We repeat a similar calculation for metal lines (using different thresholds). When calculating $C_{F}$, we do not include upper limits that exceed the adopted EW threshold. 
For \ion{H}{1} 1215, we include seven of eight of our Baseline LRGs, all nine C-LRGs, and two B-LRGs. 

%% CF calculation 
To estimate errors in the $C_{F}$, we calculate the Wilson score interval with a significance level 0.3173, corresponding to a $\sim 1 \sigma$ confidence interval.

%% HI CF
In our Baseline LRGs subsample, 4 of 8 LRGs have EW(\ion{H}{1} 1215) $> 0.3$ \AA , and one has upper limit higher than $0.3$ \AA\ (see Table \ref{tab:lines}). The derived $C_{F}$ is $0.57^{+0.17}_{-0.18}$.
For the C-LRGs we derive $C_{F}$ of $0.67^{+0.13}_{-0.17}$, which is consistent with our sample. When we combine our sample with C-LRGs and B-LRGs, we obtain a \ion{H}{1} 1215 covering fraction of $C_{F} \sim 0.56^{+0.11}_{-0.12}$ (EW $ > 0.3$ \AA ), implying that it is common to detect \ion{H}{1} in LRGs' CGM. 

% metal lines CF
We detect \ion{H}{1} in four Baseline LRG-CGM systems. Around three of these galaxies we detect metal lines. 
In most LRG-CGM systems, the strongest UV metal line that we observe is \ion{C}{3} 977.
For our Baseline sample combined with C-LRGs, the $C_{F}$ for the detection of EW(\ion{C}{3} 977) $> 0.2$ \AA\ is 0.38$^{+0.13}_{-0.11}$, which is a bit smaller than the $C_{F}$ for EW(\ion{H}{1} 1215) $> 0.3$ \AA . 
We do not find LRGs with strong \ion{H}{1} and no metal lines detected. The only LRG with \ion{H}{1} and no metal lines detected shows one of the weakest \ion{H}{1} that we detect, with only Ly$\alpha$ line detected above 3$\sigma$.

% MgII-LRGs
For the MgII-LRGs, we detect \ion{H}{1}1215 and \ion{C}{3} 977 in all cases, and obtain $C_{F}$ for EW(\ion{C}{3} 977) $> 0.3$ \AA\ of 1$^{+0}_{-0.13}$.

\subsection{EW(\ion{H}{1}) Dependence with Impact Parameter}\label{sec:hi}

\begin{figure}[h]
\centering
\includegraphics[width=0.49\textwidth]{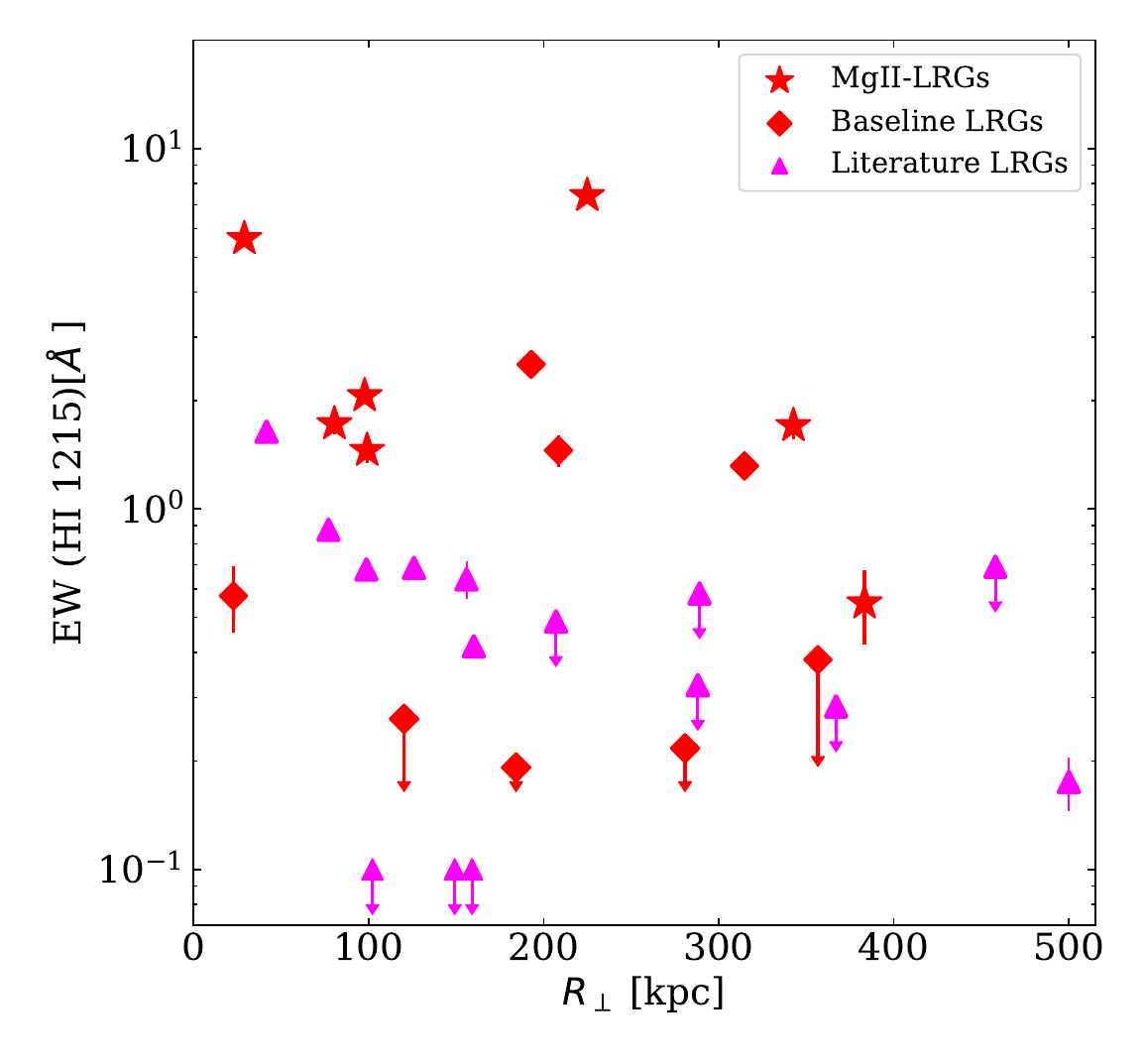}
\caption{\small 
CGM of LRGs traced by EW(\ion{H}{1} 1215) as a function of impact parameter.
Data points that lie below the y-axes limits are shown as upper limits at a fixed value of EW $= 10^{-1} \AA $.
One notes that there is no clear decrease of LRG-CGM EW(\ion{H}{1} 1215) with increasing impact parameter. 
}
\label{fig:ewhilrgs}
\end{figure}

% EW - R: previous studies
Previous studies found that EW(\ion{H}{1} 1215) from the CGM of $\sim L^{*}$ and $< L^{*}$ galaxies decreases with increasing impact parameter \citep[see e.g.][]{X11,Bordoloi18}. For LRGs, or massive $> L^{*}$ galaxies, at $z \sim 0.5$, \ion{Mg}{2}, measurements show a similar correlation. 
Motivated by these results, in Figure \ref{fig:ewhilrgs} we show EW(\ion{H}{1} 1215) vs. impact parameter for MgII-LRGs, Baseline LRGs, and literature LRGs, and we test if there is any similar correlation for LRGs. We apply the generalized Kendall's Tau test (BHK method), for which we use a code from the ASURV package \citep{Feigelson,Isobe,LaValley} that we translated into Python\footnote{\url{https://github.com/marijana777/pyasurvpart}}.

% Correlation with impact parameter for HI 
For MgII-LRGs, we obtain a correlation coefficient -0.43, implying a moderately strong (anti)correlation. On the other hand, for the general population of LRGs, we recover a weaker correction. For Baseline LRGs combined with literature LRGs, we recover a weak correlation, with a correlation coefficient of -0.30. These results are significant to $\sim 1.4 \sigma$ and $\sim 1.6 \sigma$, respectively. 
When we scale impact parameter to virial radius, we obtain similar and a bit weaker and less significant results. 
For comparison, for 24 COS-Halos blue galaxies, the generalized Kendall's Tau test gives a correlation coefficient of -0.41 and a significance level of $\sim 2.8 \sigma$. In contrast, for COS-Halos red galaxies, there is no statistically significant correlation (the correlation coefficient is 0.018, and the p-value is 0.1$\sigma$). 

In the next section, we test if there is a correlation between \ion{H}{1} column densities, $N_{HI}$, and impact parameter scaled to the virial radius for the combined sample of our and literature LRGs. 
We note that the EW(\ion{H}{1} 1215) and $N_{HI}$ measures are correlated, but are different quantities, and that we have slightly different data available for those measurements (for half of our data we measure only upper limits in $N_{HI}$, while, on the other hand, most of B-LRGs data do not have available EW(\ion{H}{1} 1215) values). For those reasons, the results for $N_{HI}$ and EW(\ion{H}{1} 1215) need not be the same.  

%% EW vs R 

\subsection{\ion{H}{1} Column Densities}\label{sec:nhi}

%% measurements - method 
The spectral resolution of our COS spectra is not high enough to calculate \ion{H}{1} column densities ($N_{HI}$) based on Voigt profile fitting to \ion{H}{1} Ly $\alpha$. Furthermore, nearly all detections lie on the flat portion of the curve of growth. Instead, we calculate $N_{HI}$ from the Lyman-limit (LL) flux decrement, using a maximum likelihood analysis \citep[see][]{X17}.
For each LRG-CGM with coverage of the LL, we estimated a continuum level just redward of the LL, by fitting the flux in an unabsorbed wavelength interval with a constant value. We use this value as a continuum level at the LL decrement. 
Then, we estimated $N_{HI}$ by applying a maximum likelihood analysis, and by using $N_{HI} = \tau_{912} /\sigma_{912} (1 {\rm Ryd}) = - \ln (F_{912}/C_{912})/\sigma_{912} (1 {\rm Ryd})$, where $F_{912}$ is flux, $C_{912}$ is continuum level, $\tau_{912}$ is the optical depth, and $\sigma_{912} (1 {\rm Ryd})$ is the \ion{H}{1} photoionization cross-section at the LL decrement. 
In QSO spectra with strong LL absorption, the flux is consistent with zero and the calculated $N_{HI}$ values represent lower limits. Similarly, when the LL level is comparable to the continuum level, the calculated $N_{HI}$ provides an upper limit. 

\begin{figure*}[t!]
\centering
\includegraphics[width=0.9\textwidth]{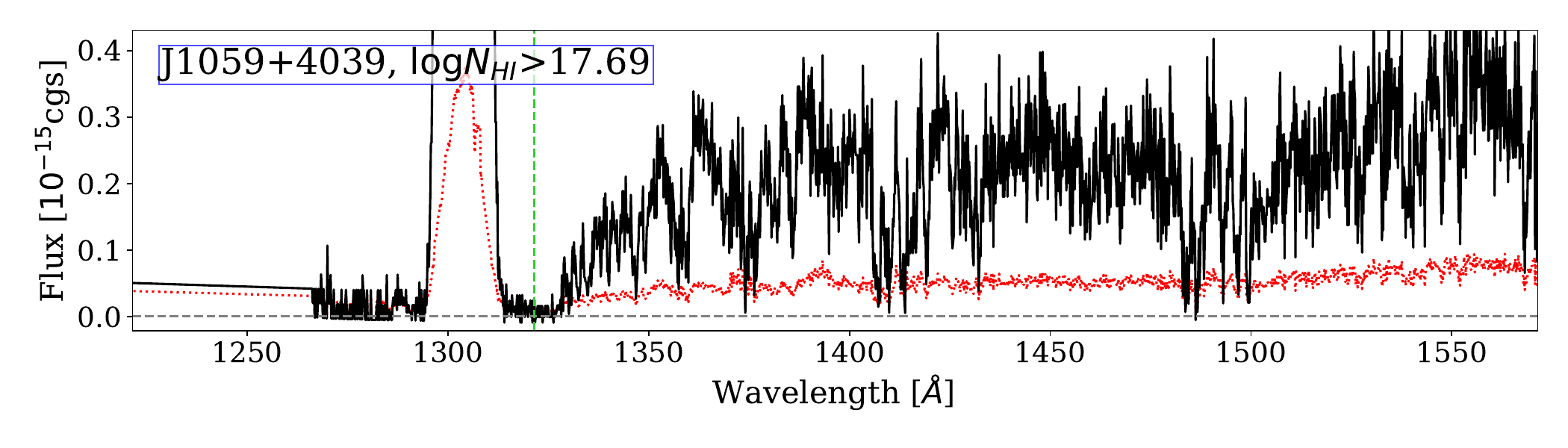}
\caption{\small 
The LL decrement in the J1059+4039 spectrum. The black line is flux, the red dotted line is the $1\sigma$ error in the flux, and the vertical green dashed line is the expected location of the LL. The measured \ion{H}{1} column density $N_{HI}$ is shown in the upper left corner. The complete figure set (15 images with the LL decrement in all our LRG-CGM systems) is 
shown in Appendix \ref{sec:appf}. % added for arxiv
%available in the online journal. % deleted for arxiv
}
\label{fig:llsfig0}
\end{figure*}

%% additional measurements - EW - NHI 
In this way, for four MgII-LRGs we obtain only lower limits in $N_{HI}$. In these cases, we set upper limits from the absence of damped Ly$\alpha$ absorption using damped regime of the curve of growth for EW(\ion{H}{1} 1215) and $N_{HI}$. 
For Baseline LRGs with \ion{H}{1} detected, we estimate lower limits from the
%\edit1{
curve of growth for for EW(\ion{H}{1} 1215) and $N_{HI}$ for Doppler parameter of $\sim 100$ km s$^{-1}$.
We obtain $\log (N_{HI} {\, \rm cm^{2}}) >$ 15.0, 17.6, 15.3, and 14.2, however, we note that these values could decrease if multiple line components are taken into account. 
In addition, using Doppler parameter of $\sim 50$ km s$^{-1}$ or less would give too high lower limits, where in 3 of 4 cases lower limits would exceed upper limits.
Note that one of these LRGs (LRG\_1102+4543)
has EW(\ion{H}{1} 1215) $\sim 2.5$ \AA , and $N_{HI} < 10^{15.77}$ cm$^{-2}$.
Based on the curve of growth, $N_{HI}$ corresponding to this EW(\ion{H}{1} 1215) are $> 10^{17.6}$ (for $b \sim 100$ km s$^{-1}$). However, multiple line components and errors in the derived $N_{HI}$ must decrease the true $N_{HI}$.
We note that one of the LRGs, LRG\_1549+0700, has a decline in the flux level at wavelength that corresponds to the LRG Lyman-limit system (912 \AA ). However, this LRG does not have detected Ly$\alpha$ or any of the \ion{H}{1} Lyman transitions. We therefore adopt the measured column density as an upper limit. 
The measurements are listed in Table \ref{tab:nhitab}. 
Figure \ref{fig:llsfig0} shows the LL decrement for one of our LRG-CGM systems. The complete figure set (15 images with the LL decrement in all our LRG-CGM systems) is available in 
Appendix \ref{sec:appf} (Figures \ref{fig:llsfig0a}-\ref{fig:llsfig14}).  % added for arxiv
% the online journal. % commented for arxiv 

% NHI table 
\begin{table}
\center
\caption{Column densities $N_{HI}$ for our LRG-CGM systems.
\label{tab:nhitab}}
\begin{threeparttable}
{
\makebox[\linewidth]
{
\hspace{-0.35in}
\centering 
\begin{tabular}{ccc}
\cutinhead{MgII-LRGs$^{a}$}
Name & $\log N_{HI}$ (LLS) & $\log N_{HI}$ (DLA)
\\
 &  (cm$^{-2}$)  & (cm$^{-2}$)  
\\
\hline
LRG\_1059+4039 & $> 17.69$ & $< 20.4$  \\
LRG\_1106-0115 & $17.26^{+0.02}_{-0.00}$ &  ...  \\
LRG\_1121+4345 & $> 17.89$ & $< 19.20$  \\
LRG\_1144+0714 & $> 17.91$ & $< 19.4$  \\ 
LRG\_1351+2753 & $17.37^{+0.03}_{-0.02}$ &  ...   \\ 
LRG\_1440-0157 & $17.02^{+0.03}_{-0.02}$ &  ...   \\ 
LRG\_1520+2534 & $> 17.86$ & $< 20.65$  \\ 
\hline
\cutinhead{Baseline LRGs$^{b}$}
Name & $\log N_{HI}$ (cog) & $\log N_{HI}$ (LLS)
\\
 &  (cm$^{-2}$)  & (cm$^{-2}$)  
\\
\hline
LRG\_0226+0014 &  ...  & $< 16.34$   \\ 
LRG\_0855+5615 & $>15.0$ &  $< 15.8$\tnote{c} \\ 
LRG\_1102+4543 & $>17.6$ & $< 15.77$   \\ 
LRG\_1217+4931 & $>15.3$ & $< 15.77$  \\ 
LRG\_1237+1106 & $>14.2$ & $< 16.64$  \\ 
LRG\_1251+3025 &  ...  & $< 16.20$  \\ 
LRG\_1306+3421 &  ...  & $< 16.66$  \\ 
LRG\_1549+0701 &  ...  & $< 16.42$  \\ 
\hline
\end{tabular}
}
\begin{tablenotes}
    \item[a] {Columns contain LRG-CGM name, $N_{HI}$ calculated from the LL flux decrement, and upper limits in $N_{HI}$ calculated from the fitted damped Ly$\alpha$ line profile.
When only lower limits in $N_{HI}$ are available from the LL flux decrement, we also calculate upper limits in $N_{HI}$  from the fitted damped Ly$\alpha$ line profile, and adopt average value between the upper and lower limit, with the error equal half of the difference between  these limits.}
    \item[b] {Columns contain LRG-CGM name, lower limits in $N_{HI}$ estimated from the curve of growth for Doppler parameter of $\sim 100$ km s$^{-1}$, and $N_{HI}$ calculated from the LL flux decrement.
    }
    \item[c] {LRG\_0855+5615 has a LLS at higher redshift, that absorbs majority of the flux blueward of it, making its $N_{HI}$ estimate less reliable.}
  \end{tablenotes}
}
\end{threeparttable}
\end{table}

\vspace{10pt}

\begin{figure}[h]
\centering
\includegraphics[width=0.49\textwidth]{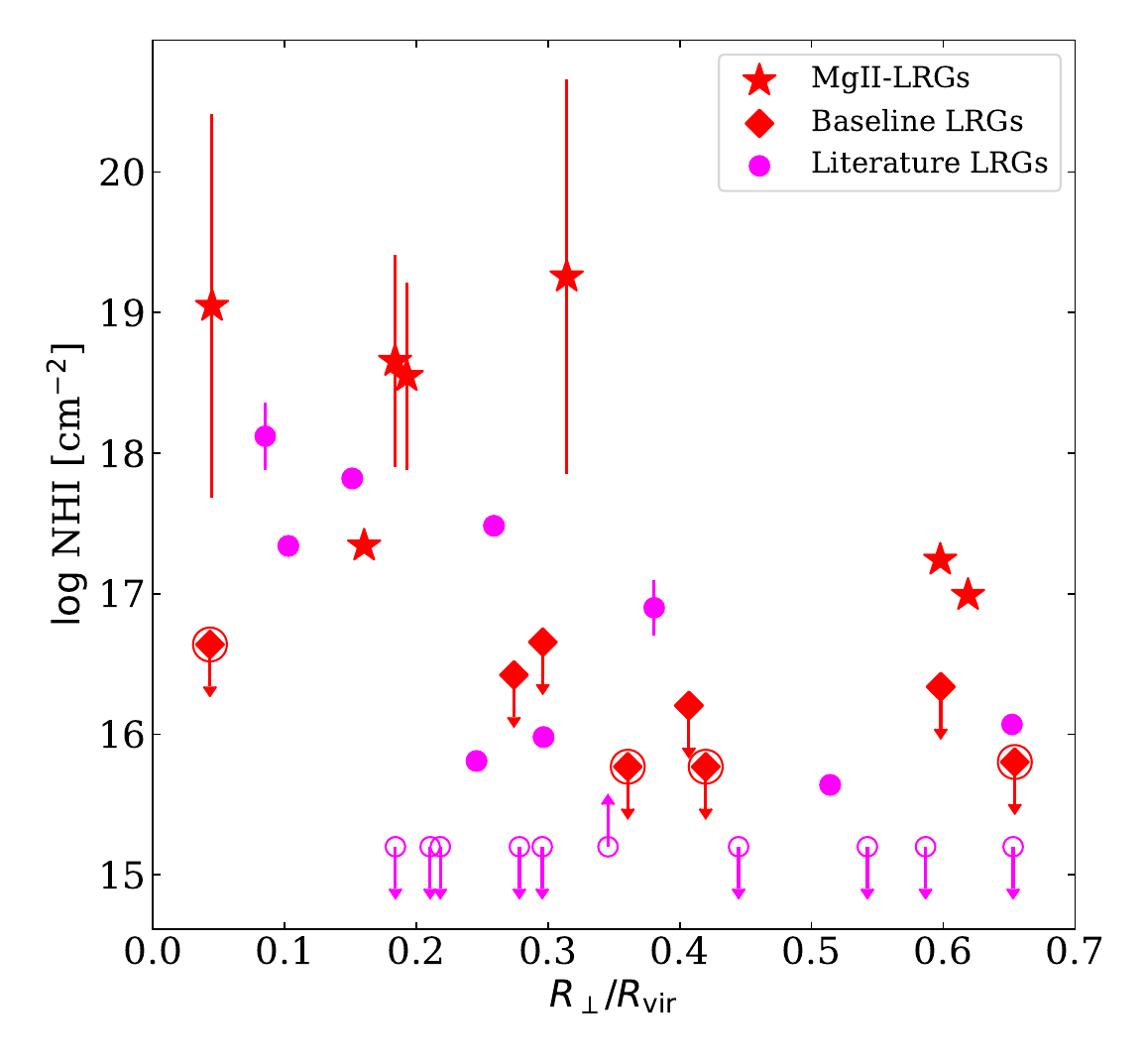}
\caption{\small 
\ion{H}{1} column densities as a function of impact parameter scaled to the virial radius, for our LRGs and literature LRGs. For the four MgII-LRGs with a lower limit from the LL decrement, the measurements represent medians of the lower limit (from the LL flux decrement) and the upper limit (from the curve of growth). 
Four Baseline LRGs with EW detection and with upper limits from the LL flux decrement are circled and shown as full diamonds upper limits.
Some of the data points would be below the y-axes limits, and these points are shown as upper limits at a fixed value $15.2$. We do not show data points beyond $R_{\perp}/R_{\rm vir} = 0.7$. 
Column density in \ion{H}{1} in LRG-CGM decreases with the impact parameter scaled to the virial radius.
}
\label{fig:nhi}
\end{figure}

% How often do we detect LLS? 
From Table \ref{tab:nhitab} one notes that we detect $N_{HI} > 10^{17}$ cm$^{-2}$ in all seven MgII-LRGs. However, for all Baseline LRGs, we measure upper limits in $N_{HI}$ that are  $\lesssim 10^{16.66}$ cm$^{-2}$ (with an exception that one Baseline LRG might not have reliable $N_{HI}$ measurement).
This result indicates that detecting optically-thick \ion{H}{1} gas around LRGs is rare, in agreement with \citet{Berg18}. 
LRG-CGM in our and \citet{Berg18} samples show lower $N_{HI}$ than in \citet{Chen18} sample, where three of the nine LRGs have $N_{HI} > 10^{17}$ cm$^{-2}$. This is likely due in part to an impact parameter effect (as we will see from Figure \ref{fig:nhi}). 

% Is MgII associated with LL? 
Our results show that the detection of high-$N_{HI}$ gas (e.g., optically-thick \ion{H}{1} gas) is often associated with the detection of \ion{Mg}{2}, or low-ionization metal ions. This implies that more dense gas is more likely to contain low-ionization metal ions. 
In addition, the B-LRG sample contains two LRGs with $N_{HI}$ greater than $10^{17.5}$ cm$^{-2}$ and associated low-ions. 
In B-LRG sample, five LRGs have measured column densities for low-ionization metal ions (such as \ion{Mg}{2} or \ion{C}{2}) that are not upper limits, and all of these five LRGs have the highest $N_{HI}$ (not taking into account one LRG with a lower limit in $N_{HI}$). 
Similarly, three C-LRGs have $N_{HI} > 10^{17}$ cm$^{-2}$, one of which does not have metals detected, and two of them show the strongest \ion{Mg}{2}.  
The other six LRGs either do not have \ion{H}{1} detected, or have EW(\ion{Mg}{2} 2796) $< 0.3 \AA$, confirming the association of strong \ion{Mg}{2} with high $N_{HI}$.

% correlation with R? 
Figure \ref{fig:nhi} shows \ion{H}{1} column density $N_{HI}$ as a function of impact parameter scaled to the virial radius, for our LRGs sample and for literature LRGs. 
We note that four of these LRGs with the highest $N_{HI}$ have impact parameter $< 0.3 R_{\rm vir}$. 
The generalized Kendall's Tau test shows a moderately strong anti-correlation between $N_{HI}$ and $R/R_{\rm vir}$ in our combined sample (with correlation coefficient -0.41)
between $N_{HI}$ and impact parameter, with significance $\sim 2.8 \sigma$, 
suggesting that a large fraction of the cool LRG CGM gas is gravitationally bound to the LRG host halos. 

%% NHI vs R 

\subsection{Kinematics}\label{sec:kin}

\begin{figure*}[t!]
\centering
\includegraphics[width=0.49\textwidth]{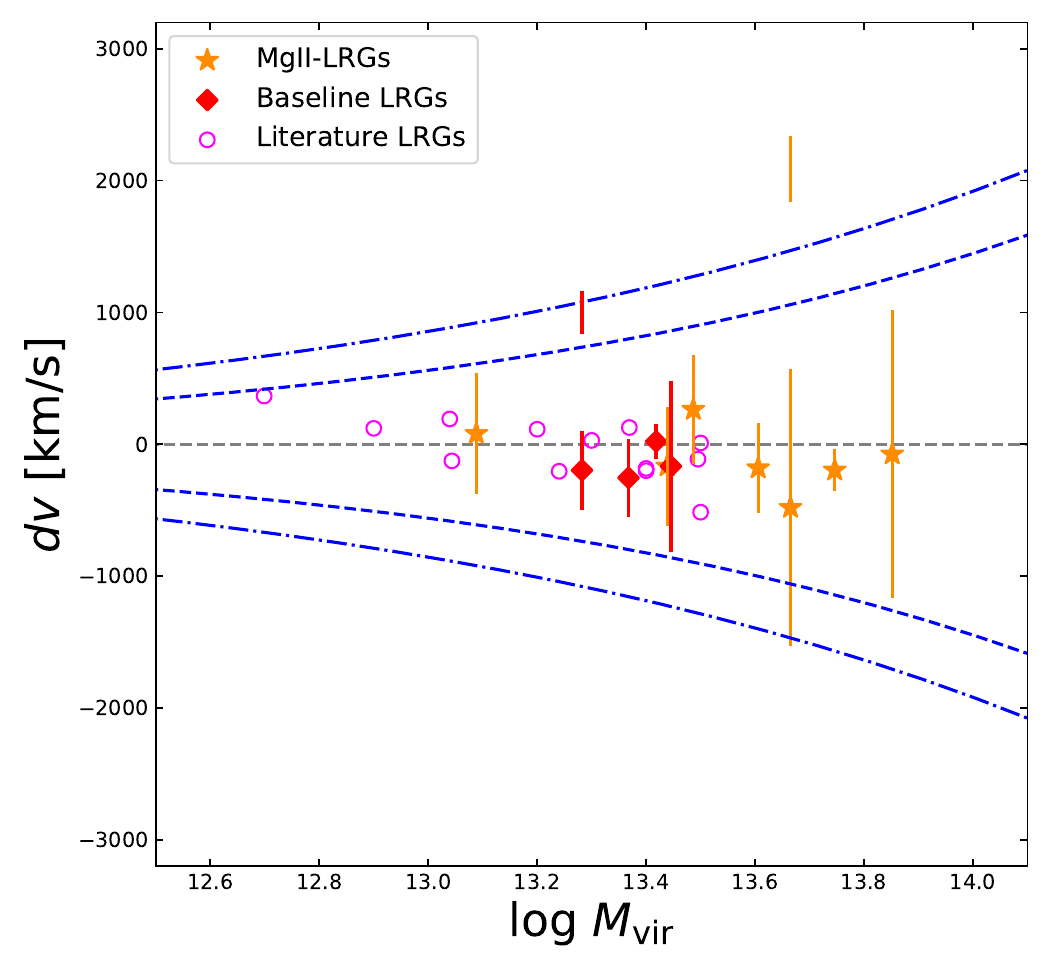} 
\includegraphics[width=0.49\textwidth]{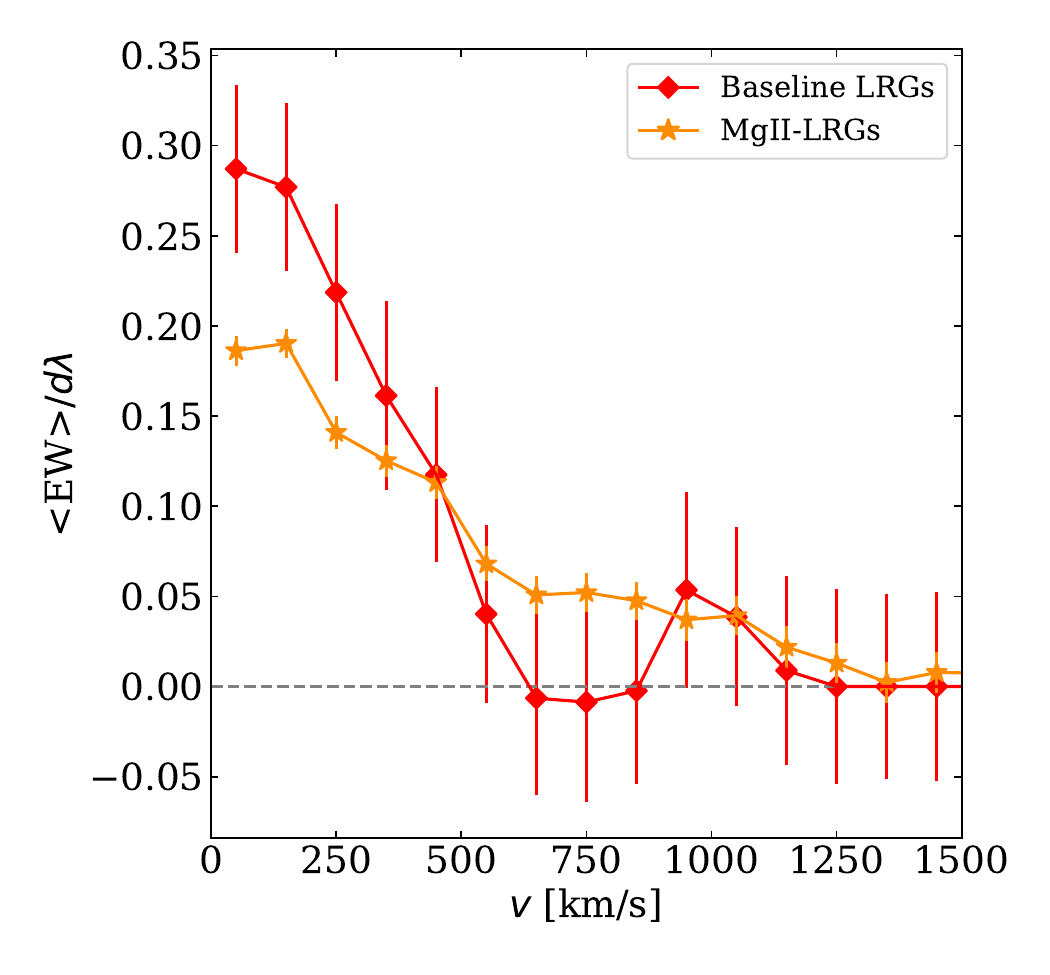} 
\caption{\small
Kinematic properties of LRG-CGM systems. 
\textit{Left panel:}
Centroid velocities ($dv$) versus virial mass for our MgII-LRGs (orange stars), our Baseline LRGs (red diamonds), and literature LRGs (magenta circles). The error bars denote velocity extents $\Delta v_{90,Ly\alpha}$. For two LRGs, we also show additional line components. For these line components we show full velocity range included in analysis instead of $\Delta v_{90,Ly\alpha}$. The gray line denotes $v = 0$ km s$^{-1}$. Blue dash-dotted lines denote escape velocity at 30 kpc and at 400 kpc.
One notes that velocities of the LRG-CGM gas are consistent with being within the escape velocity of the host halos.
\textit{Right panel:} Average rest-frame equivalent widths per wavelength interval for \ion{H}{1} 1215 for MgII-LRGs (orange line with stars) and Baseline LRGs (red line with diamonds). 
For MgII-LRGs, average rest-frame equivalent widths per wavelength interval are scaled so that the area below the function is the same for MgII-LRGs and for Baseline LRGs.
We include only data points that are inside the analyzed velocity intervals for \ion{H}{1} 1215 lines, and two additional velocity components that are also shown in the left panel. One notes that MgII-LRGs show more shallow profile than Baseline LRGs. 
}
\label{fig:kin}
\end{figure*}

%% centroid velocities, dv90 
Figure \ref{fig:kin} shows centroid velocities versus halo mass. For each LRG, the centroid velocity ($dv$, as defined in section \ref{sec:lines}) is calculated for one of the weakest reliable LRG-CGM absorption lines. The line transitions used to calculate velocity centroids are listed in Table \ref{tab:vtab}. The $\Delta v_{90}$ interval is plotted on the measurements. 
The mean centroid velocity for our sample is $-92 \pm 22$ km s$^{-1}$, and the velocity dispersion is $205$ km s$^{-1}$. For our sample combined with literature LRGs, we measure the mean centroid velocity of $-59$ km s$^{-1}$ and the velocity dispersion of $214$ km s$^{-1}$. 
One notes that the mean centroid velocity is consistent with zero within the velocity dispersion range.
One also notes that the highest velocity extents correspond to two LRGs with halo masses among the highest \citep[one presented in][]{Smailagic18} . 

% additional cmps
While previously we considered only connected pixels with the flux below the continuum when searching for CGM absorption lines, now we search for additional line components inside [-3500,3500] km s$^{-1}$. 
We find all pixels such that apparent column densities for both Ly$\alpha$ and for Ly$\beta$ exceed its errors, in at least 3 adjacent pixels. We then inspected candidate components, and selected only statistically significant and clearly identified additional components. We find such velocity components around two LRGs, one from MgII-LRGs (LRG\_1059+4039) and the other from Baseline LRGs (LRG\_0855+5615). 
The additional component around LRG\_1059+4039 shows strong \ion{H}{1} Ly$\alpha$ line, with EW (\ion{H}{1} 1215) $ = 1.33 \pm 0.08$ \AA , and other transitions (such as EW (\ion{H}{1} 1025) $ = 0.84 \pm 0.07$ \AA , EW (\ion{H}{1} 972) $ = 0.64 \pm 0.07$ \AA\ , and EW (\ion{C}{3} 977) $ = 0.40 \pm 0.07$ \AA ) .  
The additional component around LRG\_0855+5615 shows an \ion{H}{1} Ly$\alpha$ line, with EW (\ion{H}{1} 1215) $ = 0.48 \pm 0.06$ \AA , as well as EW (\ion{H}{1} 1025) $ = 0.17 \pm 0.04$  \AA\  and EW (\ion{O}{6} 1031)  $ = 0.27 \pm 0.04$ \AA . 

% profile mgii vs baseline 
The right panel of Figure \ref{fig:kin} shows the average EW per wavelength interval for MgII-LRGs and Baseline LRGs, including flux from LRGs and additional components. 
MgII-LRGs show more shallow profile than Baseline LRGs, implying that the CGM gas in MgII-LRGs and Baseline LRGs sigthlines might trace gas with different properties. 
For example, because it is expected that, due to ram-pressure drag force, the velocity of a cool cloud that is moving in a hot halo will be decreasing with time and that the decrease would be higher for less massive clouds, the CGM of the MgII-LRGs might trace cool clouds that formed more recently or that are more massive. At the same time, clouds of gas that show strong \ion{Mg}{2} absorption might contain more cool gas and hence be formed at later time or be more massive than the clouds that do not show \ion{Mg}{2}. 
In addition, \citet{Nielsen16} found that red galaxies at low redshift show smaller velocity dispersions than blue galaxies, which could be explained by more recent star formation outflows in blue galaxies. 
We conclude that in the past when LRGs or their neighboring galaxies were forming stars MgII-LRGs might on average experienced outflows more recently than Baseline LRGs. 

%% inside vesc 
From Figure \ref{fig:kin}, left panel, one notes that for all LRGs the velocity extent is inside the estimated escape velocity of LRG halos excluding additional components, implying that the CGM gas is gravitationally bound to the halos. Velocities of the two additional components (associated with LRG\_1059+4039 and LRG\_0855+5615) are similar or larger than the corresponding escape velocities. 
We found that, for LRG\_1059+4039, the additional component likely originates from the same halo as the target LRG, while for LRG\_0855+5615 we could not conclude if the additional component is associated with the target LRG halo (see Appendix \ref{sec:appe}). 
If the additional component for LRG\_1059+4039 originates from the same halo, then the CGM gas associated with this component is not gravitationally bound to the host halo. 

% kinem

% table with vcent and dv90
\begin{table}
\center
\caption{Centroid velocities and velocity extents for LRG-CGM systems$^{a}$
\label{tab:vtab}}
\begin{threeparttable}
{
 \setlength\tabcolsep{3pt} 
\makebox[\linewidth]
{
\hspace{-0.75in}
\centering 
\begin{tabular}{ccccc}
\hline
Name & $R_{\perp}$ & Transition & $dv$ & $\Delta v_{90}$
\\
 &  (kpc)  & & (km s$^{-1}$) & (km s$^{-1}$)
\\
\hline
LRG\_1059+4039 & 29 & \ion{Si}{2} 1193 & -480 & 2096  \\ 
LRG\_1106-0115 & 383 & \ion{H}{1} 919 & -196 & 319  \\ 
LRG\_1121+4345 & 80 & \ion{H}{1} 972 & 80 & 915  \\ 
LRG\_1144+0714 & 97 & \ion{Si}{2} 1193 & -165 & 901  \\ 
LRG\_1351+2753 & 99 & \ion{H}{1} 923 & -180 & 688  \\ 
LRG\_1440-0157 & 342 & \ion{H}{1} 930 & 263 & 826  \\ 
LRG\_1520+2534 & 224 & \ion{Mg}{2} 2796 & -75 & 2181  \\ 
LRG\_0226+0014 & 356 &  ...  &  ...  &  ...  \\ 
LRG\_0855+5615 & 314 & \ion{H}{1} 972 & -196 & 602  \\ 
LRG\_1102+4543 & 192 & \ion{C}{3} 977 & -167 & 1300  \\ 
LRG\_1217+4931 & 208 & \ion{H}{1} 972 & -254 & 596  \\ 
LRG\_1237+1106 & 22 & \ion{H}{1} 1215 & 22 & 268  \\ 
LRG\_1251+3025 & 280 &  ...  &  ...  &  ...   \\ 
LRG\_1306+3421 & 184 &  ...  &  ...  &  ...   \\ 
LRG\_1549+0701 & 120 &  ...  &  ...  &  ...   \\ 
\hline
\end{tabular}
}
\begin{tablenotes}
   \item[a] {Columns contain LRG-CGM name, LRG impact 
   
   parameter, transition used to derive centroid velocity, 
   
   centroid velocity ($dv$), and velocity extent ($\Delta v_{90}$). 
   
   We used \ion{H}{1} 1215 to derive $\Delta v_{90}$.} 
\end{tablenotes}
}
\end{threeparttable}
\end{table}

\vspace{10pt}

\section{Discussion: Baseline LRGs}\label{sec:discb}

\subsection{\ion{H}{1} at the Highest Masses}\label{sec:himass}

% motivation 
Recently, \citet{Bordoloi18} found that EW(\ion{H}{1} 1215) corrected for $R_{\perp}$ increases with stellar mass, and parameterized a relation between EW(\ion{H}{1} 1215), $R_{\perp}$, and stellar mass ($M_{*}$) as:
\begin{equation}
\log_{10} EW_{\rm corr} = 0.34 + 0.286\times \log (M_{*}/M_{\sun}),
\end{equation}
where 
\begin{equation}
\log_{10} EW_{\rm corr} = \log_{10} EW + 0.0026 R_{\perp}. 
\end{equation}
We now test if the \citet{Bordoloi18} relation holds for more massive galaxies. 

\begin{figure*}[h]
\centering
\includegraphics[width=0.7\textwidth]{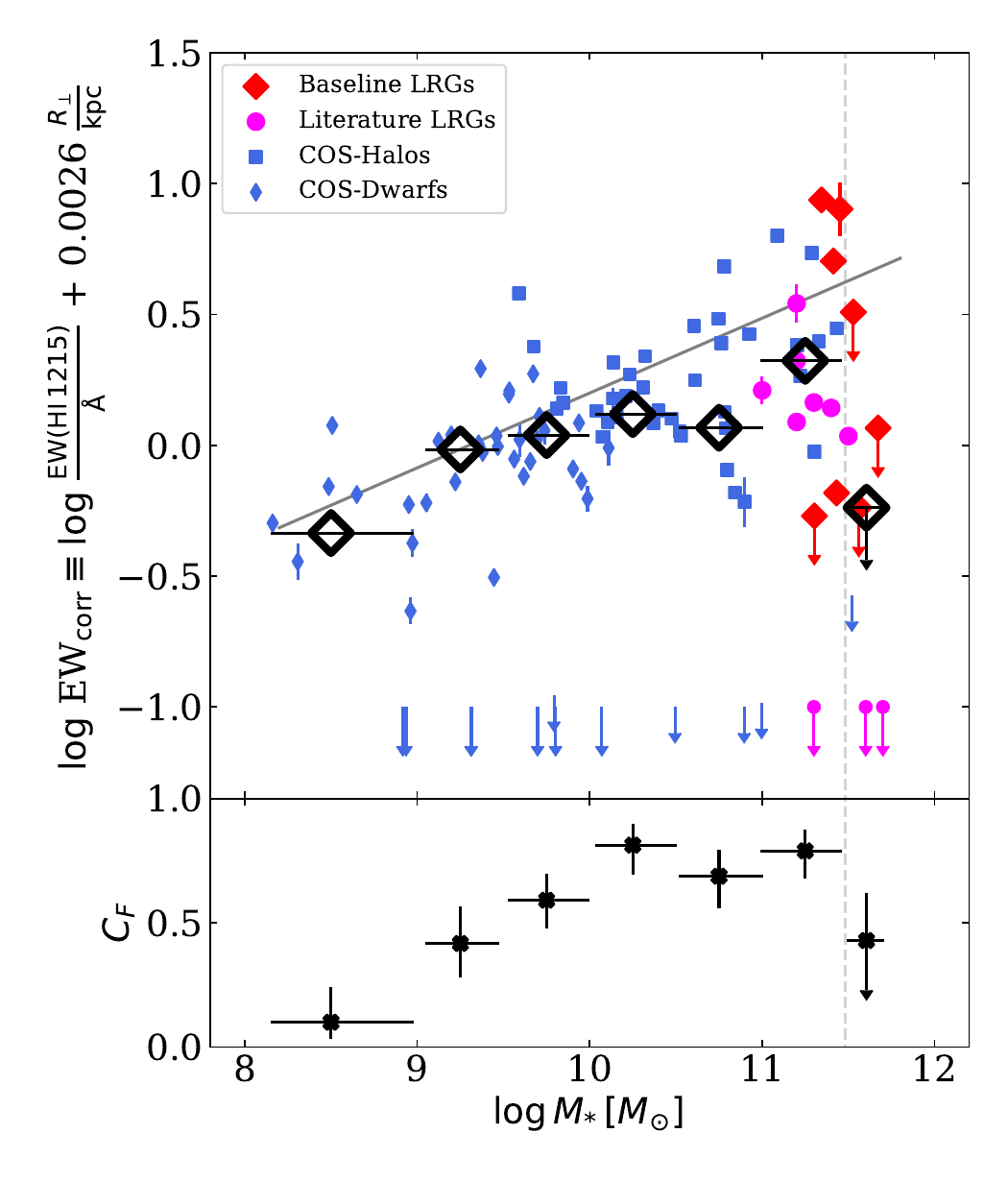}
\vskip -0.2in
\caption{\small 
CGM \ion{H}{1} as a function of stellar mass. 
\textit{Top}: $\log$ EW (\ion{H}{1} 1215) corrected for the impact parameter as a function of stellar mass, for our Baseline LRGs (red diamonds), literature LRGs (magenta circles), COS-Halos (blue squares), and COS-Dwarfs (blue diamonds). The full gray line shows relation derived by \citet{Bordoloi18}.
Black open diamonds are median values for all data points inside stellar mass bins $\log M_{*}$ of [8,9], [9,9.5], [9.5,10], [10,10.5], [10.5,11], [11,11.5], [11.5,12]. Some of the data points would be below the y-axes limits. These points are shown as upper limits at a fixed value of -1. Vertical gray dashed line corresponds to $\log M_{*} \sim 11.5$. 
\textit{Bottom}: Covering fraction for the detection of EW$_{\rm corr}$ higher than minimum EW$_{\rm corr, min} = 0$ for galaxies in stellar mass bins with width in $\log M_* $ intervals of 0.5.
While the EW$_{\rm corr}$(\ion{H}{1} 1215) increases with stellar mass for $\lesssim$ L$^{*}$ galaxies, the most massive galaxies show smaller EW$_{\rm corr}$(\ion{H}{1} 1215).
}
\label{fig:mewhi}
\end{figure*}

%% EWcorr vs mass 
In Figure \ref{fig:mewhi}, we show $\log_{10} EW_{\rm corr}$ as a function of stellar mass, for Baseline LRGs, literature LRGs, $\sim L^{*}$ galaxies and $< L^{*}$ galaxies. Literature LRGs include C-LRGs and the only one B-LRG with \ion{H}{1} 1215 significantly detected. 
One notes that, as previously found \citep[see e.g.,][]{Bordoloi18}, for $\lesssim L^{*}$ galaxies, EW$_{\rm corr}$ increases as stellar mass increases. In agreement with this increase, a few LRGs show the highest EW$_{\rm corr}$; however, the detections in LRGs also span a broader range of EW$_{\rm corr}$ than in less massive galaxies. 
One other difference is that at the highest masses, $> 10^{11.5} M_{\sun}$, the EW$_{\rm corr}$ values are, on average, smaller than at lower masses. At these masses, only one of seven galaxies has a detected cool CGM, while other six are upper limits. 

%% median 
Next, we calculate median $\log_{10} EW_{\rm corr}$ values in different stellar mass bins. Figure \ref{fig:mewhi} shows that the median $\log_{10} EW_{\rm corr}$ increases with stellar mass for masses $M_{*} < 10^{11.5} M_{\sun}$. However, for the highest-mass bin ($M_{*} > 10^{11.5} M_{\sun}$), the median $\log_{10} EW_{\rm corr}$ is below the median $\log_{10} EW_{\rm corr}$ values of nearly every other galaxy with $M_{*} > 10^{9} M_{\sun}$. 

%% EW decrease at the largest masses - statistically significant 
A decrease in $EW_{\rm corr}$ at the highest stellar masses is in agreement with predictions both from theory and from observations of clusters of galaxies \citep{Dekel06,Nelson15,FG17,Voort17,Burchett18}. To further explore the statistical significance of this result, we compare covering fractions ($C_{F}$) for the detection of EW$_{\rm corr}$ higher than minimum $\log EW_{\rm corr, min} = 0$ for galaxies in different stellar mass bins. The results are shown on the lower panel of Figure \ref{fig:mewhi}.
The $C_{F}$ and Wilson score intervals for $\sim 1 \sigma$ confidence intervals for different $\log M_{*}$ bins are listed in Table \ref{tab:ewmcf}. One notes that the upper limit on $C_{F}$ for the most massive halos ($< 0.61$) is below the $C_{F}$ corresponding to $\log M_{*} = [11,11.5]$, [10.5,11], and $[10,10.5]$. 
 
% => 
This lends further support to the notion that the average strength and incidence of \ion{H}{1} Ly$\alpha$ absorption declines in the most massive halos at $z < 1$. 

%%% table with vcent and dv90
\begin{table}
\center
\caption{Wilson score intervals for $C_{F}$ from Figure \ref{fig:mewhi}, bottom panel
\label{tab:ewmcf}}
\begin{tabular}{ccc}
\hline
$\log M_{*}$ bin & $C_{F}$ & Wilson score interval
\\
\hline
 $[8, 9]$     & 0.1      & $[0.04, 0.23]$  \\ 
 $[9, 9.5]$   & 0.42     & $[0.29, 0.56]$  \\  
 $[9.5, 10]$  & 0.59     & $[0.48, 0.69]$  \\
 $[10, 10.5]$ & 0.81     & $[0.70, 0.89]$  \\
 $[10.5, 11]$ & 0.69     & $[0.56, 0.79]$  \\  
 $[11, 11.5]$ & 0.79     & $[0.68, 0.87]$  \\  
 $[11.5, 12]$ & ... &  $< 0.61$ \\ 
\hline
\end{tabular}
\end{table}

%% not corr for R?  
We note that even if the EWs are not corrected for impact parameter, we find an even larger decrease at the highest masses. This is expected because our LRGs on average have larger impact parameters than COS-Halos and COS-Dwarfs and, hence, the difference between the corrected EWs for LRGs and $\lesssim L^{*}$ galaxies would, on average, be smaller than for uncorrected EWs.

%% vs MgII-LRGs 
On the other hand, when compared with MgII-LRGs (which are more rare and not shown on the Figure \ref{fig:mewhi}), the highest EW$_{\rm corr}$ are also found at the highest stellar masses, implying that even some of these very massive galaxies might contain significant amounts of cool CGM gas in their halos.

% EWcorr vs M* 

\subsection{Comparison with CGM Studies at Other Redshifts: Covering Fractions} \label{sec:cf}

%% LRGs CF
In section \ref{sec:cf1}, we combined our Baseline sample with literature LRGs and derived the covering fraction for EW(\ion{H}{1} 1215) $>0.3$ \AA\ for LRG-CGM, equal $C_{F} \sim 0.56^{+0.11}_{-0.12}$.

% CF evolution with z 
LRGs are massive galaxies, and are located in massive halos, at $z \sim 0.5$. 
Current theory predicts that massive halos at low redshift contain very small mass of cool gas \citep{Dekel06,Nelson15,FG17}. 
In addition, \citet{Zahedy19} found that LRG halos on average contain only $4 \times 10^{10} M_{\sun}$ in cool gas mass, which is $\sim 6-13$ percent of the expected hot CGM gas mass. 
Despite this, approximately half show significant, and in some cases strong, \ion{H}{1} absorption in the surrounding CGM. QSO host halos are massive halos at redshifts $z \sim 2 - 3$, significant fraction of which is predicted to evolve into LRG halos \citep{White12}. CGM of QSOs is known to show the strongest \ion{H}{1} and low-ionization metal CGM absorption lines \citep[e.g.][]{X13}. 
One representative example for massive halos in the local Universe is from the well studied Virgo cluster. In contrast, along lines of sight around Virgo cluster, \ion{H}{1} is rarely detected. 

\begin{figure*}[t!]
\centering
\includegraphics[width=0.49\textwidth]{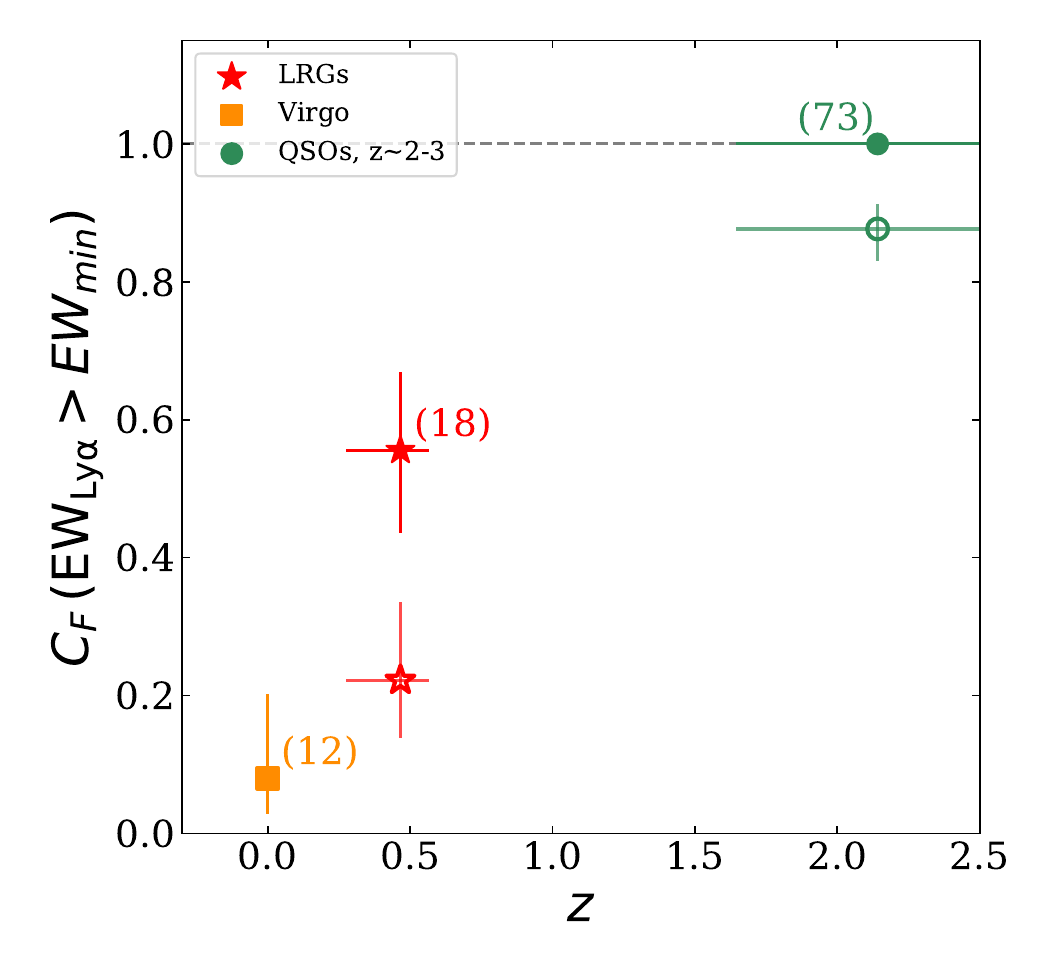} 
\includegraphics[width=0.49\textwidth]{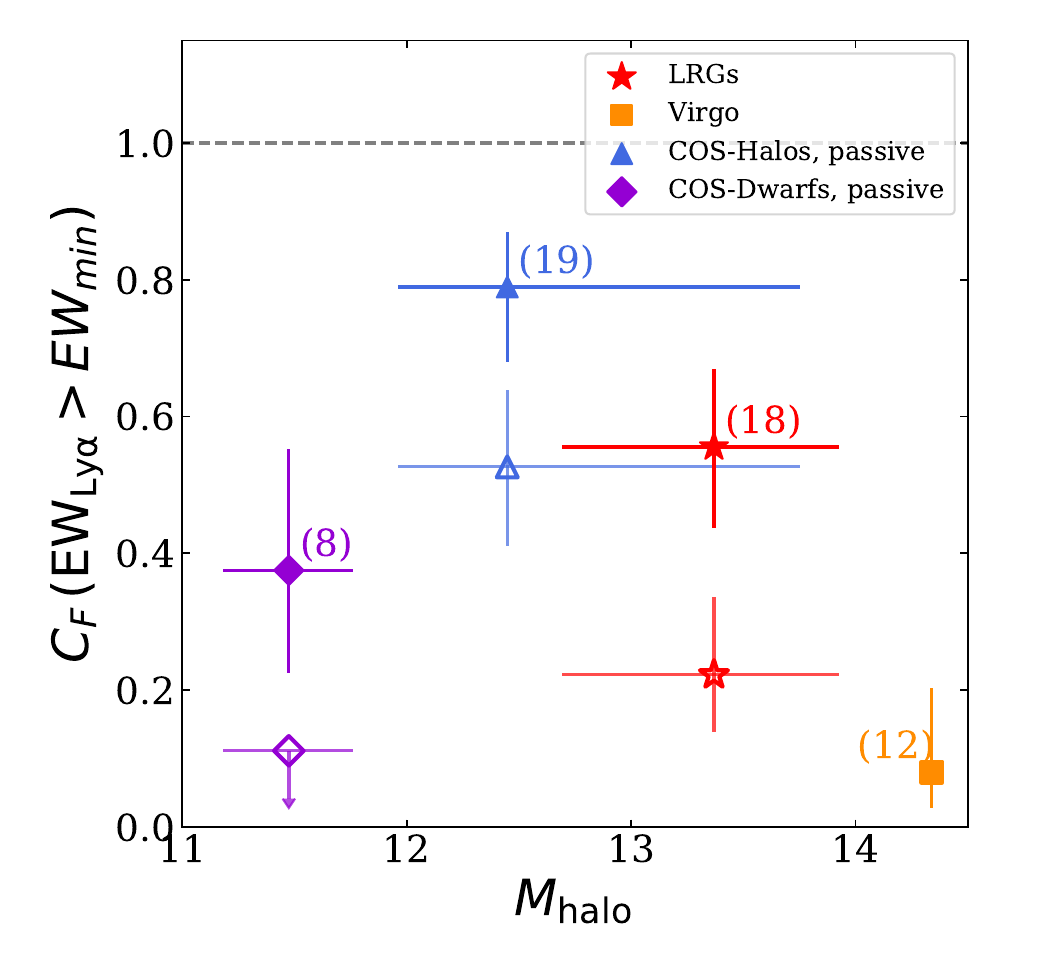} 
\caption{\small 
Covering fraction of \ion{H}{1} ($C_{F}$) in the LRG-CGM compared to the other data. The data include CGM measurements around: Virgo cluster \citep[orange square;][]{Yoon12}, LRGs (our Baseline LRGs combined with literature LRGs), QSO hosts \citep[green circle][]{X13}, non star-forming COS-Halos \citep[blue triangle;][]{Tumlinson13}, and non star-forming COS-Dwarfs \citep[violet diamond;][]{Bordoloi14}. The error bars on the $C_{F}$ correspond to the calculated Wilson score intervals.
\textit{Left}: Covering fraction versus redshift, for massive halos at different redshifts. The most massive halos show smaller $C_{F}$ at lower redshifts. Typical (average) $R_{\perp} / R_{\rm vir}$ for Virgo cluster, LRGs, and QSO hosts are 0.5, 0.3, and 1, respectively.
\textit{Right}: Covering fraction versus halo mass, for non star-forming galaxies at $z \sim 0 - 0.5.$ The typical (average) $R_{\perp} / R_{\rm vir}$ for COS-Halos and COS-Dwarfs galaxies are 0.3 and 0.5, respectively. 
All covering fractions are calculated for EW$_{\rm min}$ (\ion{H}{1} 1215) $> 0.3$ \AA\ (upper filled symbols). Covering fractions for LRGs, QSO hosts, non star-forming COS-Halos, and non star-forming COS-Dwarfs are also calculated for EW$_{\rm min}$ (\ion{H}{1} 1215) $> 1$ \AA\ (lower open symbols). 
}
\label{fig:cf}
\end{figure*}

Figure \ref{fig:cf} shows $C_{F}$ for the detection of EW(\ion{H}{1}) $> 0.3$ \AA\ in the CGM for massive halos at $z\sim 0 - 3$. For QSO host halos, we calculate $C_{F}$ for the subsample with impact parameters $<300$ kpc \citep[see ][]{X13}. The $C_{F}$ would not significantly change if instead of the subsample with impact parameters $<300$ kpc we use impact parameters $< 400$ kpc or $< 100$ kpc ($\sim 2/3$ the virial radius of the QSO host halos). 
For the Virgo cluster, we use results from \citet[][their table 6]{Yoon12}, which include absorbers within 1000 - 3000 km s$^{-1}$, with impact parameters smaller than the virial radius for Virgo cluster, and which have EW $> 0.3$ \AA . The Virgo cluster's virial radius is $\sim 1.6$ Mpc, and the $C_{F}$ might be smaller if we include absorbers only inside 400 kpc, see \citet{Yoon12}. 
The obtained covering fraction for CGM of LRGs at redshift $z\sim 0.5$ (i.e., $C_{F}(\rm LRG) \sim 0.56$) is between the $C_{F}$ of $z\sim 2-3$ QSO host galaxies ($C_{F}(\rm QSO) \sim 1$) and $z\sim 0$ Virgo cluster absorbers ($\sim 0.08$). This is expected if the amount of cold gas around massive galaxies decreases with time. 
For example, we expect that massive halos could not efficiently accrete cold gas at low redshift; and the interaction between cool gas clouds and hot halo gas lessens the mass of cool gas in the clouds that were found in massive halos at high redshift \citep[see e.g., ][]{Dekel06,Nelson15,Armillotta17,FG17,Voort17}.  

% CF vs mass 
Figure \ref{fig:cf} also compares the $C_{F}$ of non star-forming galaxies at $z\sim 0 - 0.5$ with halo masses $\sim 10^{12} - 10^{14.5} M_{\sun}$. 
Impact parameters for passive COS-Dwarfs and COS-Halos are up to $\sim 140$ kpc and $\sim 160$ kpc, which is up to $\sim 0.7$ and $\sim 0.6$ of their halo's virial radius.
$C_{F}$ increases with increasing halo mass, in agreement with previous studies for $\lesssim L^{*}$ galaxies \citep[e.g.,][]{Bordoloi18,Lan20}. For the most massive galaxies, however, we find that the $C_{F}$ decreases. The LRG $C_{F}$ is between $\sim $ local $\sim L^{*}$ galaxies and Virgo cluster, which is expected if the more massive $\gtrsim L^{*}$ galaxies contain less cold gas. 
One also notes that $C_{F}$ for LRGs and non star-forming galaxies differ by more than its 1-$\sigma$ errors.

%% CF figures (CF vs z, CF vs M)

We note that Baseline LRGs show a broad range of \ion{H}{1} EWs, from values which are similar to the average QSO hosts EWs, to the values that are below the average $\sim L^{*}$ galaxies.

\vspace{10pt}

\subsection{\ion{O}{6} in Massive Halos}\label{sec:ovi}  
% Warm-hot gas component

\begin{figure*}[t!]
\centering
\includegraphics[width=0.49\textwidth]{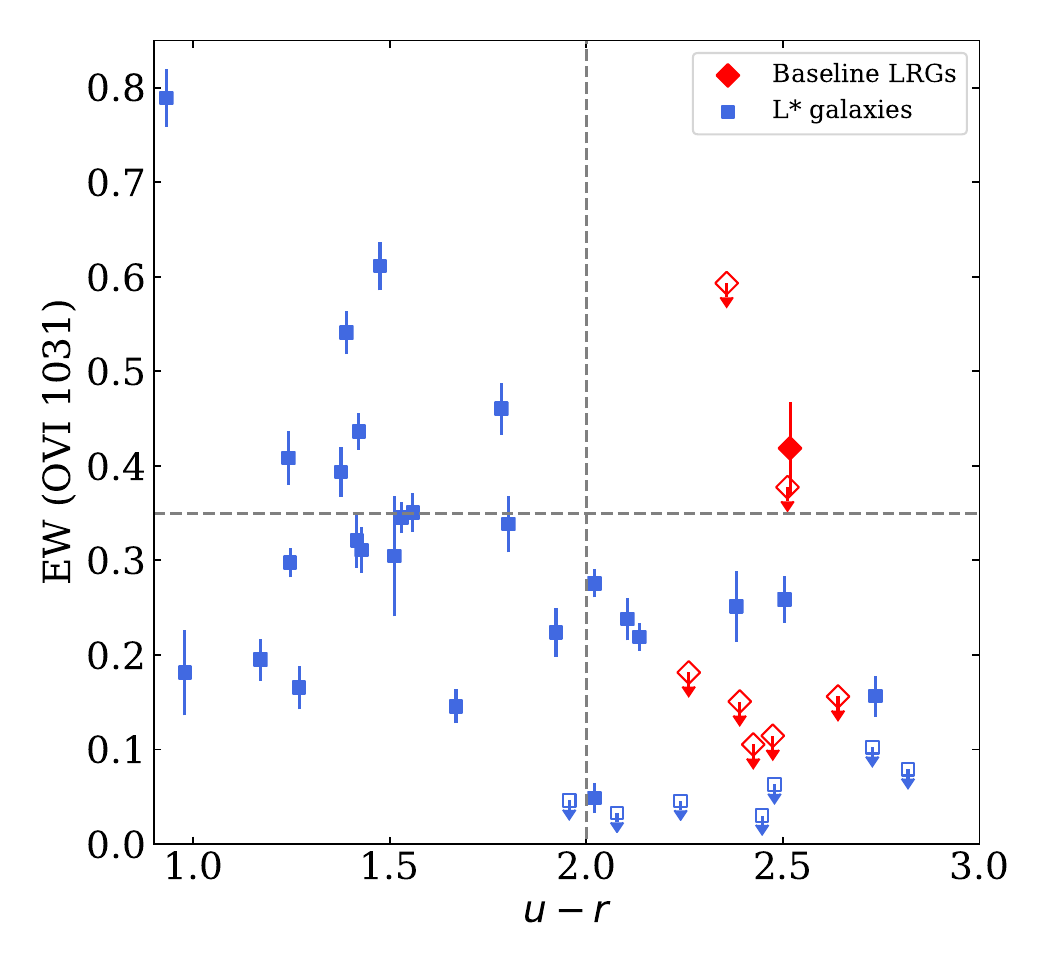} 
\includegraphics[width=0.49\textwidth]{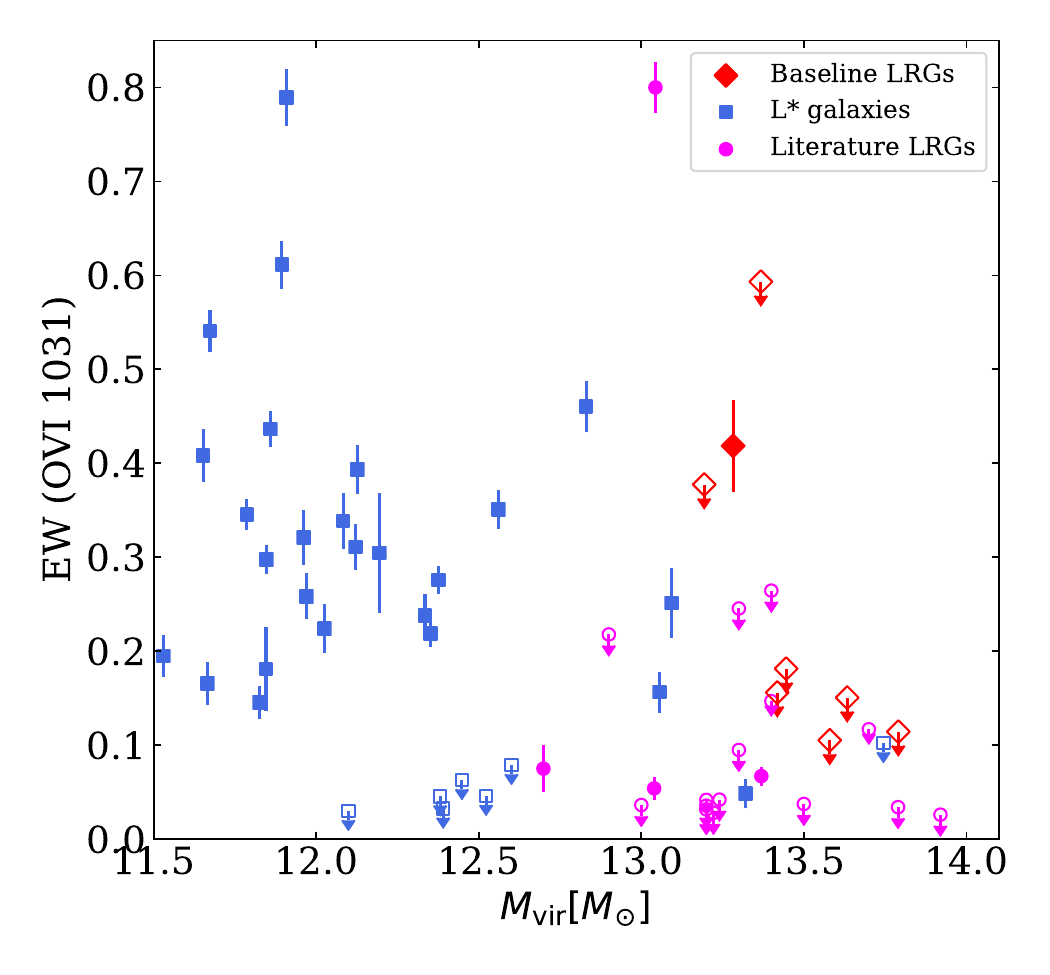} 
\caption{\small 
CGM \ion{O}{6} as a function of color and halo mass. 
\textit{Left panel:} Comparison of EW(\ion{O}{6} 1031) vs. $M_u - M_r$ color for our Baseline LRGs and COS-Halos. $M_u$ and $M_r$ are absolute SDSS magnitudes. Open symbols denote upper limits. 
\textit{Right panel:} EW(\ion{O}{6} 1031) vs. halo mass for LRGs (our Baseline LRGs combined with literature LRGs) and COS-Halos. 
We detect \ion{O}{6} 1031 in only one of eight LRG-CGM systems (LRG\_0855+5615). However, when detected, EW(\ion{O}{6} 1031) might exceed that of the CGM of red $\sim$ L$^{*}$ $z\sim 0.2$ galaxies.
}
\label{fig:ovi}
\end{figure*}

% motivation, OVI vs u-r and mass 
Previous studies showed that \ion{O}{6} 1031 is correlated with specific star-formation rate, and rarely detected in the CGM of non star-forming red galaxies \citep{Tumlinson11}. One possible explanation is that the observed red galaxies are more massive, and have higher virial temperature, such that oxygen is found in a more highly ionized state in these halos \citep{Oppenheimer16}.
In Figure \ref{fig:ovi} we show EW(\ion{O}{6} 1031) vs $M_u - M_r$ color for COS-Halos and for LRGs, where $M_u$ and $M_r$ are absolute SDSS magnitudes. 
Impact parameters for COS-Halos, our LRGs, and literature LRGs are up to $\sim 160$ kpc, 400 kpc, and 500kpc. 
One can see that, in agreement with the results in \citet[][]{Tumlinson11}, the EW(\ion{O}{6} 1031) decreases as the colors are redder, and for $M_u - M_r > 2$ all COS-Halos galaxies have EW(\ion{O}{6} 1031) $ \lesssim 0.35$ \AA . 
Most LRGs also show EW(\ion{O}{6} 1031) $ \lesssim 0.35$ \AA . 
One of the eight Baseline LRGs shows EW(\ion{O}{6} 1031) $ > 0.35$ \AA . 
% vs literature LRGs  
Consistent with these measurements, one of nine C-LRGs also shows strong \ion{O}{6}. In B-LRGs, for 11 of which we measured EW(\ion{O}{6} 1031), we do not find any similarly strong \ion{O}{6}, which is again consistent with our results. 

% OVI vs halo mass 
Figure \ref{fig:ovi} also shows EW(\ion{O}{6} 1031) vs halo mass. For COS-Halos and LRGs, EW(\ion{O}{6} 1031) decreases as halo mass increases. 
However, although LRGs have among the highest halo masses, two of the LRGs show EW(\ion{O}{6} 1031) among the highest. 
Their halo masses are among the highest in the combined sample of COS-Halos and LRGs, but among the lowest in LRGs sample.
Our results indicate that massive halos usually do not show strong \ion{O}{6}, but some contain strong \ion{O}{6}, and that the origin of \ion{O}{6} could not be explained only by different halo masses.

%% OVI vs u-r, Mhalo 

\vspace{10pt}

%%%%%%%%%%%%%%

\section{Discussion: MgII-LRGs}\label{sec:discm}

\subsection{Comparison with CGM Studies at Other Redshifts: Similarity of LRGs and QPQ for HI}

\begin{figure*}[t!]
\centering
\includegraphics[width=0.49\textwidth]{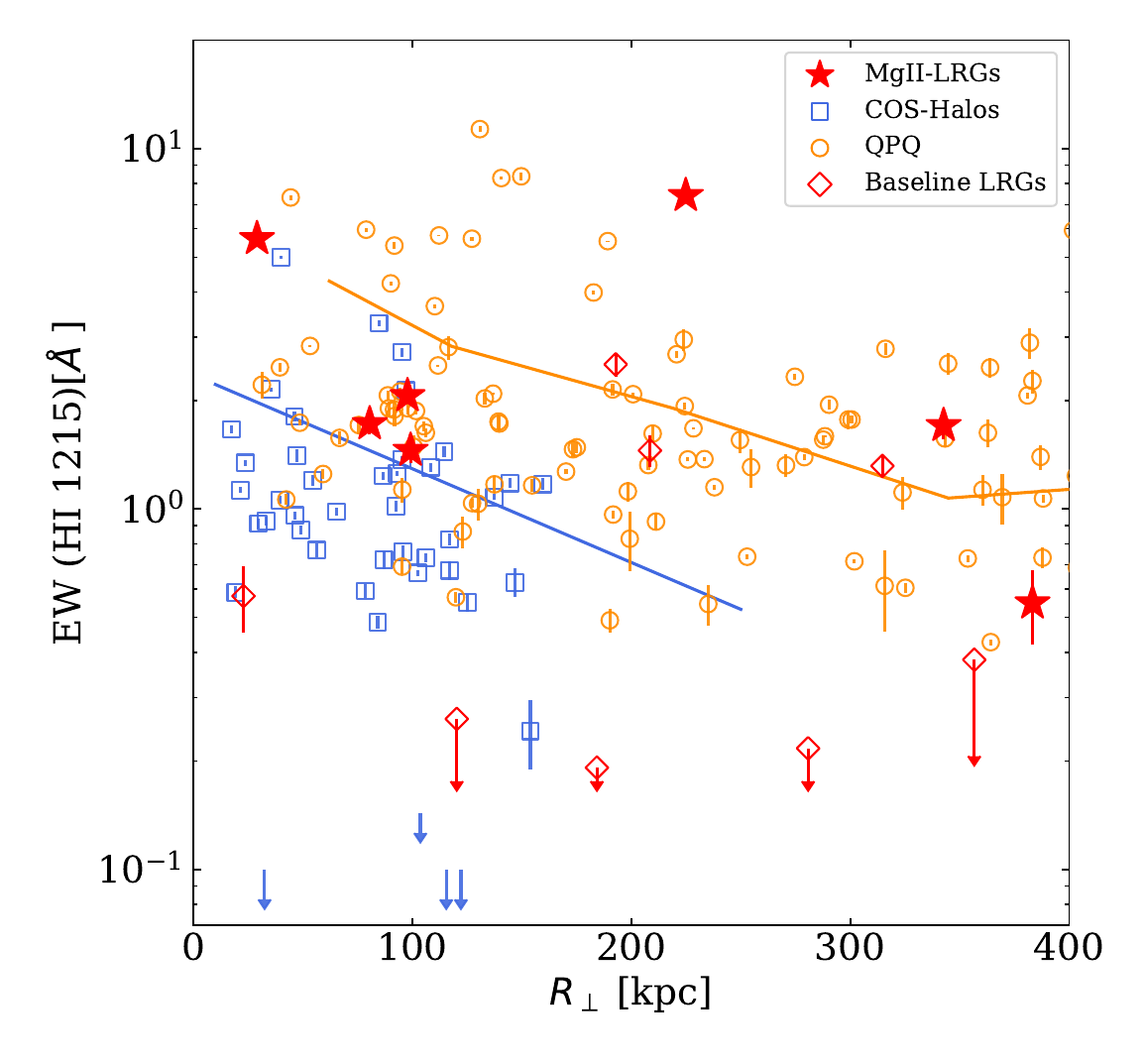}
\includegraphics[width=0.49\textwidth]{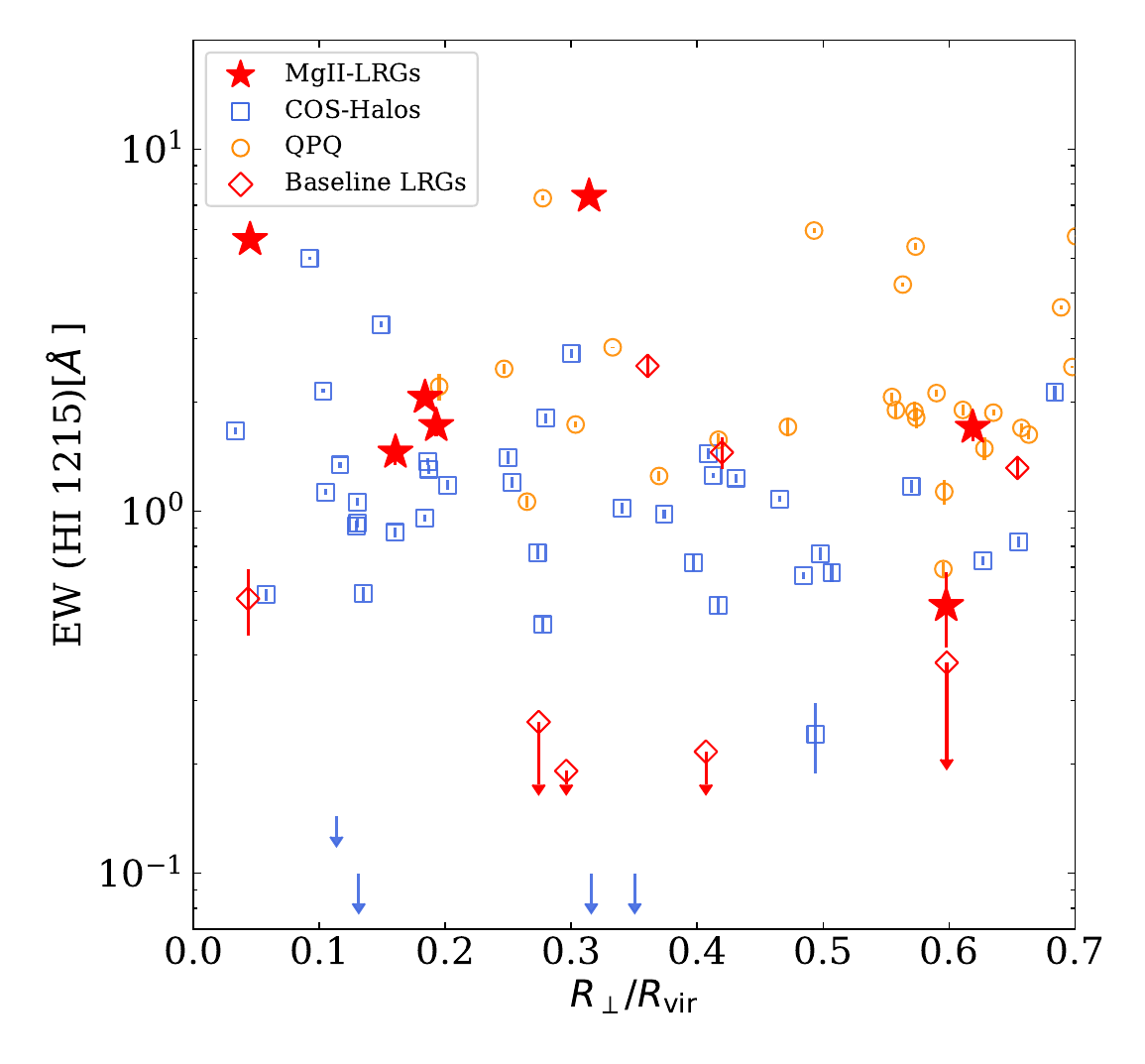}
\vskip -0.2in
\caption{\small 
EW (\ion{H}{1} 1215) as a function of impact parameter, for our MgII-LRGs (red stars), compared to other CGM data. Other CGM data shown on the Figure include: 
COS-Halos (blue open squares), QSO host galaxies (orange open circles), and our Baseline LRGs (red open diamonds). 
Blue line denotes predicted EW(\ion{H}{1} 1215) relation from \citet{Bordoloi18} for average $\log M_{*}$ for COS-Halos. Orange line denotes average EW(\ion{H}{1} 1215) for QSO hosts. 
Some of the data points would be below the y-axes limits. These points are shown as upper limits at a fixed value of -1. 
On the left panel impact parameters are not scaled, while on the right panel impact parameters are scaled to the virial radii. 
From the left panel, one notes that MgII-LRGs and QSO host galaxies show similar CGM properties, and that the CGM of LRGs and QSO host galaxies might be evolutionary connected. 
}
\label{fig:ewhimg}
\end{figure*}

% Similarity of LRGs and QPQ for HI
\citet{White12} predicted that a significant fraction of QSO hosts (massive galaxies at $z\sim 2-3$) evolve into LRGs (massive galaxies at $z\sim 0.5$). Hence, the CGM observed in QSO hosts might be evolutionary related to the CGM observed around LRGs. 
Motivated by this, in Figure \ref{fig:ewhimg} we show EW(\ion{H}{1} 1215) as a function of impact parameter for our MgII-LRGs and QSO host galaxies and, for reference, COS-Halos.  
Since EWs are expected to be correlated with impact parameter, we first correct the observed EWs for impact parameter, as in \ref{sec:himass}. 
Then, we apply a Kolmogorov-Smirnov (KS) test on the corrected EWs for our MgII-LRGs and QSO hosts within 400 kpc (the same range as for LRGs). The KS test gives statistics 0.14 and p-value 0.998, therefore we do not reject the null hypothesis that the distributions of the two samples are the same. When we applied KS test on MgII-LRGs and COS-Halos (or red COS-Halos, with $u - r > 2$ ), the obtained statistic is 0.73 (0.65), and p-value is 0.0014 (0.013), which rejects the hypothesis that our MgII-LRGs and COS-Halos (either whole sample or only red galaxies) originate from the same distribution. 

% => 
This supports the idea that the QSO-CGM and LRG-CGM might be evolutionary connected. 
One possible explanation is, as briefly discussed in \citet{Smailagic18}, that the cool gas around QSOs is heated with time, and the largest cool gas clouds might survive until $z\sim 0.5$ (and be observed as MgII-LRGs CGM), while the smaller clouds might be heated and transformed into hot gas (and be observed as Baseline LRGs without \ion{H}{1} detected), see e.g. \citet{Armillotta17}, \citet{Mandelker20}, \citet{GronkeOh18} and discussion in \citet{Armillotta17}. 
\citet{Butsky20} found that cosmic rays pressure could prevent cool CGM gas from falling into the central galaxy without changing the temperature of this gas \citep[see also][]{Hopkins20}. The authors found that this effect may explain abundant cool gas in the CGM around quenched galaxies.   
The same effect could prevent some of the cool gas clouds around QSOs from falling into the central galaxy.
%%% 
Another possibility is that the strongest QSO-CGM systems evolve into the CGM of MgII-LRGs and that when the strongest QSO-CGM systems evolve to the redshift $z\sim 0.5$, then the distribution of the \ion{H}{1} gas in their CGM is the same as in the overall population of QSOs at $z\sim 2-3$. 

% but: vs R/Rvir 
We note that if impact parameters are scaled to the virial radius, we do not obtain statistically significant results when we compare the distribution of the CGM around MgII-LRGs with the CGM around QSO hosts (and around $\sim$ local $\sim L^{*}$ galaxies). However, if the QSOs' CGM are gravitationally bound, then we might expect that a large fraction of this gas will stay at similar or lower distance from the halo center. 
Since we find that the distributions of the cool CGM around MgII-LRGs and QSO hosts as a function of impact parameter are the same, our results support this scenario, where cool gas clouds stay at approximately the same distance from the central galaxy over a few Gyr. 

% SMGs
We note that SMGs might also evolve into LRGs \citep[see][]{White12}. \citet{Fu16} studied CGM around three SMGs at $\sim 100 - 200$ kpc, and found that their EWs are $\sim 2-3$ \AA , consistent with QSO hosts and our measurements. However, these SMGs do not show metal absorption or optically thick \ion{H}{1} gas in their CGM. In contrast, our LRGs often show metal absorptions in their CGM.

% MgII-LRGs: EW vs R, vs R/Rvir 

\subsection{\ion{O}{6} in MgII-LRG CGM Systems}\label{sec:ovi}  
% Warm-hot gas component

% motivation, OVI vs u-r and mass 
In our sample of MgII-LRGs, five of seven CGM show \ion{O}{6} 1031, three of which have EW(\ion{O}{6} 1031)$ > 0.35$ \AA\ \citep[among which are the two LRGs discussed in][]{Smailagic18}. One further notes that the highest EW(\ion{O}{6} 1031) are detected in the lowest halo masses, for MgII-LRGs, as well as for Baseline LRGs. 

% probability 
Among COS-Halos with EW(\ion{Mg}{2} 2796) $> 0.3$ \AA , five galaxies have $u - r > 2$, and all of them have EW(\ion{O}{6} 1031)$ \lesssim 0.35$ \AA . 
This gives a probability to detect EW(\ion{O}{6} 1031) $ \gtrsim 0.35$ \AA\ along one line of sight for a sample with the same \ion{O}{6} properties as COS-Halos, which is $< 1/5$. If LRGs and COS-Halos originate from the same distribution, a probability to detect EW(\ion{O}{6} 1031)$ > 0.35$ \AA\ in three of seven MgII-LRGs is less than 0.28, which is close to 1 - $\sigma$.
We conclude that the warm-hot gas phases in the CGMs of LRGs and red COS-Halos might have different properties, but we need higher number of galaxies to obtain statistically significant results. One possible explanation is that some of LRGs are surrounded with star-forming neighboring galaxies, and that CGM \ion{O}{6} originates from groups containing both a LRG and a blue star-forming neighboring galaxy.
We discuss this possibility in the next section.

\vspace{10pt}

\section{Environment and Neighboring Galaxies}\label{sec:ng}

\subsection{Selection and Colors}\label{sec:ngs0}

% motivation, introduce the topic 
In this section, we explore the relation between neighboring galaxies and the observed CGM absorption around LRGs. 
All galaxies near the QSO sightline might contribute to the observed CGM absorption. 
\citet{Smailagic18} explored the same question and performed similar analysis; however, the authors discussed only results that were relevant to the two LRGs with extreme CGM, and did not provide results for the rest of the sample. Here we briefly describe the method, which is similar as in \citet{Smailagic18} but with small modifications, and provide results for the whole sample of 15 LRGs. 

% method: search
We searched for all galaxies in SDSS that are projected within 200 kpc of our QSO sightlines. Most of these galaxies have only photometric redshifts available, which we adopted. 
We selected all of these galaxies with photometric redshift consistent with the LRG redshift, i.e., for which $|z_{\rm LRG} - z_{\rm phot}| < \sigma (z_{\rm phot})$, where $z_{\rm phot}$ is the photometric redshift, $z_{\rm LRG}$ is the LRG redshift, and $\sigma (z_{\rm phot})$ is the error in the photometric redshift. 
We will refer to these galaxies as neighboring galaxies (NGs). We found 0 - 6 NGs around each LRG. 

% method: k-correction
Next, we calculated k-corrected magnitudes and its errors by using the function {\it sdss\_kcorrect} from Michael Blanton's IDL package {\it kcorrect}\footnote{\url{http://kcorrect.org/}} \citep[version 4.3;][]{Blanton03}. As input, we use SDSS Petrosian {\it ugriz} fluxes, its inverse variances, and extinctions in {\it ugriz} bands, and we assume that all NGs around some LRG-QSO sightline have redshifs the same as the LRG's redshift. 

% method: colors
For each galaxy we calculate its color. Since LRGs are typically found in dense environments \citep{Padmanabhan07}, such as groups of galaxies, and are typically quenched \citep{Banerji10}, we expect that most of their NGs will have red colors, but blue colors would not be unexpected \citep[see e.g.,][]{Kovac14}. We define blue and red galaxies as defined in \citet{Gavazzi10}. The authors studied galaxies in the local Universe, and found that on the $g - i$ color - $i$ magnitude diagram red and blue galaxies could be separated with a line: $g - i = -0.0585 (M_{i} + 16) + 0.78$, where $M_{i}$ is absolute $i$ magnitude. 
We define $C_{gi}$ color as:
\begin{equation}
C_{gi} = -0.0585 (M_{i} + 16) + 0.78 - (g-i).
\end{equation}
Here, we use k-correction output absolute $g$ and $i$ magnitudes. 
When $C_{gi}$ color is greater (smaller) than zero, galaxies are defined as blue (red).
We confirm that in all cases when LRGs satisfy our selection criteria for NGs, LRGs will have red $C_{gi}$ colors.  

% method: colors: motivation
We choose to use the described definition to separate blue and red galaxies because $u$ magnitudes often have large errors.
However, errors in $C_{gi}$ colors are still large for some NGs. 
Specifically, redder galaxies are more likely to have faint bluer magnitudes, and hence larger errors in the $C_{gi}$ color. For this reason, in the further text, we define galaxies as blue only if the $C_{gi}$ color is greater than twice the error in the $C_{gi}$ color. We mark other galaxies as red. 

\vspace{10pt} 

\begin{figure}[h]
\centering
\includegraphics[width=0.49\textwidth]{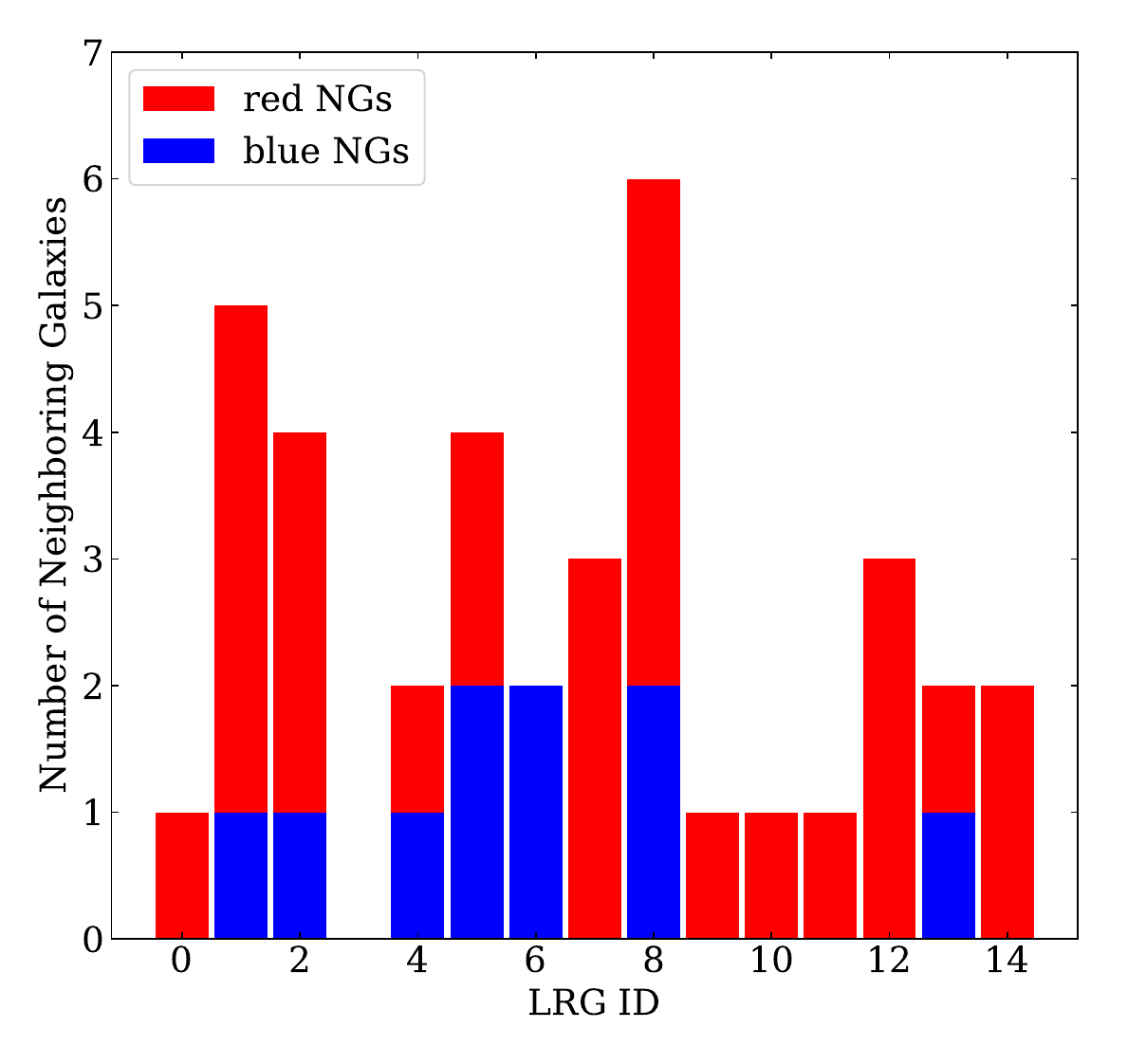} 
\caption{\small
Number of neighboring galaxies (NGs) inside $< 200$ kpc from the QSO sightlines. 
Red and blue NGs are labeled in red and blue color, respectively (see text for the definition that we use). Different bins correspond to different LRGs. LRGs are sorted from the maximum to the minimum EW(\ion{H}{1} 1215). 
ID numbers from 0 to 14 correspond to the following LRGs: LRG\_1520+2534 (0), LRG\_1059+4039 (1), LRG\_1102+4543 (2), LRG\_1144+0714 (3), LRG\_1121+4345 (4), LRG\_1440-0157 (5), LRG\_1351+2753 (6), LRG\_1217+4931 (7), LRG\_0855+5615 (8), LRG\_1237+1106 (9), LRG\_1106-0115 (10), LRG\_0226+0014 (11), LRG\_1549+0701 (12), LRG\_1251+3025 (13), LRG\_1306+3421 (14).
NGs do not include LRGs.
In our sample, LRGs have 0-6 NGs, which includes both blue and red galaxies. It could be also seen from the figure that there is no a correlation between EW(\ion{H}{1} 1215) and the number of NGs or blue NGs.
}
\label{fig:ngs}
\end{figure}

% number of blue and red NGs
Figure \ref{fig:ngs} shows a histogram with the number of red and blue NGs within 200 kpc from the QSO sightlines. A full list of NGs is shown in Appendix \ref{sec:appc}. 
The Figure shows that around most of the LRGs we find at least one NG, and the number of NGs around each of our LRGs is $\leq 6$. The number of NGs that are blue is $0-3$. We note that in some cases the number of blue NGs exceeds the number of red NGs. It could be also seen from the figure that there is no a correlation between EW(\ion{H}{1} 1215) and the number of NGs or blue NGs.

% note about NGs with spectroscopic redshift
Next, we searched for NGs with spectroscopic redshifts, and found two such galaxies, both of which are LRGs:

1) LRG\_1059+4039: another LRG was found at $\sim 200$ kpc, and its redshift is offset from the target LRG by $200$ km s$^{-1}$; 

2) LRG\_1549+0701: another LRG was found at 1793 km s$^{-1}$ from the targeted LRG. 
This  another LRG does not have photometric redshift consistent with the target LRG, and was not included in our sample of NGs. Due to a relatively large velocity difference, we do not include it in our analysis. 

% additional NG 
\citet{Smailagic18} found one additional galaxy close to the QSO\_1059+4039 sightline. Based on Panoramic Survey Telescope and Rapid Response System (Pan-STARRS) images, we assume that this galaxy has red color.

% NGs figure

\subsection{The Role of Blue NGs}

% OVI: motivation 
We now investigate connection between the environment of LRGs and the detection of \ion{O}{6} in the CGM. 
Blue and red galaxies are typically star-forming and passive galaxies, respectively.
Previous work by \citet{Tumlinson11} found that the CGM of star-forming $\sim L^{*}$ $z\sim 0.2$ galaxies often shows strong \ion{O}{6} absorption, while passive galaxies rarely have \ion{O}{6} detected in their CGM, and that \ion{O}{6} column densities are correlated with specific star-formation rates in host galaxies.  
Later, theory work by \citet{Oppenheimer16} found that one possible explanation is that different halo masses could explain this correlation.  
These authors found that the host halos of star-forming galaxies have the virial temperature at which the \ion{O}{6} ionization fraction reaches maximum, while passive galaxies are typically located in more massive halos. These more massive halos have oxygen mostly in more highly ionized states.  
However, in contrast to these results, we find \ion{O}{6}, even strong, in the CGM around some of massive quenched galaxies, or LRGs, in our sample. 
This finding motivated us to study if such LRGs have more blue galaxies in their surroundings, than the ones without \ion{O}{6}. 

% OVI:results: numbers 
In our sample, LRGs with blue NGs show \ion{O}{6} in 5 of 7 cases. In contrast, only one of 8 LRGs without blue NGs shows \ion{O}{6}. In addition, this one LRG without blue NGs that shows \ion{O}{6} has the lowest \ion{O}{6} among all detections in our sample, and it does not have red NGs either. 
In addition, we note that all red COS-Halos galaxies with EW(\ion{O}{6} 1031) greater than 0.24 \AA\ have blue NGs. 
These results indicate that there is a possible correlation between NGs and the LRG-CGM \ion{O}{6}. 

\begin{figure*}[t!]
\centering
\includegraphics[width=0.7\textwidth]{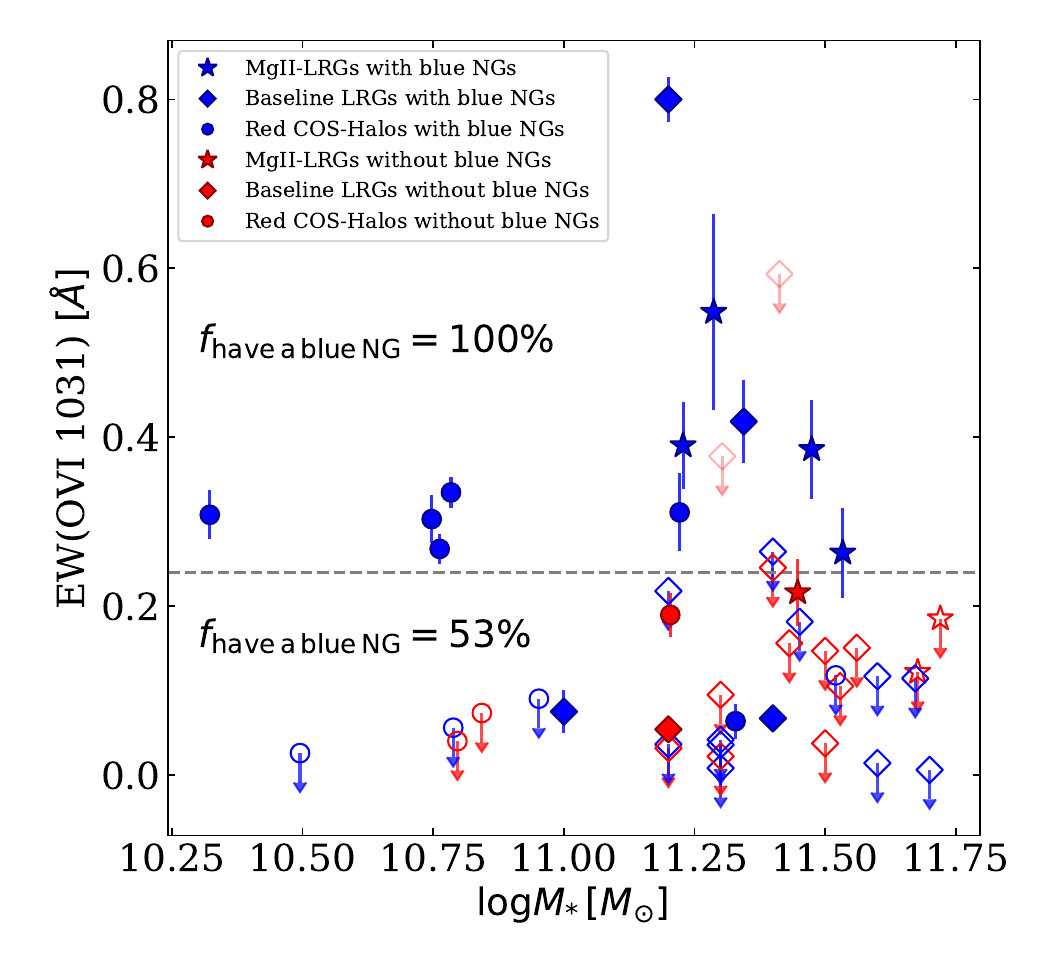} 
\includegraphics[width=0.29\textwidth]{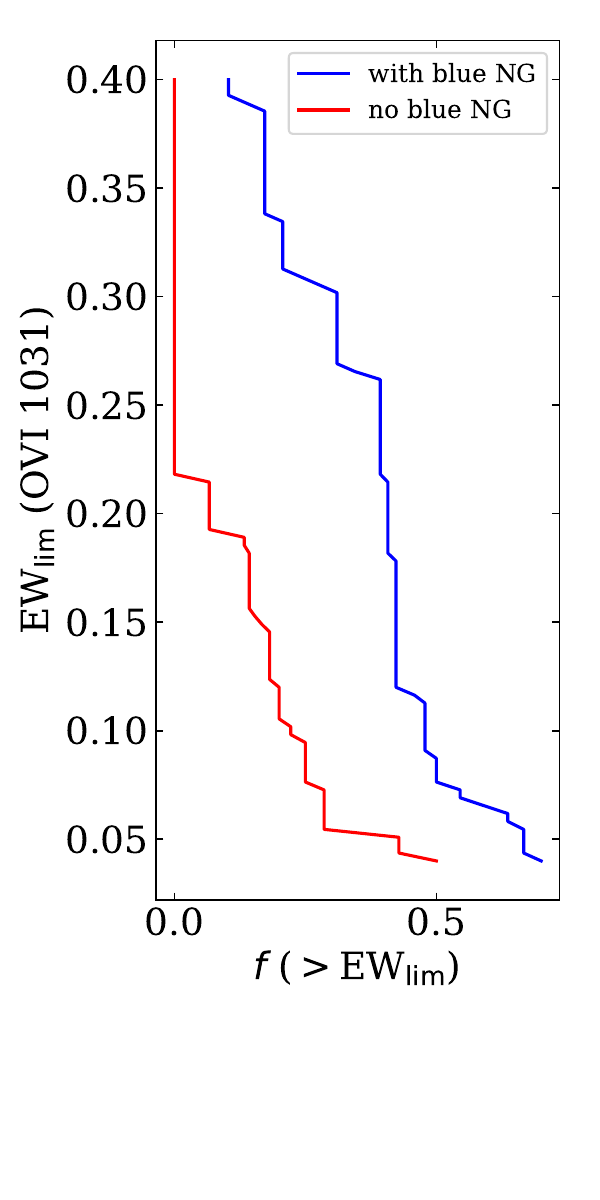} 
\caption{\small
{\textit Left:} Equivalent widths EW(\ion{O}{6} 1031) versus stellar mass for LRGs and red COS-Halos galaxies, with and without at least one blue NG.
Our Baseline and literature LRGs are marked with diamonds, while MgII-LRGs are marked with stars, and COS-Halos are marked with circles. Blue and red symbols correspond to galaxies with at least one and without blue NGs, respectively. 
Measurements for blended \ion{O}{6} transitions are shown in fainter colors. 
Typically, galaxies with strong CGM \ion{O}{6} absorption have a blue NG. 
All ($100 \%$) galaxies with measured EW(\ion{O}{6} 1031) $> 0.24$ \AA\ that are not upper limits have a blue NG, while for galaxies with EW(\ion{O}{6} 1031) $< 0.24$ \AA\ the percent of galaxies that have a blue NG is $52 \%$, or twice smaller. 
Gray dashed line denotes EW(\ion{O}{6} 1031) $= 0.24$ \AA .  
{\textit Right:} Fraction of galaxies with (blue line) and without (red line) blue NGs that show EW(\ion{O}{6} 1031) greater than EW$_{\rm lim}$, for different EW$_{\rm lim}$, in the combined sample of LRGs and red COS-Halos galaxies. One notes that for all EW$_{\rm lim}$ between 0.05 and 0.4, the fraction of galaxies with EW(\ion{O}{6} 1031)$ \, > \rm EW_{\lim}$ is higher for galaxies with blue NGs than for galaxies without blue NGs.
}
\label{fig:ngsblue2}
\end{figure*}

% OVI: what we did
We combine our sample of 15 LRGs with literature LRGs and red COS-Halos galaxies and divide the combined sample in two subsamples, one with blue NGs, and the other without. In the combined sample, 29 of 47 galaxies have a blue NG inside 200 kpc, and in our sample of 15 LRGs, 7 LRGs have blue NGs.  
We note that two of blue NGs around one of COS Halos galaxies (J0928+6025) have a confirmed spectroscopic redshift in \citet{Werk12}.
In the left panel of Figure \ref{fig:ngsblue2}, we show EW(\ion{O}{6} 1031) versus stellar mass for LRGs (including our sample and literature data) and red COS-Halos, with and without blue NGs. One notes that 11 galaxies have the strongest \ion{O}{6} detected, with EW(\ion{O}{6} 1031) greater than 0.24 \AA , and that in all of these cases there is a blue NG close to the QSO sightline. 
In contrast, among the galaxies with EW(\ion{O}{6} 1031) smaller than 0.24 \AA , approximately half have a blue NG. 
If EW(\ion{O}{6} 1031) distribution is the same for galaxies with and without blue NGs, the probability that all of the 11 galaxies that show the strongest EW(\ion{O}{6} 1031) have blue NGs is 0.0020.
The probability that at least one of these galaxies does not have a blue nearby NG would be 0.9980, or $\sim 3 \sigma$. 
In the right panel of Figure \ref{fig:ngsblue2}, we show the fraction of galaxies with and without blue NGs that show EW(\ion{O}{6} 1031) greater than EW$_{\rm lim}$, for different values of EW$_{\rm lim}$, in the combined sample of LRGs and red COS-Halos galaxies. For all values of EW$_{\rm lim}$ between 0.05 and 0.4, the fraction of galaxies with EW(\ion{O}{6} 1031)$ \, > \rm EW_{\lim}$ is higher for galaxies with blue NGs than for galaxies without blue NGs.
These results support the scenario in which the observed \ion{O}{6} around quenched galaxies originates from 
groups containing both a (massive) quenched galaxy and a blue star-forming neighboring galaxy. The \ion{O}{6} might originate from the blue star-forming neighboring galaxy or from environmental effects in such groups.
% Groups - similar 
We find a similar conclusion for a sample of 12 sightlines that probe gas inside $z\sim 0$ groups \citep{Stocke19}: in this sample one moderately strong \ion{O}{6} system is detected, and that sightline contains the brightest composite galaxy (a galaxy that shows both emission and absorption lines) inside 200 kpc from the sightline. 

Our results are also in agreement with a recent work of \citet{Tchernyshyov22}, where the authors found that among galaxies with the same stellar or halo mass, \ion{O}{6} is more commonly detected in the CGM around star-forming galaxies than around passive galaxies. \citet{Tchernyshyov22} discussed two possible explanations for the detection of the \ion{O}{6} in the CGM around two passive galaxies from their sample, one was that the \ion{O}{6} is associated with another star-forming interloping/neighboring galaxy, and the other was that galaxies that show \ion{O}{6} in their CGM were quenched more recently and their CGM was still not transformed. Our results show that the former scenario might be more common.   
%blue NGs  

\vspace{12pt}

%% blue ngs figure

% HI and MgII results
For comparison, for EW(\ion{H}{1} 1215) and EW(\ion{Mg}{2} 2796) we do not find similarly clear trends. For example, in our sample of 15 LRGs, the number of LRGs with and without blue NGs that show \ion{H}{1} 1215 are comparable, and are 6 of 7 (86\%) and 5 of 8 (63\%), respectively. 
The same numbers for \ion{Mg}{2} 2796 are also comparable, and are 4 of 7 (57\%) and 3 of 8 (38\%). 
In contrast, for \ion{O}{6} we obtain 71\% and 13\% . 
In the combined sample of LRGs and red COS-Halos, among 11 galaxies which show the strongest EW(\ion{H}{1} 1215) (and EW(\ion{Mg}{2} 2796)), 7 have blue NGs. 
If the EW(\ion{H}{1} 1215) distribution is the same for galaxies with and without blue NGs, the probability that among 11 galaxies with the strongest EW 7 have blue NGs is 0.28, much higher than we obtained for \ion{O}{6}. 
However, despite the lack of statistically significant results, we note that the strongest two \ion{Mg}{2} absorbers in our sample of 15 LRGs are found around LRGs with blue NGs, and that two of three LRGs with more than one red NGs and no blue NGs do not show any CGM lines detected. A larger sample (and spectroscopic redshifts) might show more statistically significant results if \ion{H}{1} and \ion{Mg}{2} are correlated with blue NGs.

\subsection{The Role of the Overall NGs Population}

% previous work 
\citet{Smailagic18} found extreme \ion{C}{3} absorption in the CGM around two LRGs. The authors found that the measured EW and velocity extent are in agreement with their predictions for the combined EW and the combined velocity extent for LRG and its NGs, if they are seeing a superposition of the CGM systems. In addition, the authors found that the two LRGs with extreme CGM are two out of three LRGs with the highest number of NGs inside 100 kpc. 

% CIII 
We calculate the probability that the two LRGs with extreme CGM traced by strong \ion{C}{3} absorption would be located in the two of three regions with the highest number of galaxies in our sample of 15 LRGs. We found that the probability is 0.029, indicating that more overdense regions might contain more metals or have higher Doppler parameter. However, we note that when we applied generalized Kendall's Tau test (BHK method), we do not find a statistically significant correlation between the number of NGs inside 100 or 200 kpc, and EW(\ion{C}{3} 977), EW(\ion{H}{1} 1215), or EW(\ion{Mg}{2} 2976). 

% NGs: multiple components 
We also note that the two LRGs with identified separate components at large $\delta v$ (LRG\_1059+4039 and LRG\_0855+5615) are two of three LRGs with the highest number of NGs inside 100 kpc, and the probability for this is again 0.029, which is $\sim 1 - 2 \sigma$, indicating that more overdense regions might show multiple separate components from their CGM, and that multiple galaxies might contribute to the observed CGM absorption, which supports the \citet{Smailagic18} conclusion. 

% predicted EW (from LRG and NGs)
We also predicted EW(\ion{H}{1} 1215), based on the combined EW from all NGs from \citet{Bordoloi18}, as \citet{Smailagic18} did for two LRG-CGM systems. However, we do not find a clear correlation, and in some cases we obtain higher, and in other lower EW(\ion{H}{1} 1215) than observed. For example, inside 200 kpc around the QSO-LRG with the highest EW(\ion{H}{1} 1215), we find only one red NG, and the predicted EW(\ion{H}{1} 1215) $\sim 1.1$ \AA , which is few times smaller than the observed EW(\ion{H}{1} 1215) $\sim 7.4$ \AA . This indicates that LRG-CGM systems might have more complex origins. 

\vspace{10pt}

\section{Conclusions}\label{sec:conc}

In this work, we study the circumgalactic gas around 15 LRGs at redshift $z\sim 0.5$, and with impact parameters up to 400 kpc, 7 of which have already detected \ion{Mg}{2} in their CGM (MgII-LRGs) and 8 without \ion{Mg}{2} (Baseline LRGs).  
We obtained HST COS FUV spectra of background QSOs. The available spectral lines include transitions in \ion{H}{1}, \ion{C}{3}, \ion{C}{2}, \ion{O}{6}, \ion{Si}{2}, \ion{Si}{3}, and other. 
This is the first study of \ion{H}{1} and UV metal lines in a sample of LRGs with strong \ion{Mg}{2} already detected (previous studies focused on a general population of LRGs, without previous knowledge about the detection of \ion{Mg}{2}); and the first presentation of our full sample of the CGM around 15 LRGs. 
We found that the covering fraction of \ion{H}{1} with EW(\ion{H}{1} 1215)$> 0.3$ \AA\ is $\sim 0.56^{+11}_{-12}$ and that metals are commonly detected (three of eight Baseline LRG-CGM systems show a metal absorption line). 

Our main results are summarized below:

1) While for $\sim L^{*}$ and $< L^{*}$ galaxies, CGM \ion{H}{1} 1215 absorption is generally stronger for more massive galaxies, cool CGM traced by \ion{H}{1} 1215 is suppressed above stellar masses of $\sim 10^{11.5} M_{\sun}$.  

2) CGM around MgII-LRGs and CGM around QSO host galaxies show similar EW(\ion{H}{1}) absorptions at similar impact parameters, and we could not reject the null hypothesis that the distributions of the two samples are the same. This implies that the CGM of LRGs and QSO host galaxies might be evolutionary connected. 

3) On average, LRG-CGM systems show weak \ion{O}{6}, if detected. However, some fraction of LRGs shows \ion{O}{6} 1031 higher than around red $\sim$ L$^{*}$ $z\sim 0.2$ galaxies. While EW(\ion{O}{6}) is $< 0.35$ \AA\ around all red COS-Halos galaxies, we find EW(\ion{O}{6}) greater than 0.35 \AA\ around one Baseline LRG and three MgII-LRGs. 

4) The \ion{O}{6} detected in the CGM around LRGs and other quenched galaxies might originate from groups containing both a LRG and a blue (star-forming) neighboring galaxy (NG).
In the combined sample of our LRGs, literature LRGs, and red COS-Halos, CGMs around 11 galaxies show the strongest \ion{O}{6}, with EW(\ion{O}{6} 1031) greater than 0.24 \AA , and in all of these cases there is a blue NG close to the QSO sightline. 
In contrast, among the galaxies with EW(\ion{O}{6} 1031) smaller than EW$_{\rm lim} = 0.24$ \AA , approximately half have a blue NG. If EW(\ion{O}{6} 1031) distribution is the same for galaxies with and without blue NGs, the probability that all of the 11 galaxies which show the strongest EW(\ion{O}{6} 1031) have blue NGs is 0.0020. The probability that at least one of these galaxies does not have a blue nearby NG would be 0.9980, or $\sim 3 \sigma$. For all EW$_{\rm lim}$ between 0.05 and 0.4, the fraction of galaxies with EW(\ion{O}{6} 1031)$ \, > \rm EW_{\lim}$ is higher for galaxies with blue NGs than for galaxies without blue NGs.

5) The velocity extent of the CGM absorptions is within the estimated escape velocity of LRG halos, implying that the CGM gas is gravitationally bound to the halos. However, some of LRG-CGM systems have additional components with velocity similar to or larger than the escape velocity.

6) The $N_{HI}$ is anti-correlated with the impact parameter scaled to the virial radius, implying that the \ion{H}{1} is associated with LRGs halos. The generalized Kendall’s Tau test gives correlation coefficient -0.41 with significance $\sim 2.8 \sigma$. 

7) Strong \ion{Mg}{2} in LRG-CGM systems is associated with high $N_{HI}$.

\vspace{10pt}

\begin{acknowledgments}

 \begin{center}
\textit{Acknowledgments}     
 \end{center}

 Support for program GO-14171 was provided by NASA through a grant from the Space Telescope Science Institute, which is operated by the Association of Universities for Research in Astronomy, Inc., under NASA contract NAS 5-26555. We would like to thank Brice M{\' e}nard and Sanchayeeta Borthakur for contribution to this project and G{\' a}bor Worseck for assisting with data reduction. We would also like to thank Jessica Werk, Kirill Tchernyshyov, Matthew Wilde, Nicolas Tejos, Tobias Westmeier and Glenn Kacprzak for helpful discussion. 
\end{acknowledgments}

\facilities{HST(COS), Sloan} 
\software{
Astropy \citep{astropy13, astropy18}}, 
CALCOS \citep[2.21;][]{calcos},
COS-REDUX \citep{cosredux},
Kcorrect \citep[4.3;][]{Blanton03},
Linetools \citep{linetools},
Matplotlib \citep{Hunter07},
Numpy \citep{Oliphant06},
Pyasurvpart \citep{pyasurvpart},
Pyigm \citep{pyigm},
Scipy \citep{Virtanen20}, 
Statsmodels \citep{statsmodels}

%%%%%%%%%%%%%%%%%%%%%%%%%%%%%%%%%%%%%%%%

\clearpage

\appendix

\section{Customized Data Reduction}\label{sec:app1}

% 0) General information
We reduced the data by using CALCOS v2.21 and our own custom Python codes COS\_REDUX. 
% \edit1{COS\_REDUX}
%\deleted{\footnote{\url{https://github.com/pypit/COS_REDUX}, or in the future \url{https://github.com/pypeit/COS_REDUX}}} 
Below we describe our customized data reduction in more details. We use the same notation as it was used in the COS Data Handbook (\url{https://hst-docs.stsci.edu/cosdhb}).

Based on \citet{Worseck16} and our private communication with G{\' a}bor Worseck, our custom Python code combined with CALCOS performs a customized data reduction, which includes customized dark subtraction and co-addition of sub-exposures. 
CALCOS v3.1. uses a new TWOZONE extraction algorithm for data taken at Lifetime Position 3, while CALCOS v2.21 used a boxcar extraction. 
The previously used boxcar algorithm rejects every wavelength bin that contains a bad pixel in the extraction region. The TWOZONE algorithm divides the extraction region into an inner and an outer zone and rejects a wavelength bin only if a bad pixel is located in the inner zone. The boundaries of these two zones are defined such that each includes a fraction of the total flux, and are wavelength dependent.
However, since the TWOZONE extraction is difficult to apply to dark frames, we chose, as did \citet{Worseck16}, to use CALCOS v2.21. 
We applied our data reduction steps separately to segments A and B. 

As a part of our customized data reduction, we found object trace and defined a rectangular extraction region in the combined corrtag file for a given segment. We did not apply flat fielding. We restricted pulse height amplitudes (PHA) to the range of [2,11]. 

We found darks taken within 90 days around the date of our science observations, with the detector voltages (HVLEVELA and HVLEVELB) the same as in the science corrtag files, and defined background regions in corrtag files and in darks. 
Next, for each dark, we performed Kolmogorov-Smirnov (KS) test on the cumulative PHA histograms for the science and the dark frame, and selected all darks with the KS test probability PK\_S $>$ 0.1. When the number of the selected darks was too low ($< 4$), we lowered the criterion. 
Finally, we coadded all of the selected darks, and made a histogram with $16385$ bins of the shifted XFULL (shifts are defined in the science file header as SHIFT1A or SHIFT1B), for the pixels where YFULL is inside the aperture of the extraction region, with good data quality ($\rm DQ = 0$), and with PHA between the minimum and the maximum selected value for the science frames. 
The final dark spectrum was obtained by scaling the obtained spectrum to the number of points in the background regions of the science frame, and smoothing by 500 pixels. 

We coadded x1d sub-exposures, and binned them by 2 pixels. 
First, we selected pixels where $\rm DQ=0$, flux is greater than 0, and wavelength is less than 2100 \AA . For these pixels, we found WAVELENGTH and CALIB, where $\rm CALIB = NET / FLUX$, and NET and FLUX are the net count rate and the (default calculated and not precise) CALCOS pipeline flux.
Then, we concatenated the data for all sub-exposures, and sorted them. 
Next, we interpolated CALIB over the wavelength. 
We defined total counts (TOTALCOUNTS), as the sum of counts from the sub-exposures, multiplied by DQ\_WGT. 
Here, DQ\_WGT are data quality weights, that are equal 0 or 1. 
We defined TOTALDARK in a similar way.    
Finally, we calculated flux as 
\begin{equation}
\rm FLUX = \frac{TOTALCOUNTS-TOTALDARK}{CALIB \times TOTALTIME},
\end{equation}
where TOTALTIME is the sum of DQ\_WGT $\times$ EXPTIME over all subexposures.
We also provided a simple estimate of the error in the flux as
\begin{equation}
\rm ERROR = \frac{TOTALCOUNTS^{1/2}}{CALIB \times TOTALTIME}.
\end{equation}

Finally, we coadded FUVA and FUVB spectra. 
We note that for one of the LRGs (see Table \ref{tab:qsotab}) there are two visits. However, since one of these visits has much better $S/N$ than the other, the combined spectrum would marginally improve the $S/N$, and we decided to use only the visit with the better $S/N$. 

\vspace{20pt}

\section{Further analysis for less reliable lines}\label{sec:appb} 

% OVI - motivation
In 8 LRG-CGM systems, we find an absorption line at wavelength that corresponds to \ion{O}{6} 1031 at the LRG-CGM's redshift. However, such lines could also be interloping lines from another redshift, and not associated with LRGs. In addition, previous studies found that \ion{O}{6} is not commonly detected around red galaxies. In most cases when a potential \ion{O}{6} 1031 is detected, \ion{O}{6} 1037 is not detected at a significance level higher than $3 \sigma$. 

\begin{figure*}[t!]
\centering
\includegraphics[width=0.49\textwidth]{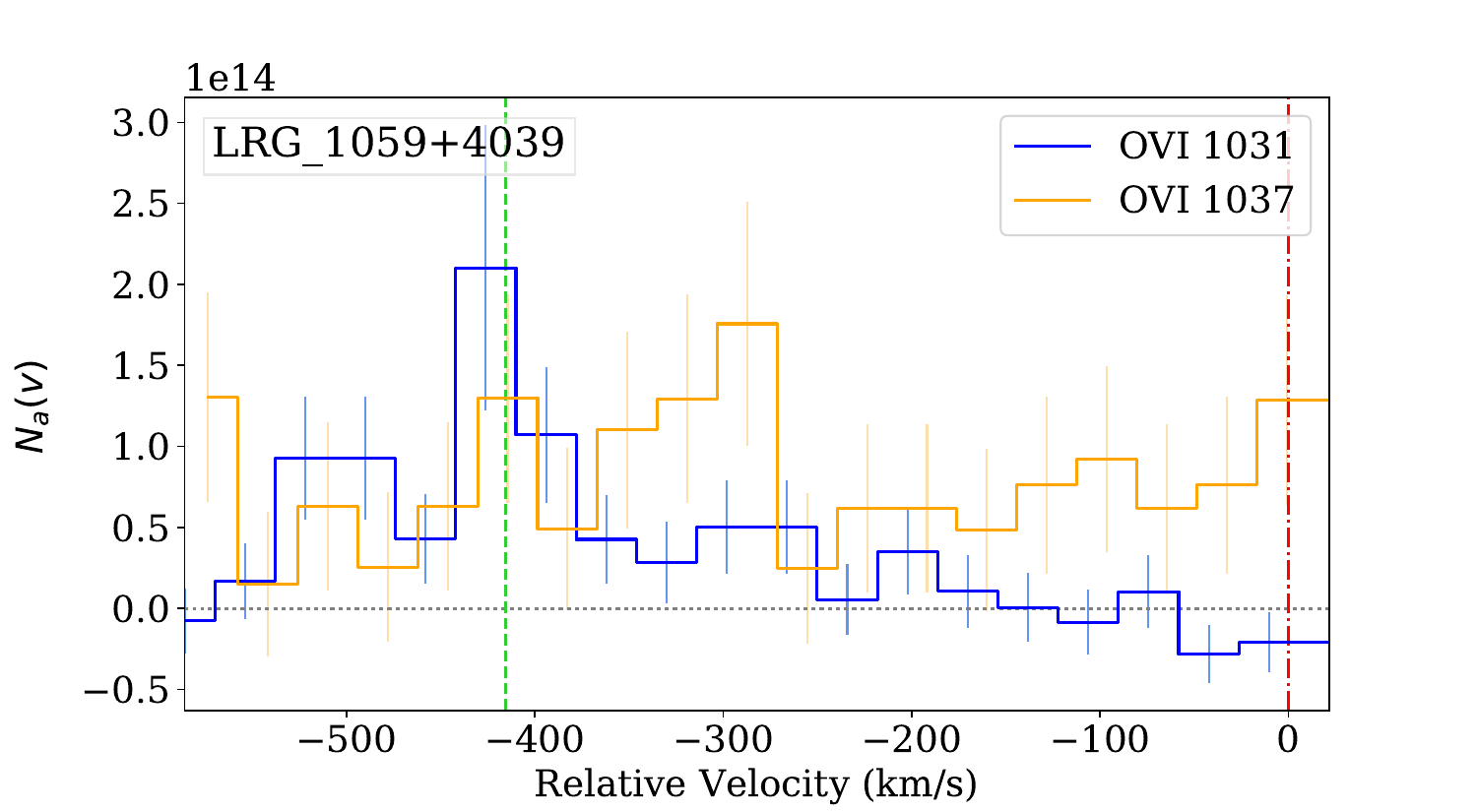}
\includegraphics[width=0.49\textwidth]{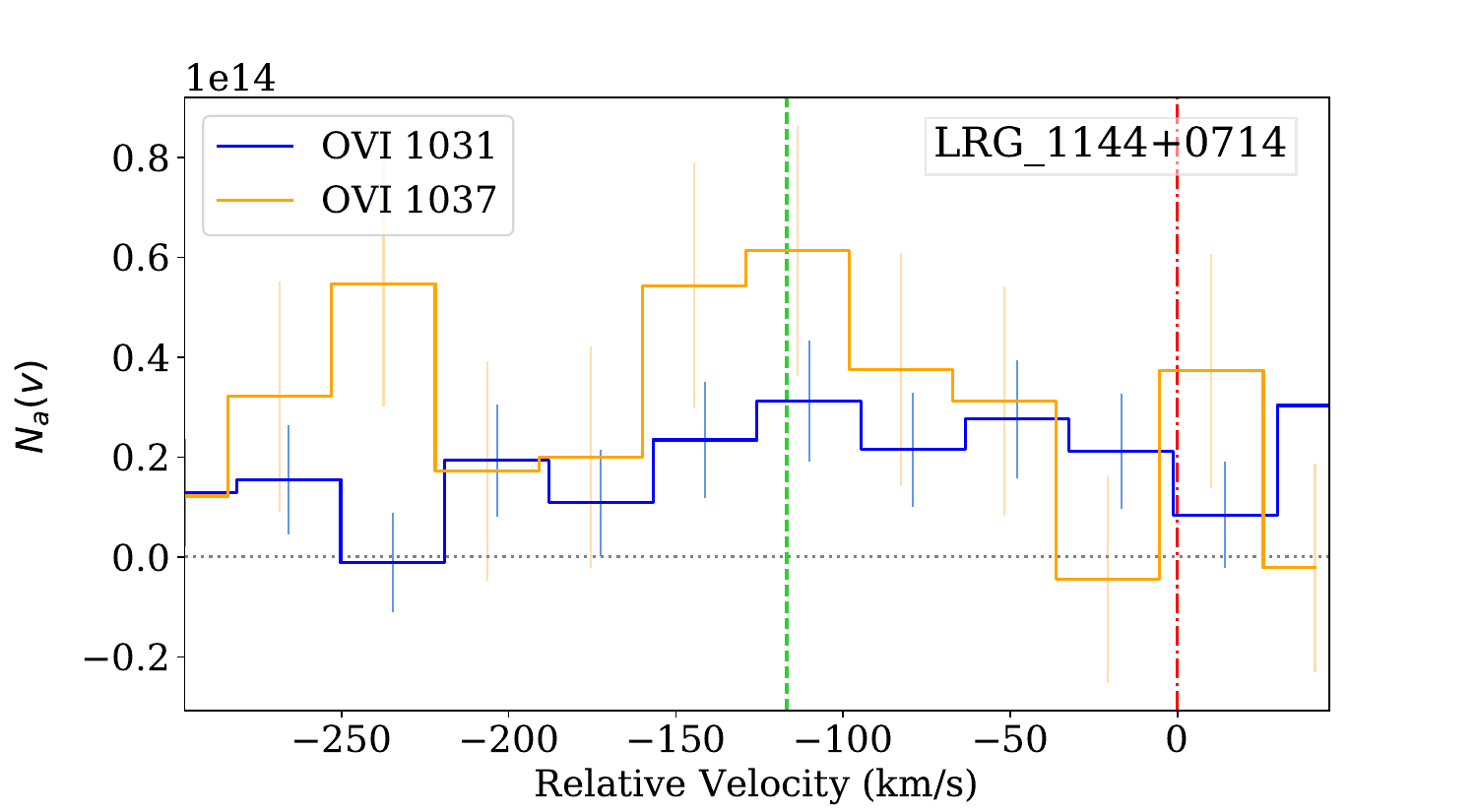}
\includegraphics[width=0.49\textwidth]{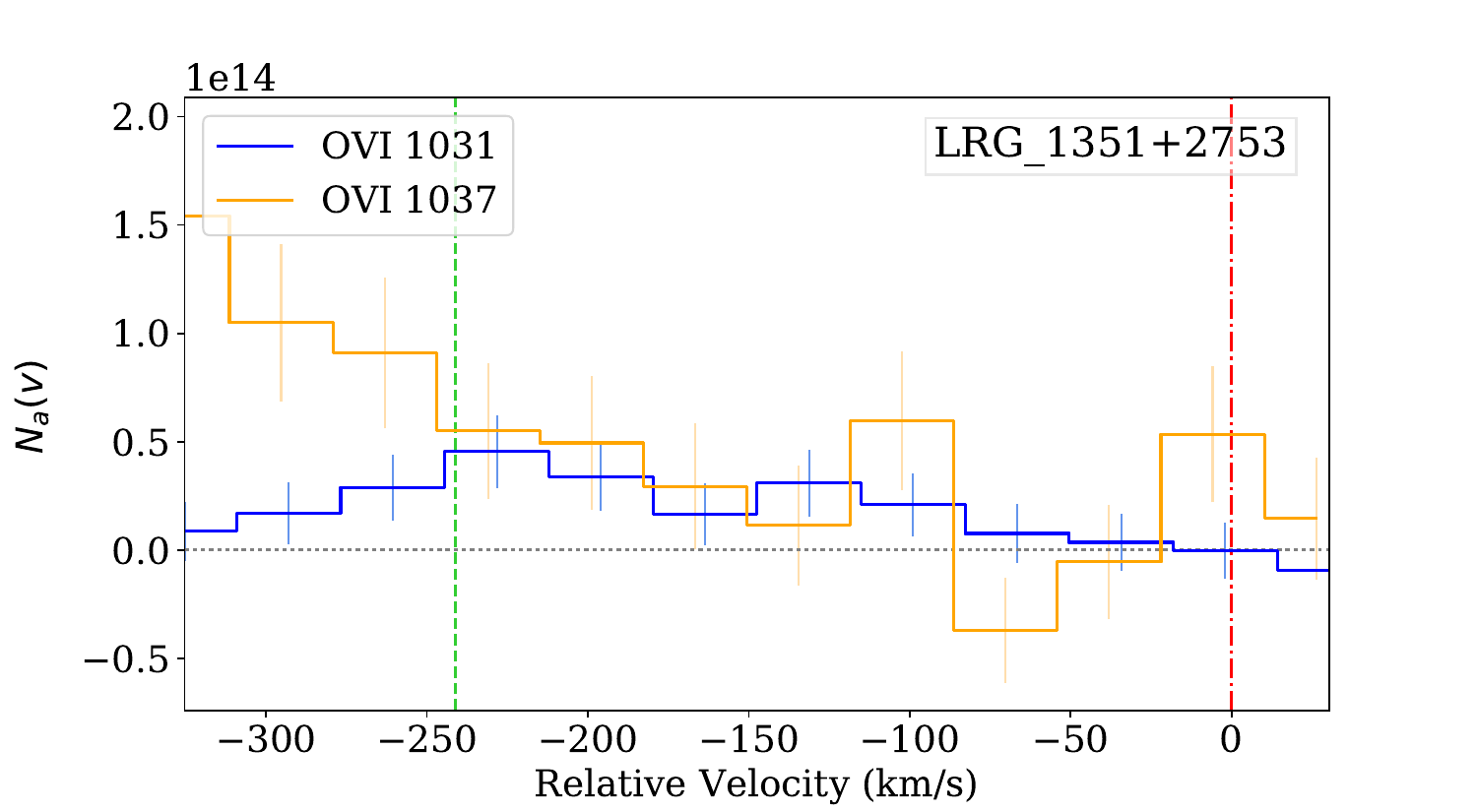}
\includegraphics[width=0.49\textwidth]{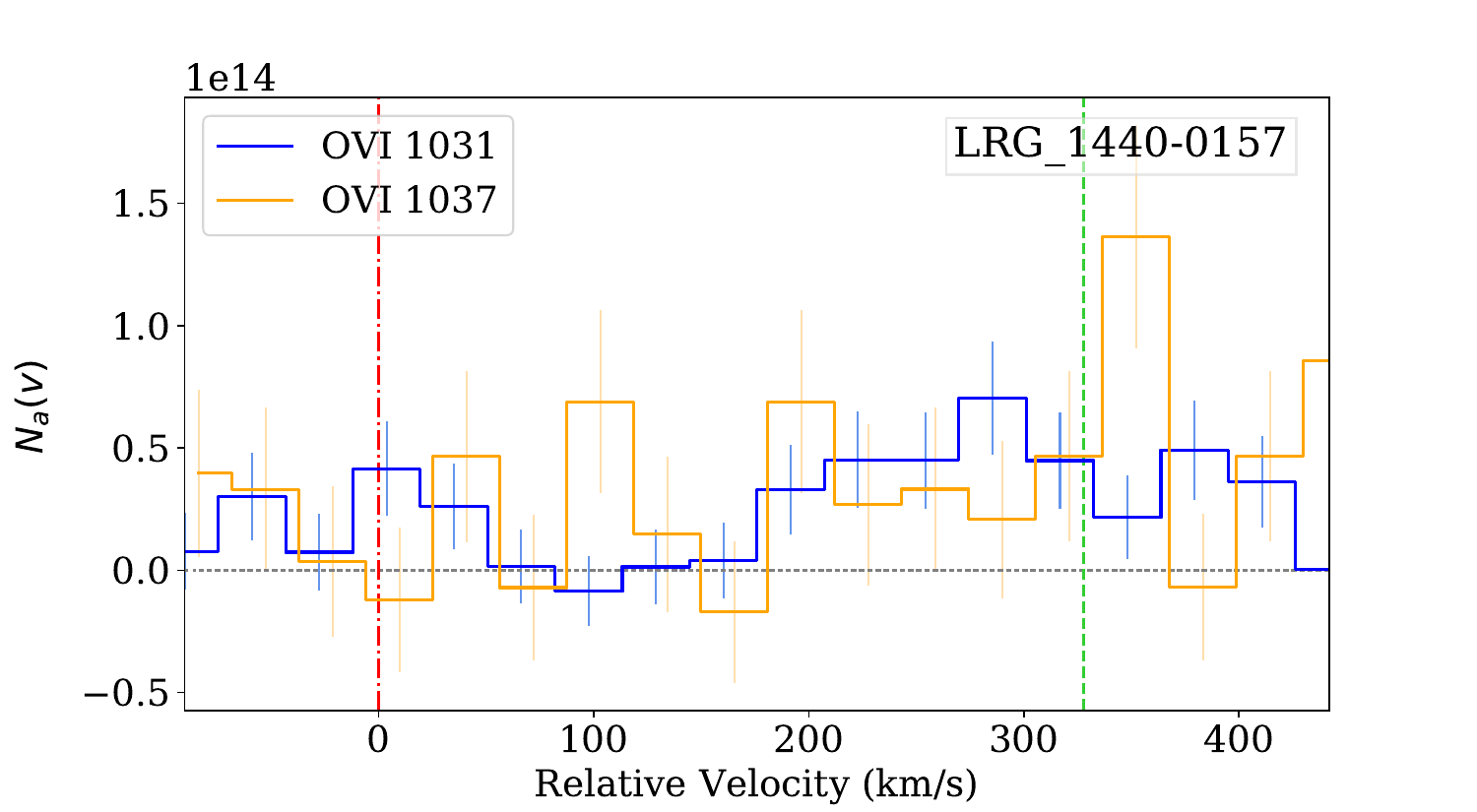}
\includegraphics[width=0.49\textwidth]{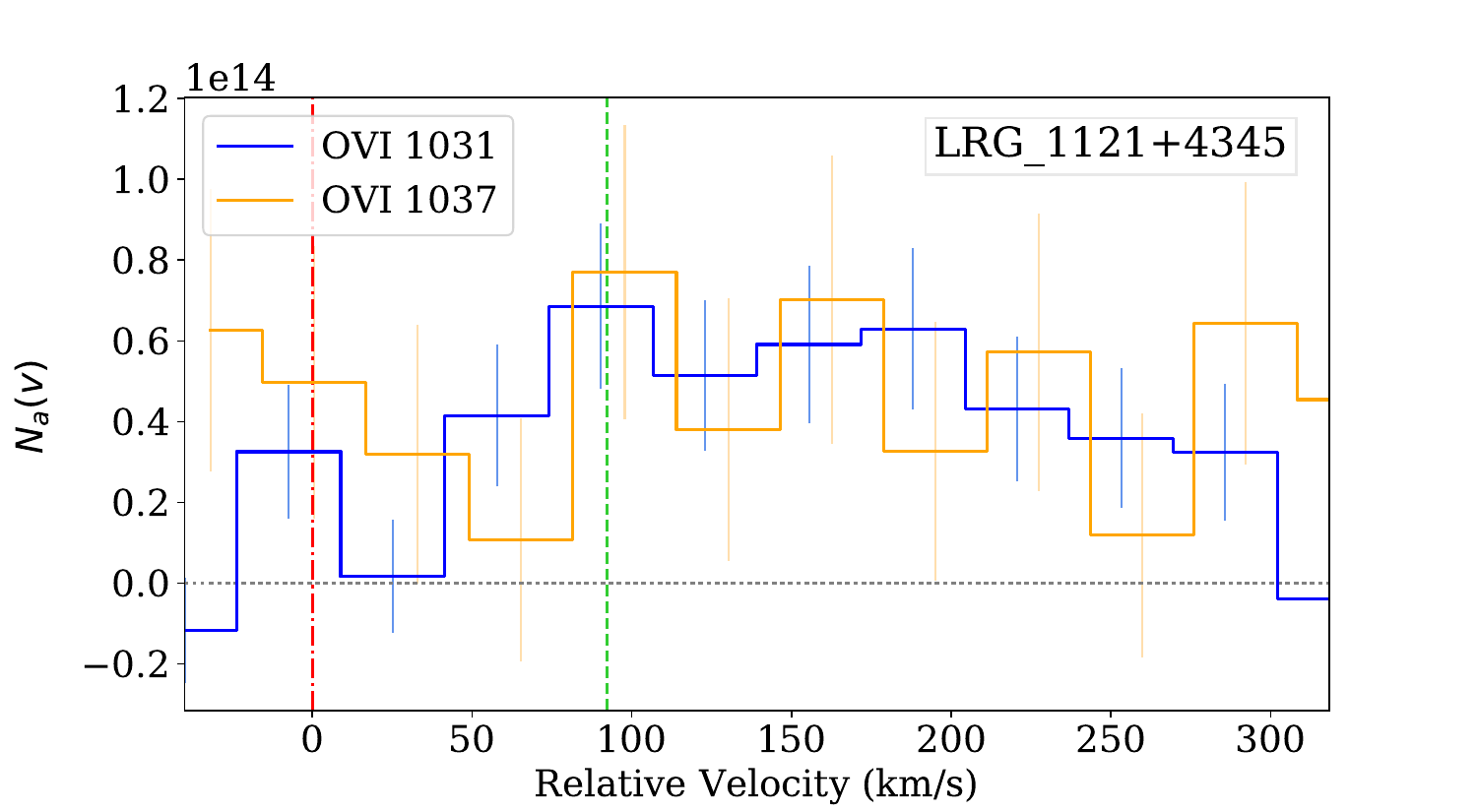}
\includegraphics[width=0.49\textwidth]{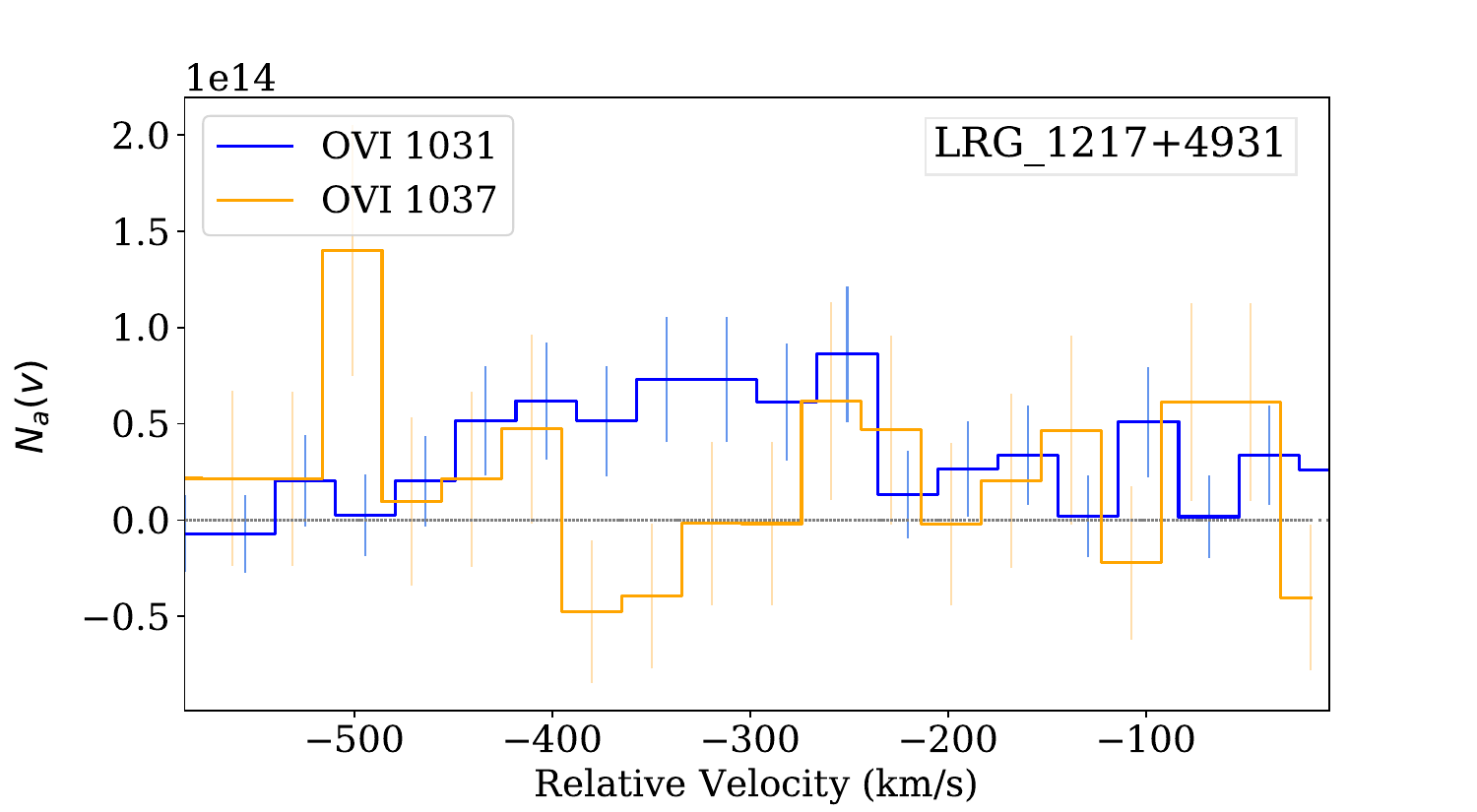}
\includegraphics[width=0.49\textwidth]{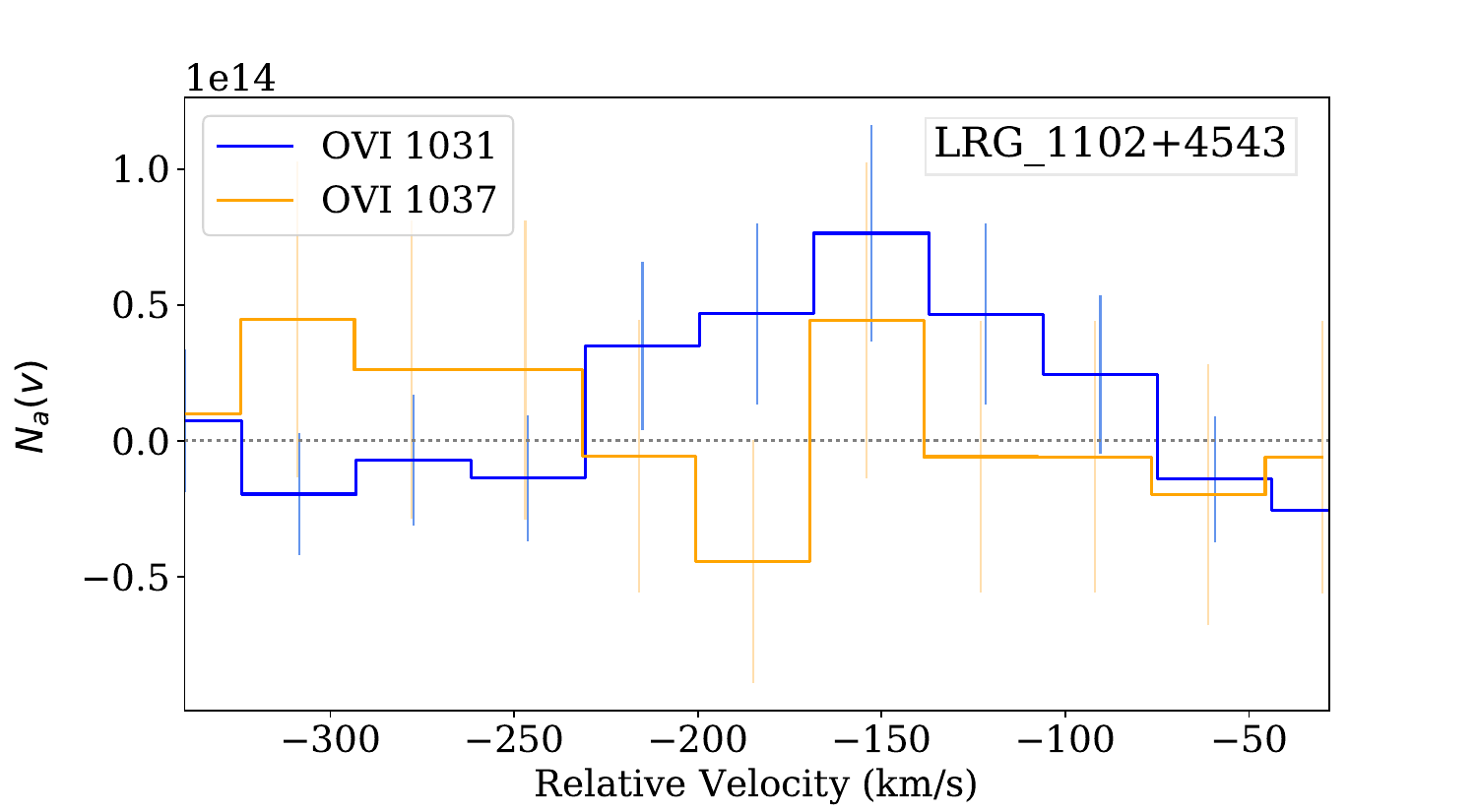}
\includegraphics[width=0.49\textwidth]{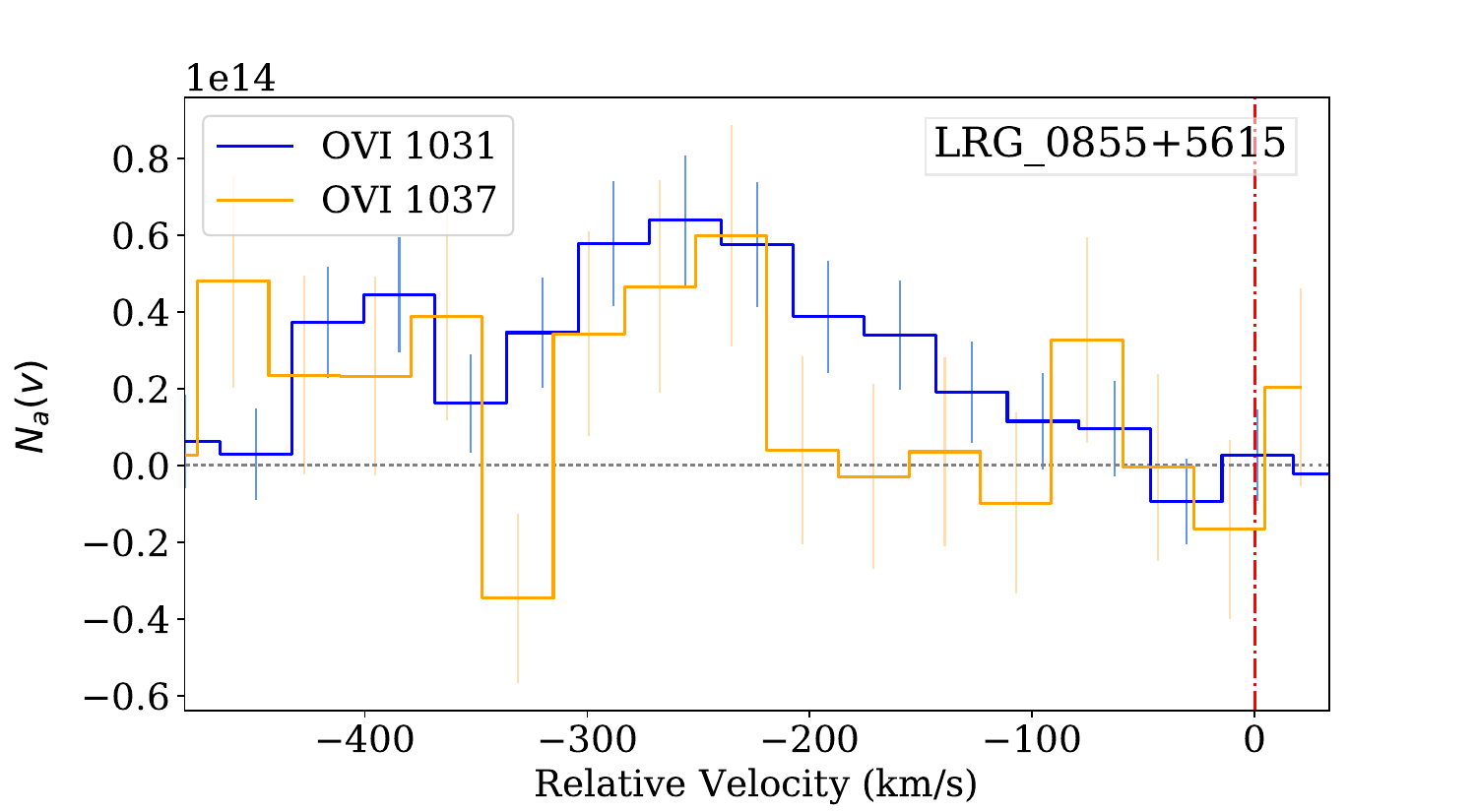}
\caption{\small 
Apparent column density profiles for a few LRGs with possible \ion{O}{6} 1031 detection. Blue and orange line represent \ion{O}{6} 1031 and \ion{O}{6} 1037 apparent column densities multiplied by the length of the velocity intervals corresponding to the velocity bins. 
Red and green vertical line correspond to the LRG's redshift and \ion{Mg}{2} LRG-CGM system's redshift, respectively. LRG name is shown in the upper right corner. 
}
\label{fig:oviaodm}
\end{figure*}

% OVI - method 
To confirm if we detect \ion{O}{6}, we calculate the apparent column density ($N_a (v)$) multiplied by the velocity interval as in \citet{SavageSem91}:   % eq. 8
$N_a (v) = \frac{\tau (v) dv}{2.654\times 10^{15} f \lambda}$, where  $\lambda$ is rest-frame wavelength in angstroms, $f$ is oscillator strength, and $v$ is velocity. Next, we compare $N_a (v)$ profiles between \ion{O}{6} 1031 and \ion{O}{6} 1037, and if the profiles are consistent within their 1 $\sigma$ error, we consider the line as a detection, otherwise we place upper limits. The results are shown on Figure \ref{fig:oviaodm}. 

% OVI - results 
We consider that we detect \ion{O}{6} in LRG-CGM systems in all cases, except LRG\_1217+4931 and LRG\_1102+4543. In the first case, near the \ion{O}{6} 1031 line center (peak in the blue line), in two pixels the flux in \ion{O}{6} 1037 is lower by more than 1$\sigma$ than the lower limit in the \ion{O}{6} 1031 flux. In the second case, we find the same, but for one pixel only. 
In LRG\_1059+4039, LRG\_1121+4345, and LRG\_0855+5615, line profiles for the two \ion{O}{6} transitions show a similar shape, and are in agreement within 1$\sigma$ errors. 
For LRG\_0855+5615, we also compare $N_a (v)$ profiles of \ion{O}{6} 1031 and \ion{H}{1} 972. 
For LRG\_1351+2253, the blue side of \ion{O}{6} 1037 is blended, but the red sides of the two transitions have similar profile and agree within 1$\sigma$ errors.
For LRG\_1144+0714 and LRG\_1440-0157, we find that fluxes from the two \ion{O}{6} transitions are in agreement in each pixel (see also stack plots in Appendix \ref{sec:appd}). % and in Figure Set 3). 

\begin{figure*}[t!]
\centering
\includegraphics[width=0.49\textwidth]{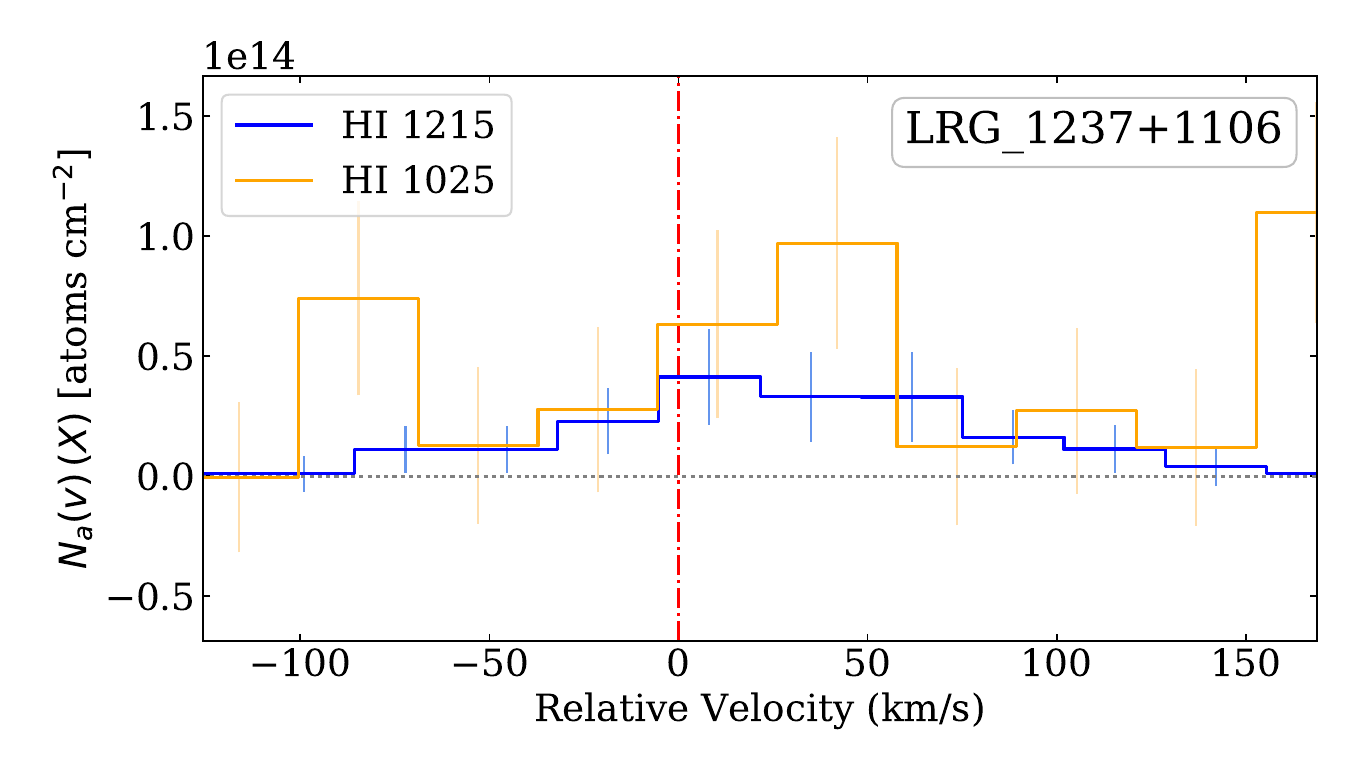}
\includegraphics[width=0.49\textwidth]{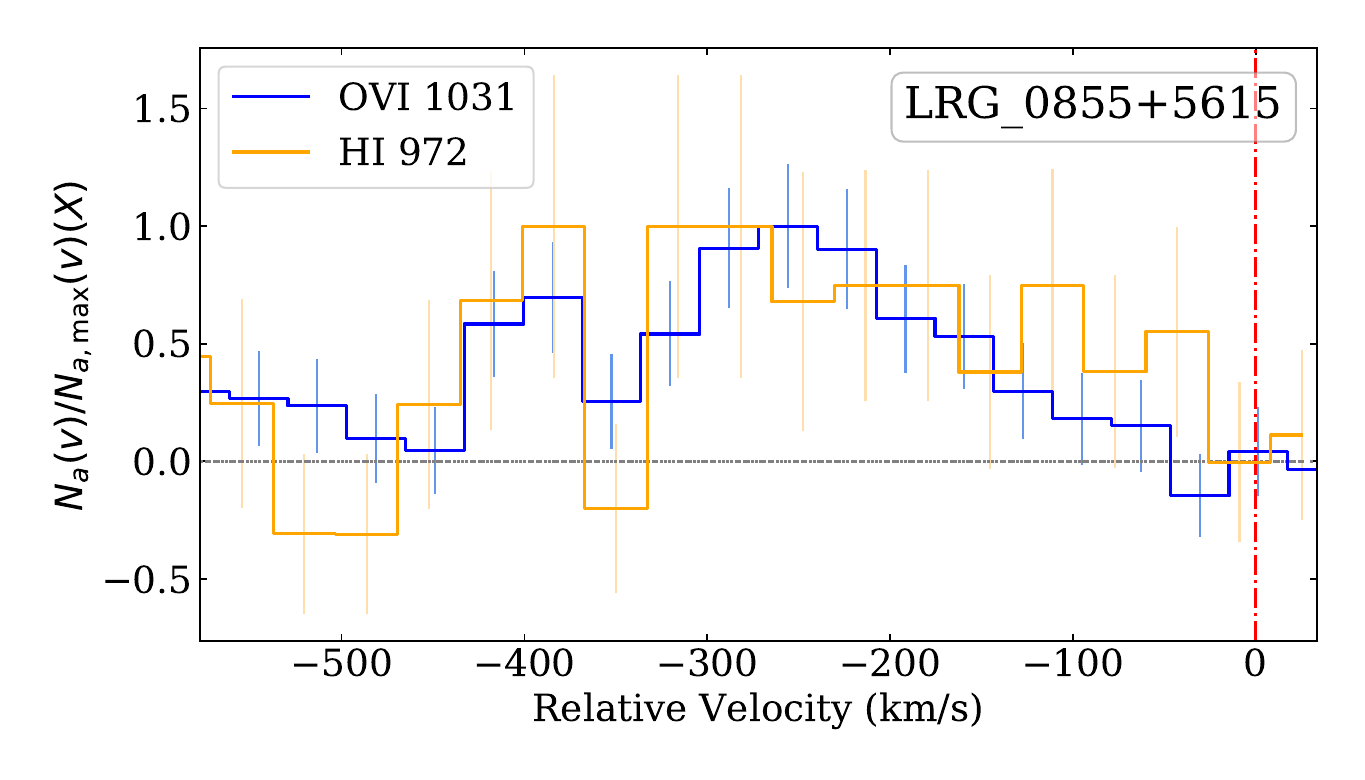}
\caption{\small 
The same as Figure \ref{fig:oviaodm}, but for confirmation of \ion{H}{1} in LRG\_1237+1106, and for a comparison between \ion{O}{6} 1031 and \ion{H}{1} 972 in LRG\_0855+5615.
}
\label{fig:hiaodm}
\end{figure*}

% Other transitions - HI 
We apply the same method to \ion{H}{1} 1215 and \ion{H}{1} 1025 transitions in the CGM of LRG\_1237+1106, because only \ion{H}{1} 1215 was detected at a significance level greater than $3 \sigma$. We show the results on Figure \ref{fig:hiaodm}, and conclude that we detect \ion{H}{1} in the CGM of LRG\_1237+1106.

% OVI AODM

% HI AODM

\section{Neighboring galaxies around LRG sightlines}\label{sec:appc}

In Tables \ref{tab:ngtab07} and \ref{tab:ngtab0}, we list properties of neighboring galaxies (NGs, see section \ref{sec:ngs0}) inside 200 kpc from the QSO sightline for our LRGs sample.

\begin{table*}
\begin{rotatetable*}
\center
\caption{Properties of neighboring galaxies (NGs) around LRG-QSO sightlines$^{a}$  
\label{tab:ngtab0}}
\begin{threeparttable}
{
\makebox % [\lineheight] 
{
\hspace{-2.9 in} 
\centering 
\begin{tabular}{cccccccccccc}
\hline
LRG & $R_{\rm LRG}$ & $z_{\rm LRG}$ & $R_{\rm gal}$ & $z_{\rm phot,\, gal}$ & 
$u - r$ & $r$ & $C_{gi}$ & $r$-mag & $g$-mag & $i$-mag &  R/B 
\\
 &  (kpc)  & & (kpc) & & & & & & & &  
\\
\hline
J1059+4039 & 29 & 0.449 & 94 & $0.460 \pm 0.108$ & 3.2 & 21.69 & $-0.6 \pm 1.1$ & $-21.30 \pm 0.10$ & $-20.05 \pm 1.10$ & $-21.75 \pm 0.05$ & R \\
J1059+4039 & 29 & 0.449 & 117 & $0.498 \pm 0.083$ & 0.5 & 21.13 & $1.0 \pm 0.3$ & $-21.10 \pm 0.02$ & $-20.38 \pm 0.18$ & $-20.43 \pm 0.21$ & B \\
J1059+4039 & 29 & 0.449 & 137 & $0.395 \pm 0.097$ & 1.9 & 22.42 & $-0.4 \pm 7.0$ & $-19.70 \pm 0.28$ & $-18.44 \pm 6.99$ & $-19.87 \pm 0.30$ & R \\
J1059+4039$^{c}$ & 29 & 0.449 & 198 & $0.436 \pm 0.017$ & 3.7 & 19.59 & $-0.1 \pm 0.0$ & $-23.18 \pm 0.00$ & $-22.32 \pm 0.05$ & $-23.63 \pm 0.00$ & R$^{b}$ \\
J1351+2753 & 99 & 0.432 & 88 & $0.364 \pm 0.073$ & 1.5 & 21.02 & $1.4 \pm 0.3$ & $-20.61 \pm 0.02$ & $-20.31 \pm 0.04$ & $-19.88 \pm 0.25$ & B \\
J1351+2753 & 99 & 0.432 & 161 & $0.351 \pm 0.150$ & 2.0 & 21.44 & $0.4 \pm 0.1$ & $-20.66 \pm 0.03$ & $-20.23 \pm 0.04$ & $-20.95 \pm 0.04$ & B \\
J1520+2534 & 225 & 0.538 & 157 & $0.430 \pm 0.153$ & 3.1 & 21.90 & $-2.5 \pm 389.4$ & $-22.02 \pm 0.10$ & $-18.25 \pm 389.40$ & $-21.86 \pm 0.21$ & R \\
J1106-0115 & 383 & 0.611 & 146 & $0.552 \pm 0.143$ & 1.1 & 21.89 & $0.3 \pm 0.8$ & $-21.35 \pm 0.18$ & $-20.48 \pm 0.75$ & $-21.26 \pm 0.14$ & R \\
J1440-0157 & 343 & 0.480 & 31 & $0.434 \pm 0.070$ & 2.7 & 21.18 & $-1.2 \pm 5579.6$ & $-22.32 \pm 31.74$ & $-20.34 \pm 5579.61$ & $-22.71 \pm 15.81$ & R \\
J1440-0157 & 343 & 0.480 & 77 & $0.451 \pm 0.108$ & 1.4 & 21.19 & $0.4 \pm 0.0$ & $-21.22 \pm 0.02$ & $-20.87 \pm 0.03$ & $-21.55 \pm 0.02$ & B \\
J1440-0157 & 343 & 0.480 & 87 & $0.428 \pm 0.087$ & 1.0 & 21.96 & $1.0 \pm 0.2$ & $-19.77 \pm 0.09$ & $-19.82 \pm 0.08$ & $-19.80 \pm 0.14$ & B \\
J1440-0157 & 343 & 0.480 & 154 & $0.432 \pm 0.109$ & -0.1 & 22.08 & $1.2 \pm 0.8$ & $-19.34 \pm 0.33$ & $-19.25 \pm 0.26$ & $-18.99 \pm 0.72$ & R \\
J1237+1106 & 23 & 0.473 & 165 & $0.523 \pm 0.088$ & 0.0 & 21.21 & $-0.1 \pm 0.3$ & $-21.25 \pm 0.05$ & $-20.30 \pm 0.25$ & $-21.52 \pm 0.04$ & R \\
J1121+4345 & 80 & 0.422 & 125 & $0.394 \pm 0.065$ & 2.4 & 21.18 & $-0.2 \pm 0.1$ & $-21.29 \pm 0.02$ & $-20.41 \pm 0.11$ & $-21.75 \pm 0.01$ & R \\
J1121+4345 & 80 & 0.422 & 162 & $0.457 \pm 0.113$ & 1.8 & 21.09 & $0.3 \pm 0.0$ & $-21.16 \pm 0.03$ & $-20.78 \pm 0.04$ & $-21.60 \pm 0.03$ & B \\
J1549+0701 & 120 & 0.500 & 61 & $0.562 \pm 0.154$ & -0.2 & 21.99 & $0.2 \pm 0.4$ & $-20.11 \pm 0.26$ & $-19.61 \pm 0.23$ & $-20.41 \pm 0.27$ & R \\
J1549+0701 & 120 & 0.500 & 151 & $0.519 \pm 0.042$ & 3.2 & 21.78 & $-2.2 \pm 19.4$ & $-21.86 \pm 0.08$ & $-19.14 \pm 19.42$ & $-22.48 \pm 0.02$ & R \\
J1549+0701$^{d}$ & 120 & 0.500 & 178 & $0.591 \pm 0.120$ & 2.5 & 22.63 & $-0.3 \pm 2.5$ & $-21.00 \pm 0.70$ & $-19.92 \pm 2.45$ & $-21.35 \pm 0.58$ & R \\
\hline
\end{tabular}
}
\hspace{-1.9 in} 
\begin{tablenotes} 
    \item[a] {First three columns contain name of the LRG-CGM system, LRG impact parameter, LRG redshift. Other columns contain information 
    
    about NGs: impact parameter between the galaxy and the QSO ($R_{\rm gal}$); photometric redshift and its error ($z_{\rm phot,\, gal}$; note that these errors 
    
    are likely underestimated); $u - r$ color; apparent $r$ magnitude ($r$); $C_{gi}$ color (see section \ref{sec:ng}); absolute $r$, $g$, $i$ magnitudes (k-corrected); 
    
    and indicator if NG is red or blue (R/B). All colors are derived from k-corrected absolute magnitudes. We note that errors in colors 
    
    are in some cases very big.}
    \item[b] {This galaxy has spectroscopic redshift measured by SDSS, $z\sim 0.450$, and is actually another LRG.}
    \item[c] {J1059+4039 has one more NG, that was identified in Pan-STARRS and does not have SDSS data. This NG is not 

    included in this table.}
    \item[d] {There is also one another galaxy close to LRG\_1549+0701, which is also a LRG. This galaxy has spectroscopic redshift $z\sim 0.509$, and 
    
    impact parameter 125 kpc. Its photometric redshift is not consistent with the targeted LRG.}
  \end{tablenotes}
}
\end{threeparttable}
\end{rotatetable*}
\end{table*}

\begin{table*}
\begin{rotatetable*}
\center
\caption{Properties of neighboring galaxies around LRG-QSO sightlines - continuation   
\label{tab:ngtab07}}
\begin{threeparttable}
{
\makebox %[\lineheight] 
{
\hspace{-2.5in}
\centering 
\begin{tabular}{cccccccccccc}
\hline
LRG & $R_{\rm LRG}$ & $z_{\rm LRG}$ & $R_{\rm gal}$ & $z_{\rm phot,\, gal}$ & 
$u - r$ & $r$ & $C_{gi}$ & $r$-mag & $g$-mag & $i$-mag &  R/B 
\\
 &  (kpc)  & & (kpc) & & & & & & & &   
\\
\hline
J1306+3421 & 184 & 0.469 & 181 & $0.366 \pm 0.140$ & 2.7 & 22.12 & $-5.0 \pm 19453.0$ & $-20.78 \pm 0.12$ & $-14.85 \pm 19452.98$ & $-20.91 \pm 0.14$ & R \\
J1306+3421 & 184 & 0.469 & 199 & $0.460 \pm 0.030$ & 1.9 & 21.01 & $-0.2 \pm 0.9$ & $-21.69 \pm 0.02$ & $-20.50 \pm 0.88$ & $-21.85 \pm 0.04$ & R \\
J1217+4931 & 208 & 0.523 & 134 & $0.570 \pm 0.052$ & 1.3 & 22.80 & $-0.9 \pm 2.1$ & $-20.92 \pm 0.48$ & $-19.26 \pm 2.07$ & $-21.27 \pm 0.41$ & R \\
J1217+4931 & 208 & 0.523 & 180 & $0.508 \pm 0.080$ & 1.1 & 22.59 & $-0.5 \pm 2.9$ & $-21.37 \pm 0.12$ & $-19.78 \pm 2.81$ & $-21.42 \pm 0.50$ & R \\
J1217+4931 & 208 & 0.523 & 189 & $0.445 \pm 0.133$ & 1.8 & 22.57 & $-0.1 \pm 0.5$ & $-20.42 \pm 0.16$ & $-19.86 \pm 0.45$ & $-21.07 \pm 0.17$ & R \\
J1102+4543 & 193 & 0.487 & 94 & $0.565 \pm 0.114$ & 0.7 & 22.17 & $0.9 \pm inf$ & $-19.55 \pm 0.14$ & $-19.40 \pm inf$ & $-19.52 \pm 0.41$ & R \\
J1102+4543 & 193 & 0.487 & 109 & $0.438 \pm 0.094$ & 0.5 & 20.59 & $0.8 \pm 0.2$ & $-21.66 \pm 0.03$ & $-20.99 \pm 0.15$ & $-21.26 \pm 0.12$ & B \\
J1102+4543 & 193 & 0.487 & 142 & $0.444 \pm 0.160$ & 0.2 & 22.43 & $-0.4 \pm 0.4$ & $-18.51 \pm 22.37$ & $-19.94 \pm 0.36$ & $-21.44 \pm 0.13$ & R \\
J1102+4543 & 193 & 0.487 & 172 & $0.497 \pm 0.041$ & 2.9 & 21.51 & $-0.7 \pm 1.2$ & $-22.09 \pm 0.06$ & $-20.79 \pm 1.24$ & $-22.67 \pm 0.02$ & R \\
J1251+3025 & 281 & 0.513 & 73 & $0.554 \pm 0.144$ & 1.5 & 22.27 & $0.7 \pm 0.1$ & $-19.95 \pm 0.05$ & $-19.93 \pm 0.05$ & $-20.31 \pm 0.05$ & B \\
J1251+3025 & 281 & 0.513 & 145 & $0.648 \pm 0.153$ & 1.2 & 22.27 & $0.5 \pm 1.2$ & $-20.23 \pm 0.08$ & $-19.66 \pm 1.20$ & $-20.15 \pm 0.11$ & R \\
J0226+0014 & 357 & 0.473 & 102 & $0.461 \pm 0.154$ & 1.8 & 22.12 & $-0.1 \pm 0.2$ & $-21.27 \pm 0.14$ & $-20.58 \pm 0.16$ & $-21.82 \pm 0.08$ & R$^{e}$ \\
J0855+5615 & 315 & 0.440 & 70 & $0.507 \pm 0.074$ & 0.2 & 21.35 & $0.4 \pm 0.4$ & $-20.68 \pm 0.05$ & $-20.00 \pm 0.13$ & $-20.62 \pm 0.43$ & R \\
J0855+5615 & 315 & 0.440 & 76 & $0.424 \pm 0.066$ & 2.0 & 20.53 & $0.1 \pm 0.4$ & $-22.02 \pm 0.17$ & $-21.31 \pm 0.44$ & $-22.39 \pm 0.03$ & R \\
J0855+5615 & 315 & 0.440 & 84 & $0.526 \pm 0.111$ & 1.3 & 21.83 & $-3.4 \pm 365.6$ & $-20.79 \pm 0.10$ & $-16.76 \pm 365.63$ & $-21.23 \pm 0.05$ & R \\
J0855+5615 & 315 & 0.440 & 84 & $0.411 \pm 0.104$ & 1.9 & 21.20 & $1.6 \pm 0.7$ & $-20.50 \pm 0.12$ & $-20.67 \pm 0.18$ & $-20.10 \pm 0.70$ & B \\
J0855+5615 & 315 & 0.440 & 160 & $0.446 \pm 0.131$ & 1.7 & 20.42 & $0.4 \pm 0.0$ & $-21.81 \pm 0.01$ & $-21.35 \pm 0.03$ & $-22.13 \pm 0.01$ & B \\
J0855+5615 & 315 & 0.440 & 199 & $0.376 \pm 0.128$ & 0.2 & 20.85 & $-5.4 \pm 19790.8$ & $-21.42 \pm 0.09$ & $-14.99 \pm 19790.77$ & $-21.51 \pm 0.08$ & R \\

\hline
\end{tabular}
}
\hspace{-1.4 in} 
\begin{tablenotes} 
    \item[e] {Absolute $r$-magnitudes of two NGs around J1251+3025 and of the NG around J0226+0014 significantly differ,
    
    although all three NGs
    have similar apparent $r$-magnitudes. This is because of significantly different k-corrections
    
    of these sources, which are -0.13, 0.15 and 
    1.25, respectively. Larger k-corrections correspond to older stellar
    
    populations. Here, the source with the lowest k-correction is a blue 
    NG while two other sources are red NGs,
    
    as expected.    }
\end{tablenotes}
}
\end{threeparttable}
\end{rotatetable*}
\end{table*}

\section{Notes about individual absorbers}\label{sec:appd}

% adding paragraph and figures for arxiv  (all before section D.1.)

\begin{figure*}[t!]
\centering
\includegraphics[width=0.55\textwidth]{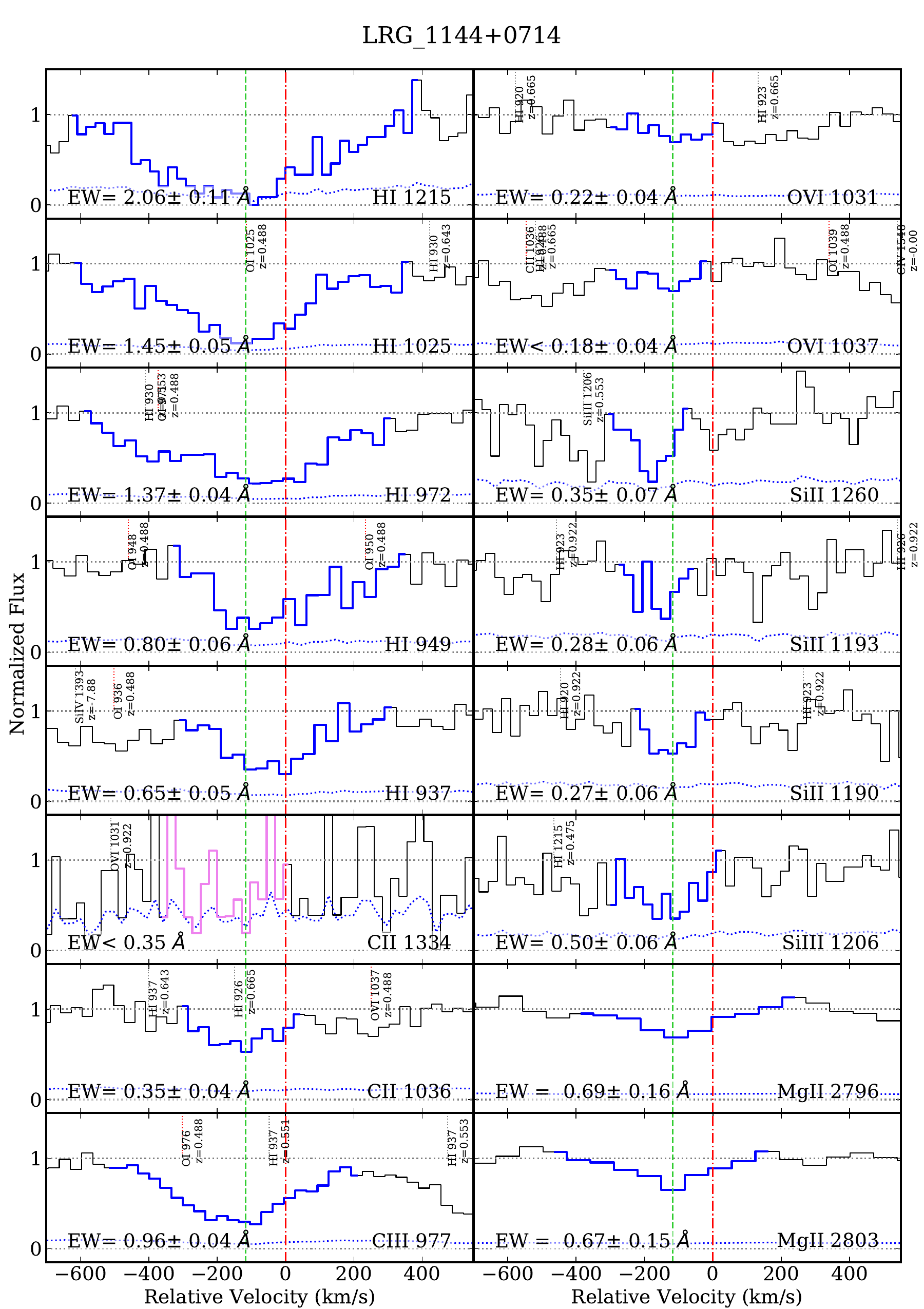}
\caption{\small 
The same as Figure \ref{fig:stacksmgii1}, but for LRG\_1144+0714. Line profiles, shown as continuum normalized flux vs. velocity in the LRG rest frame. Transitions are labeled in the lower right corner, and rest-frame EWs are shown in the lower left corner. The analyzed velocity ranges are marked in blue color for reliable transitions, and in red color for blended or uncertain detections. 
% magenta and cyan - ADDED  
Velocity ranges of additional components are marked in cyan, and analyzed velocity ranges for some of non detections are labeled in magenta.  
Red dot-dashed and green dashed vertical lines denote LRG redshift and velocity offset of the corresponding LRG-CGM system (for MgII-LRGs determined from the central velocity of \ion{Mg}{2} absorption), respectively. The blue dotted line shows the flux uncertainty. Interloping transitions are marked with red and gray vertical dotted lines, if they originate from the CGM around the LRG, or from absorption systems found at other redshifts, respectively. We consider that transitions within 1200 km s$^{-1}$ of the LRG redshift are associated with the CGM around the LRG. 
Name and redshift of these interloping transitions are labeled alongside of the vertical lines. 
}
\label{fig:stacksmgii2}
\end{figure*}

\begin{figure*}[t!]
\centering
\includegraphics[width=0.55\textwidth]{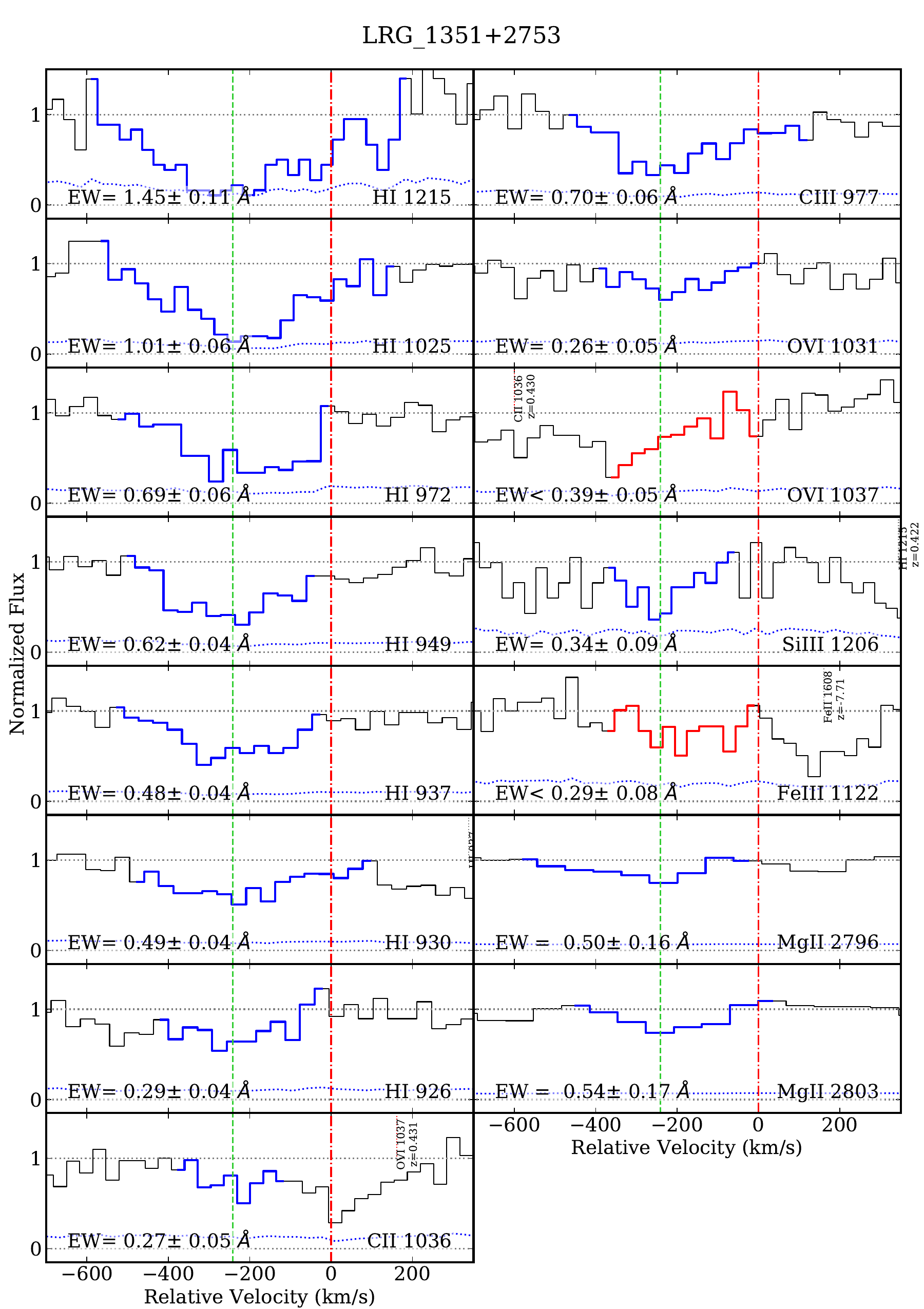}
\caption{\small 
The same as Figure \ref{fig:stacksmgii2}, but for LRG\_1351+2753. 
}
\label{fig:stacksmgii3}
\end{figure*}

\begin{figure*}[t!]
\centering
\includegraphics[width=0.55\textwidth]{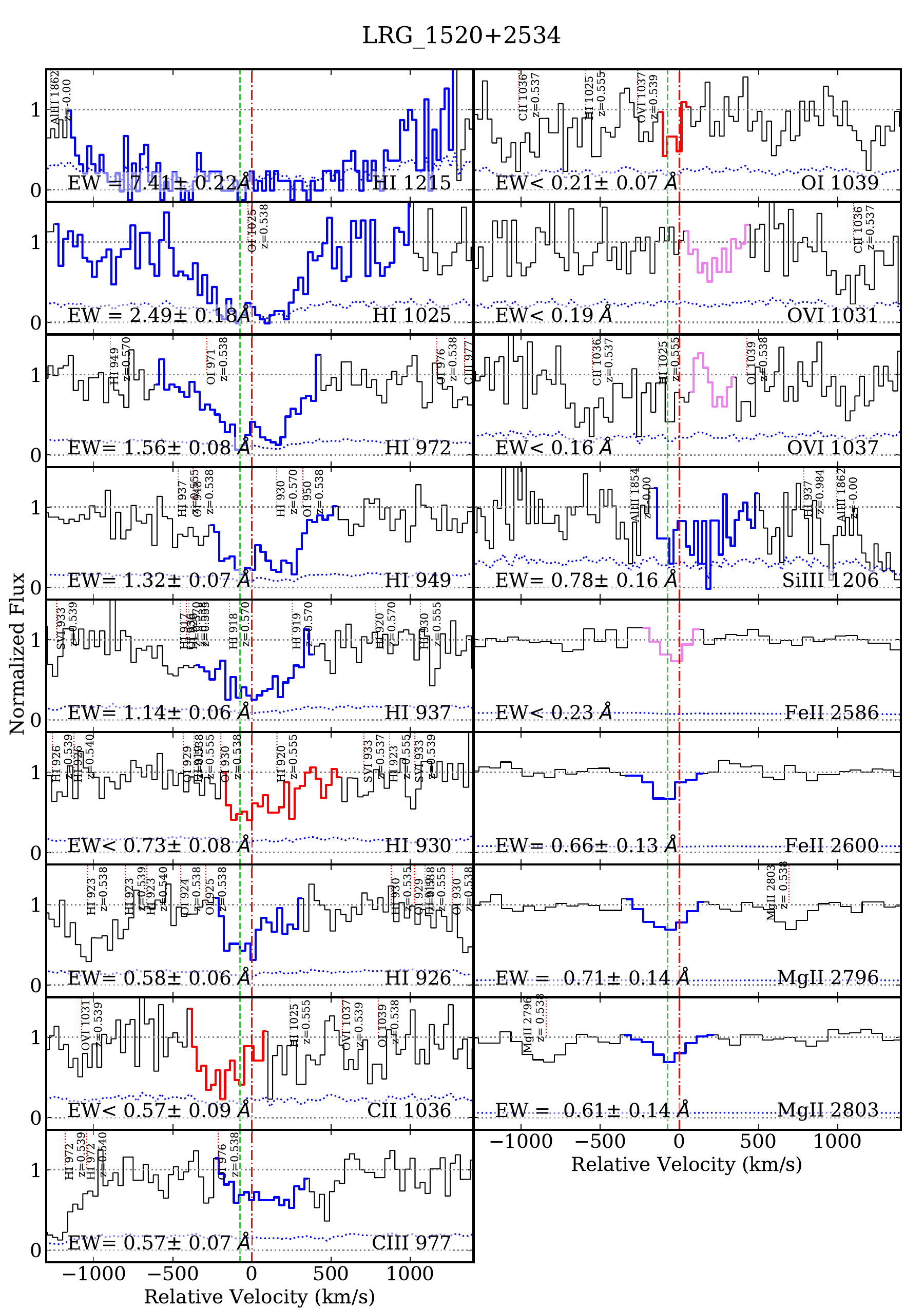}
\caption{\small 
The same as Figure \ref{fig:stacksmgii2}, but for LRG\_1520+2534.
}
\label{fig:stacksmgii4}
\end{figure*}

\begin{figure*}[t!]
\centering
\includegraphics[width=0.55\textwidth]{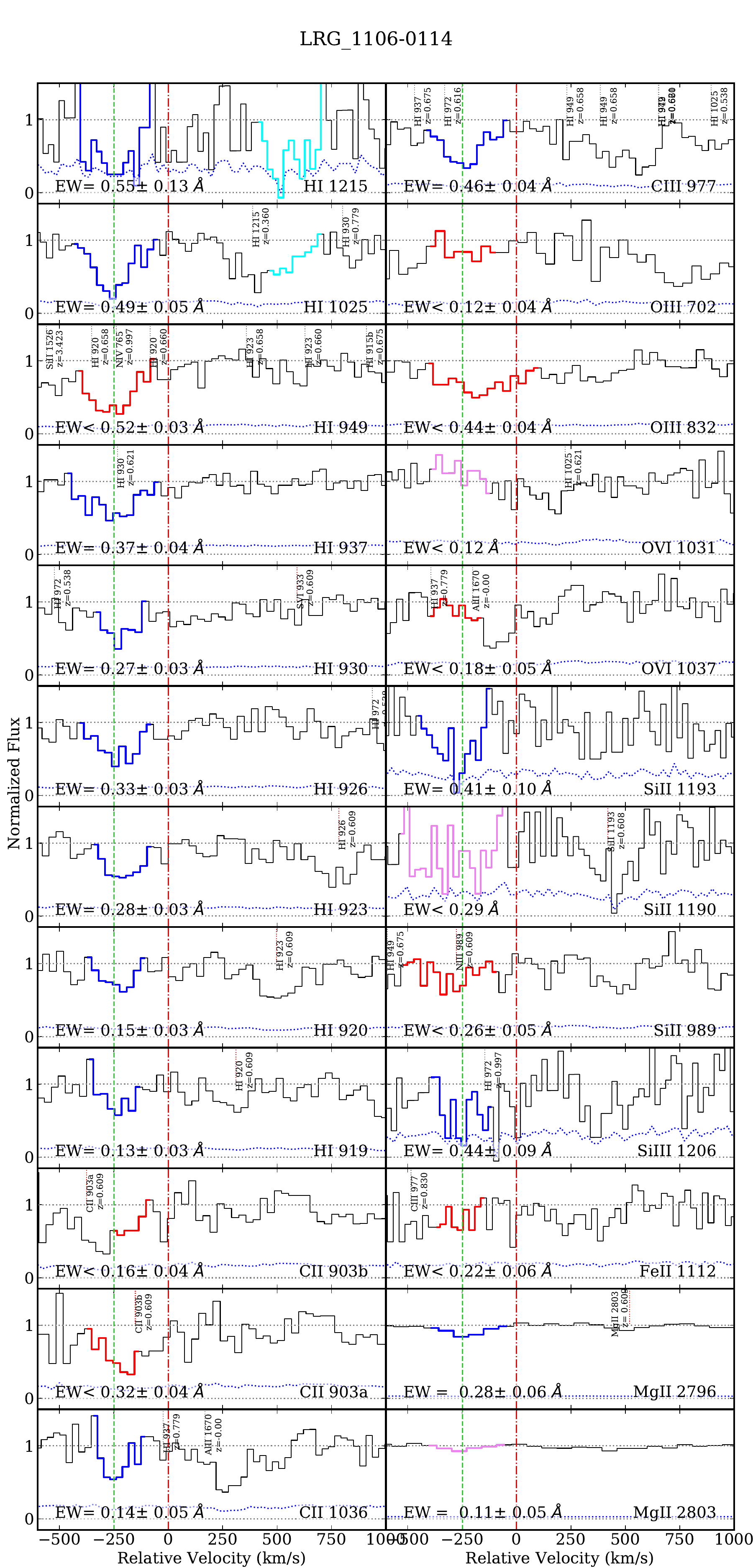}
\caption{\small 
The same as Figure \ref{fig:stacksmgii2}, but for LRG\_1106-0115.
}
\label{fig:stacksmgii5}
\end{figure*}

\begin{figure*}[t!]
\centering
\includegraphics[width=0.55\textwidth]{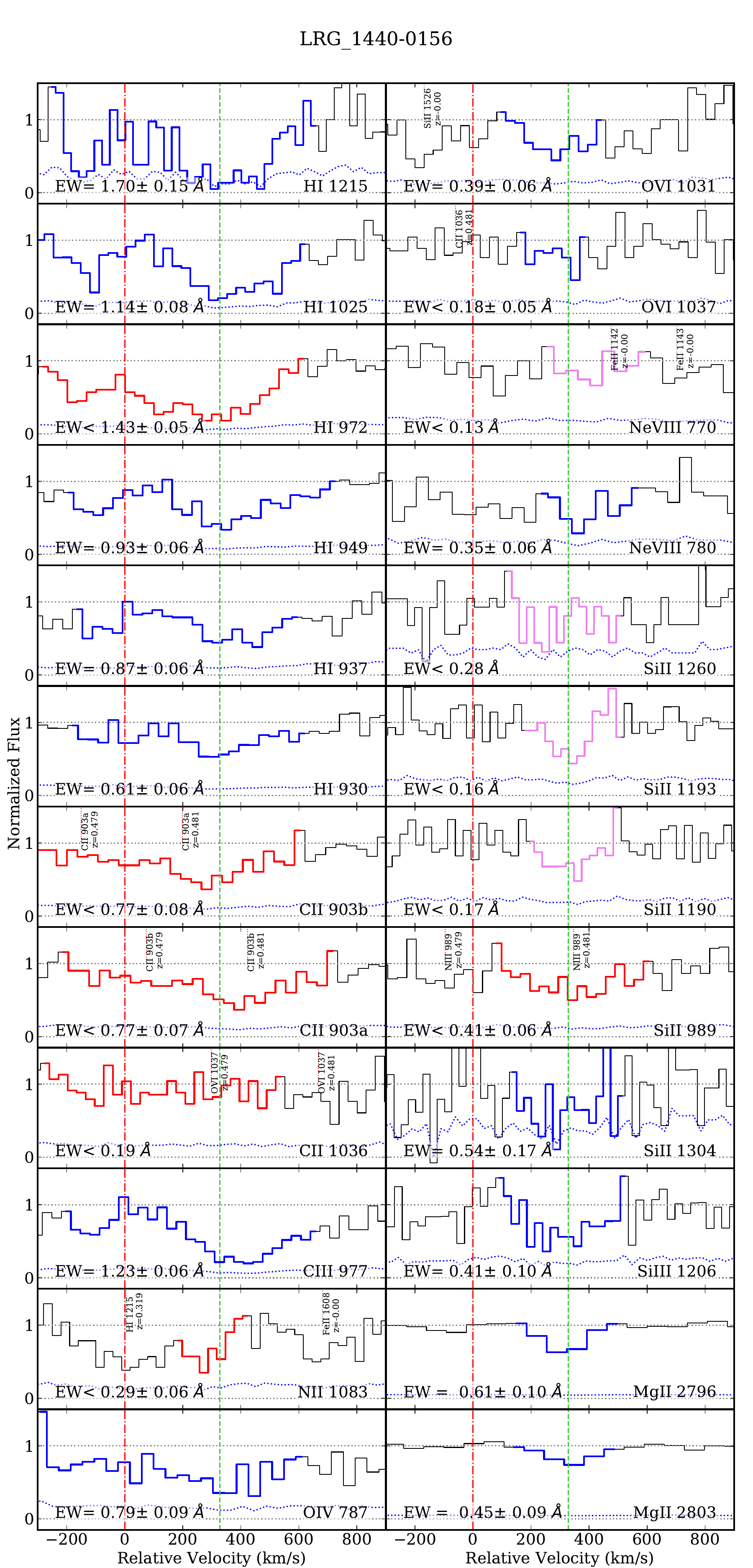}
\caption{\small 
The same as Figure \ref{fig:stacksmgii2}, but for LRG\_1440-0157.
}
\label{fig:stacksmgii6}
\end{figure*}

\begin{figure*}[t!]
\centering
\includegraphics[width=0.55\textwidth]{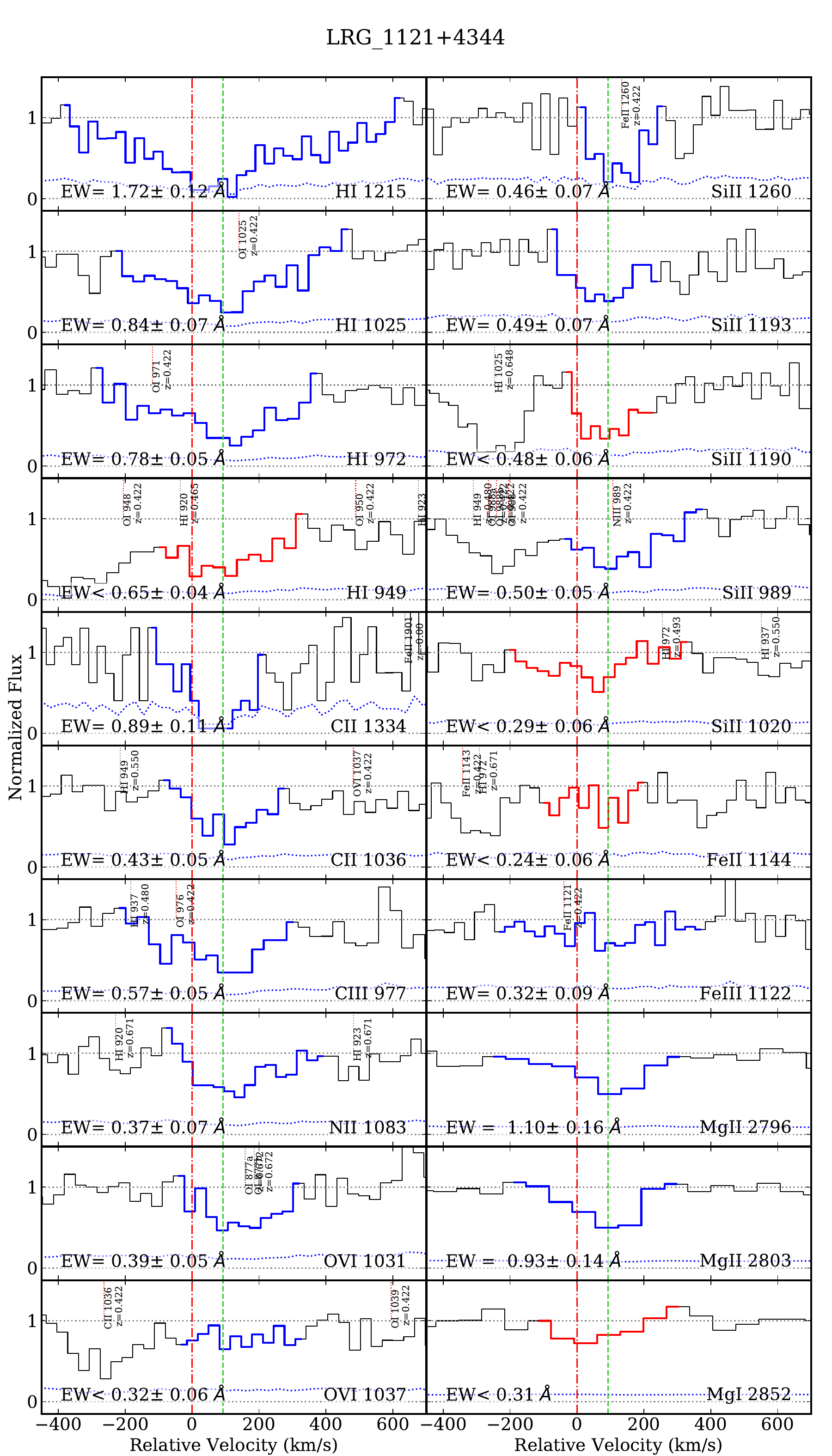}
\caption{\small 
The same as Figure \ref{fig:stacksmgii2}, but for LRG\_1121+4345.
}
\label{fig:stacksmgii7}
\end{figure*}

\begin{figure*}[t!]
\centering
\includegraphics[width=0.49\textwidth]{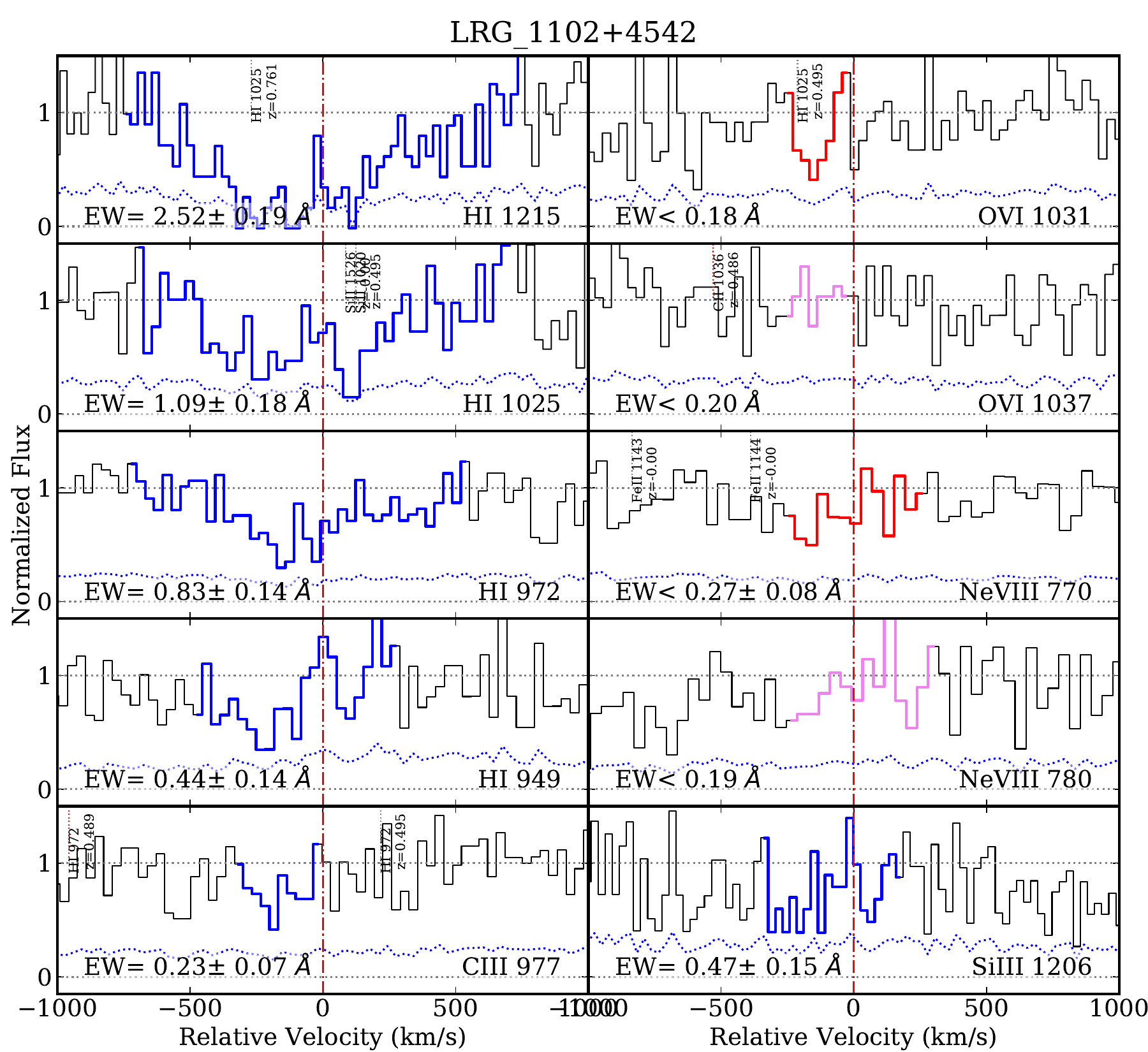}
\includegraphics[width=0.49\textwidth]{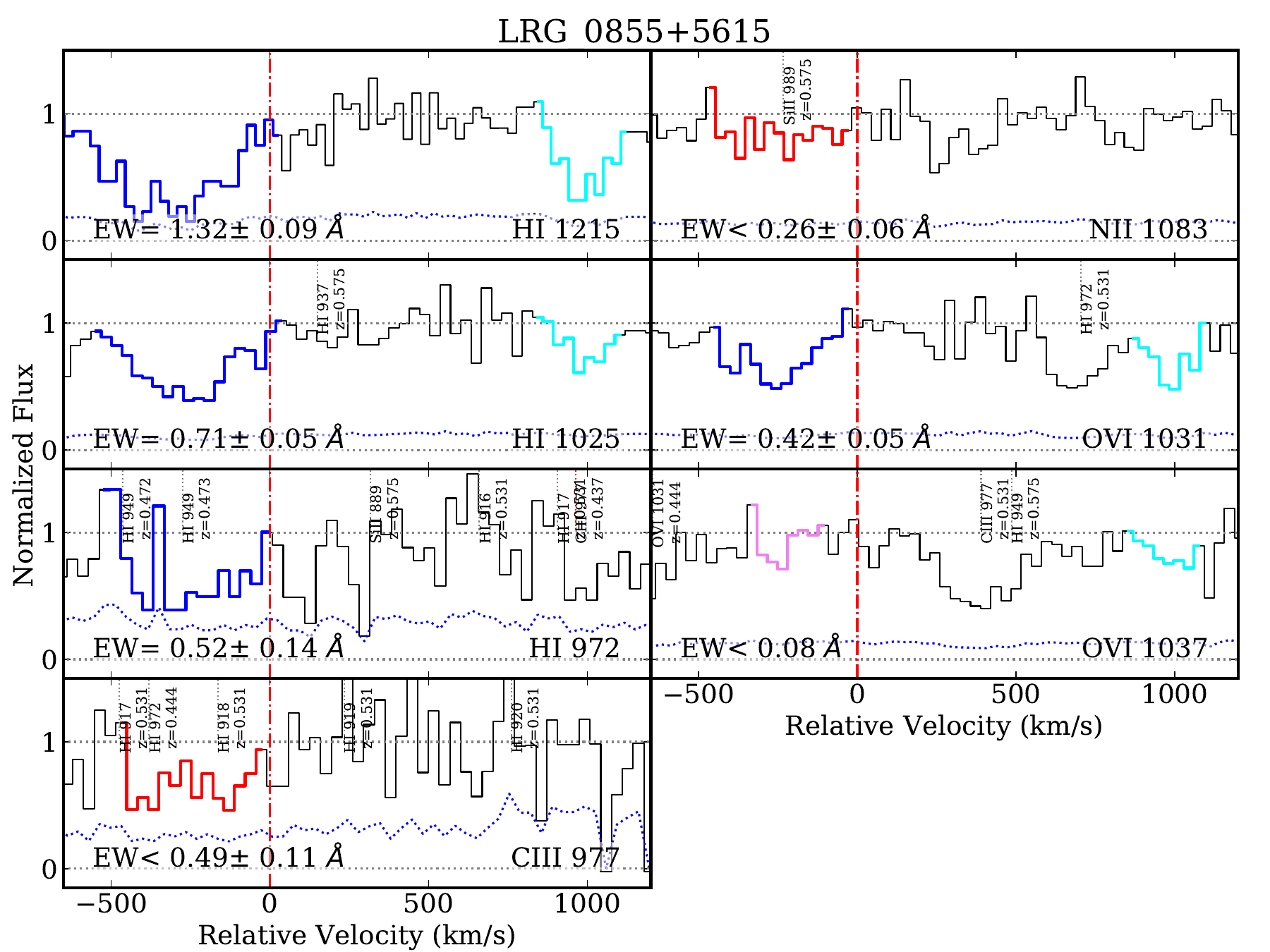} 
\includegraphics[width=0.49\textwidth]{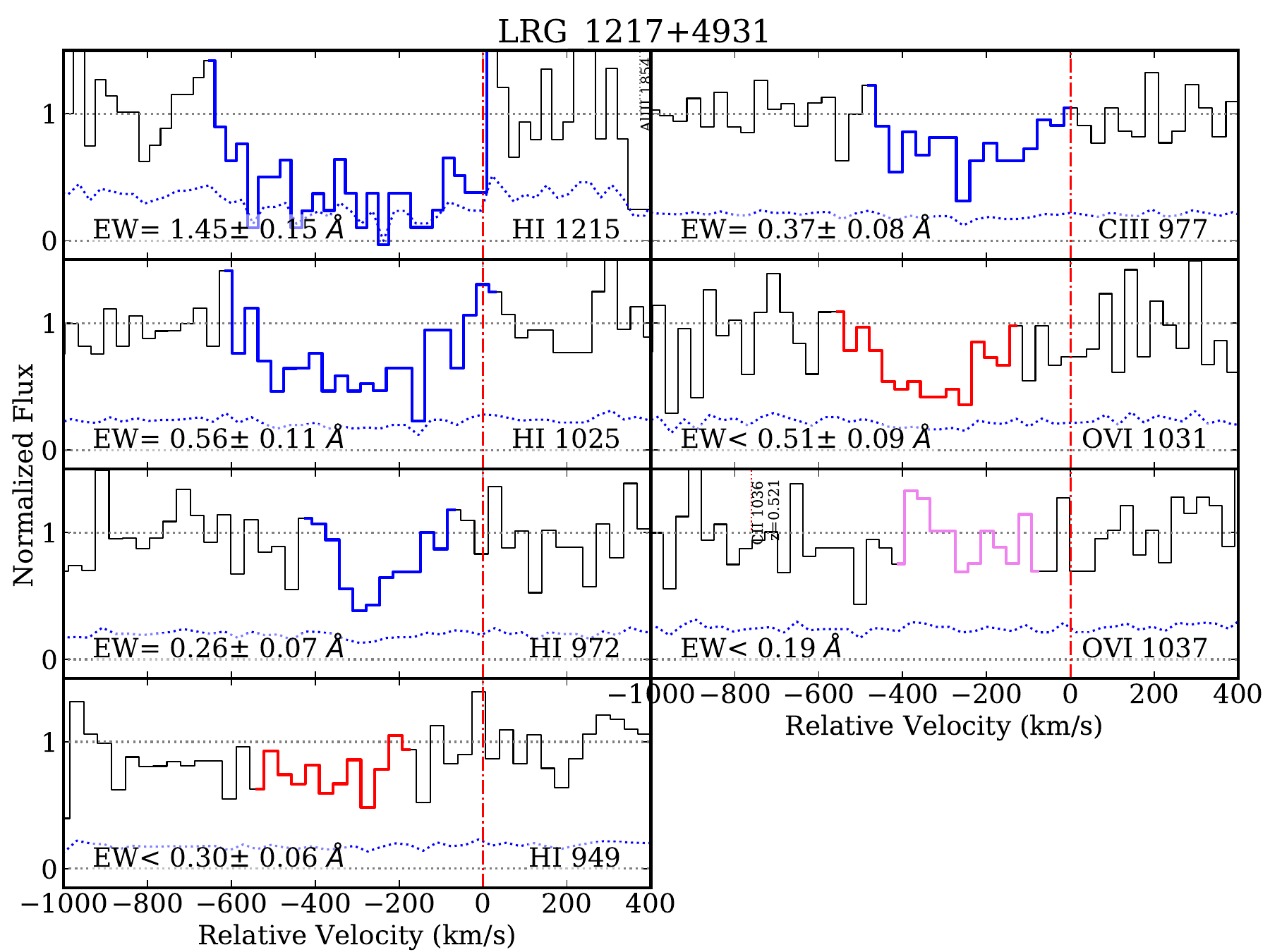}
\caption{\small 
The same as Figure \ref{fig:stacksmgii2}, but for Baseline LRGs with two or more \ion{H}{1} transitions significantly detected.
}
\label{fig:stacksnomg}
\end{figure*}

\begin{figure*}[t!]
\centering
\includegraphics[width=0.30\textwidth]{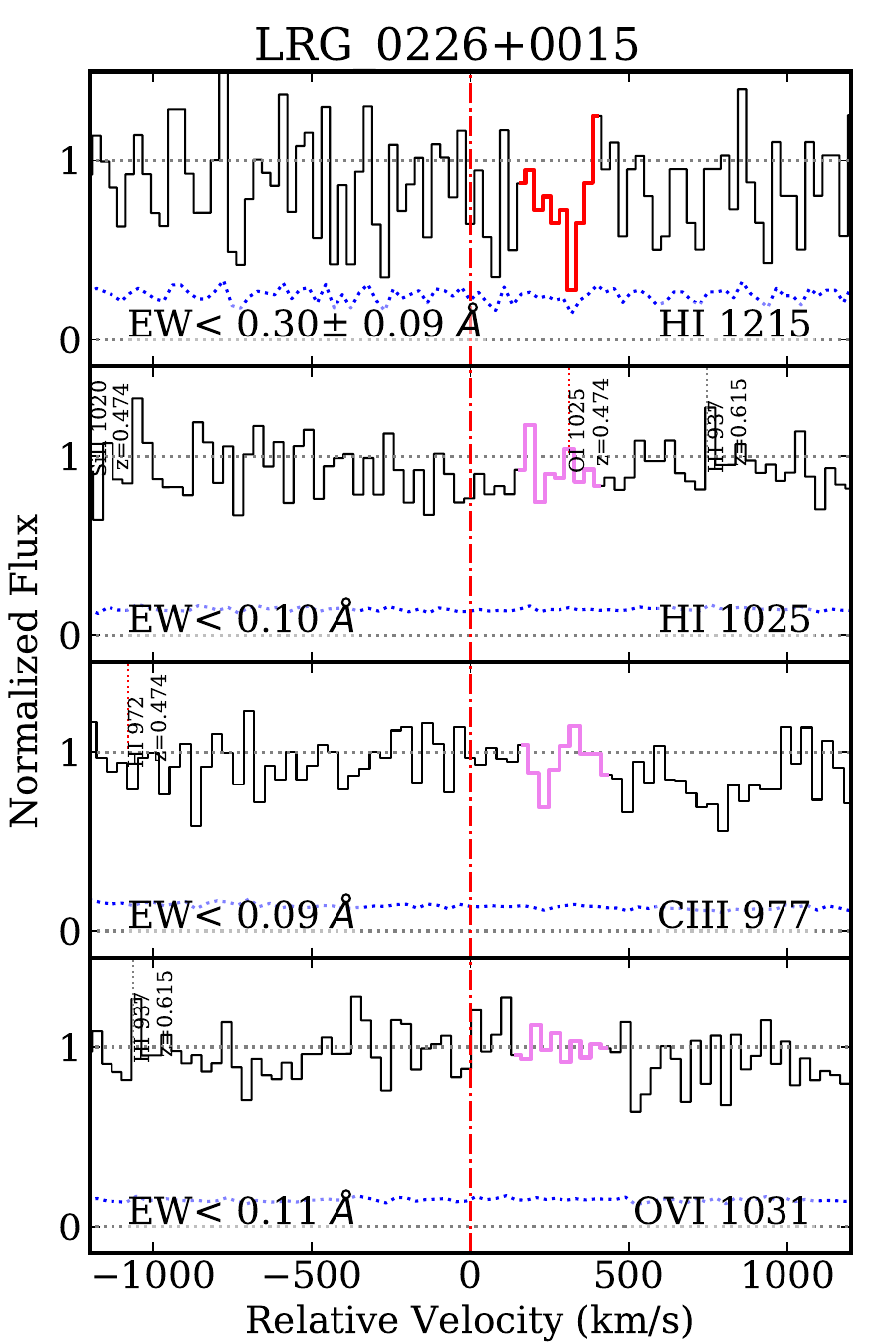}
\includegraphics[width=0.30\textwidth]{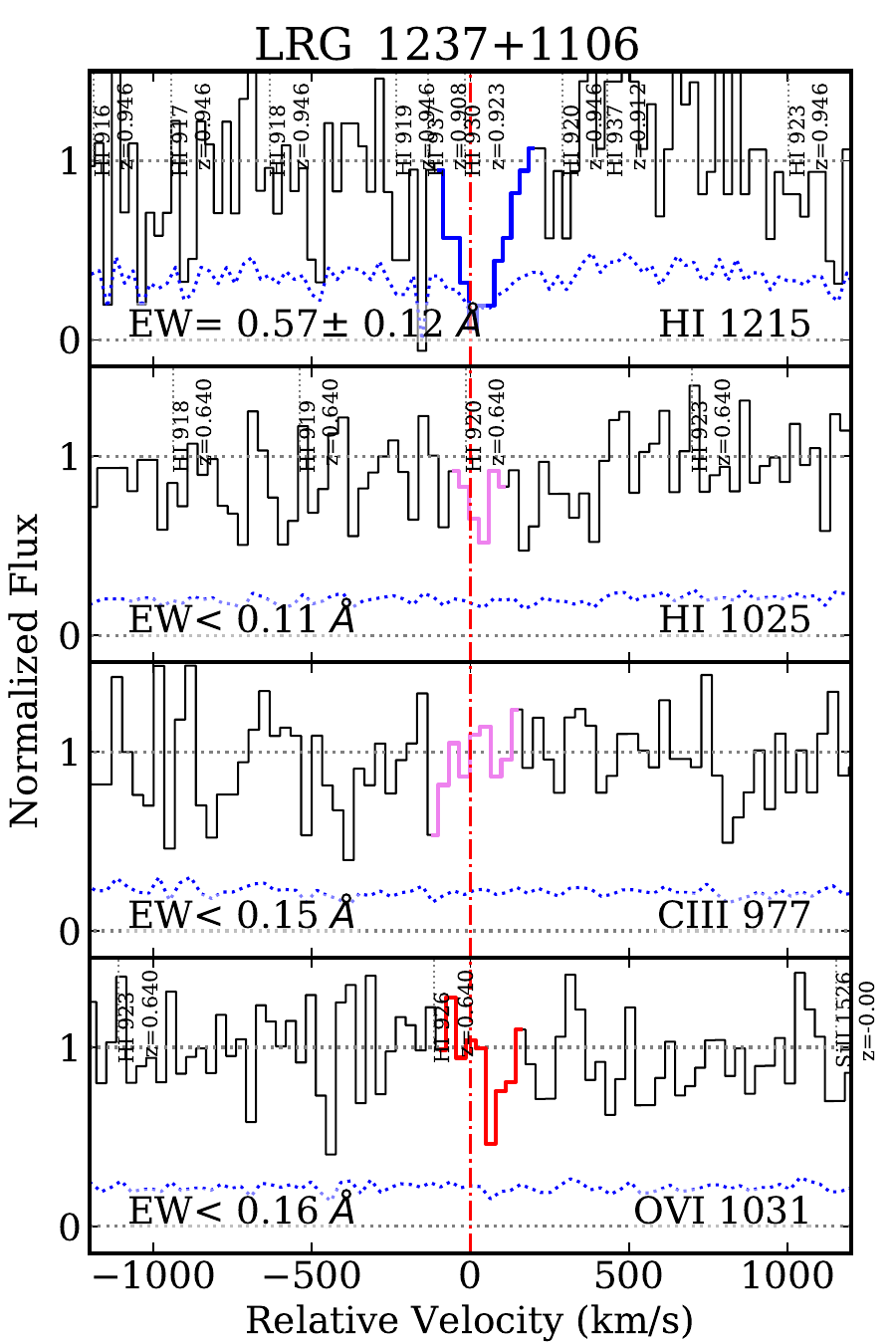}
\includegraphics[width=0.30\textwidth]{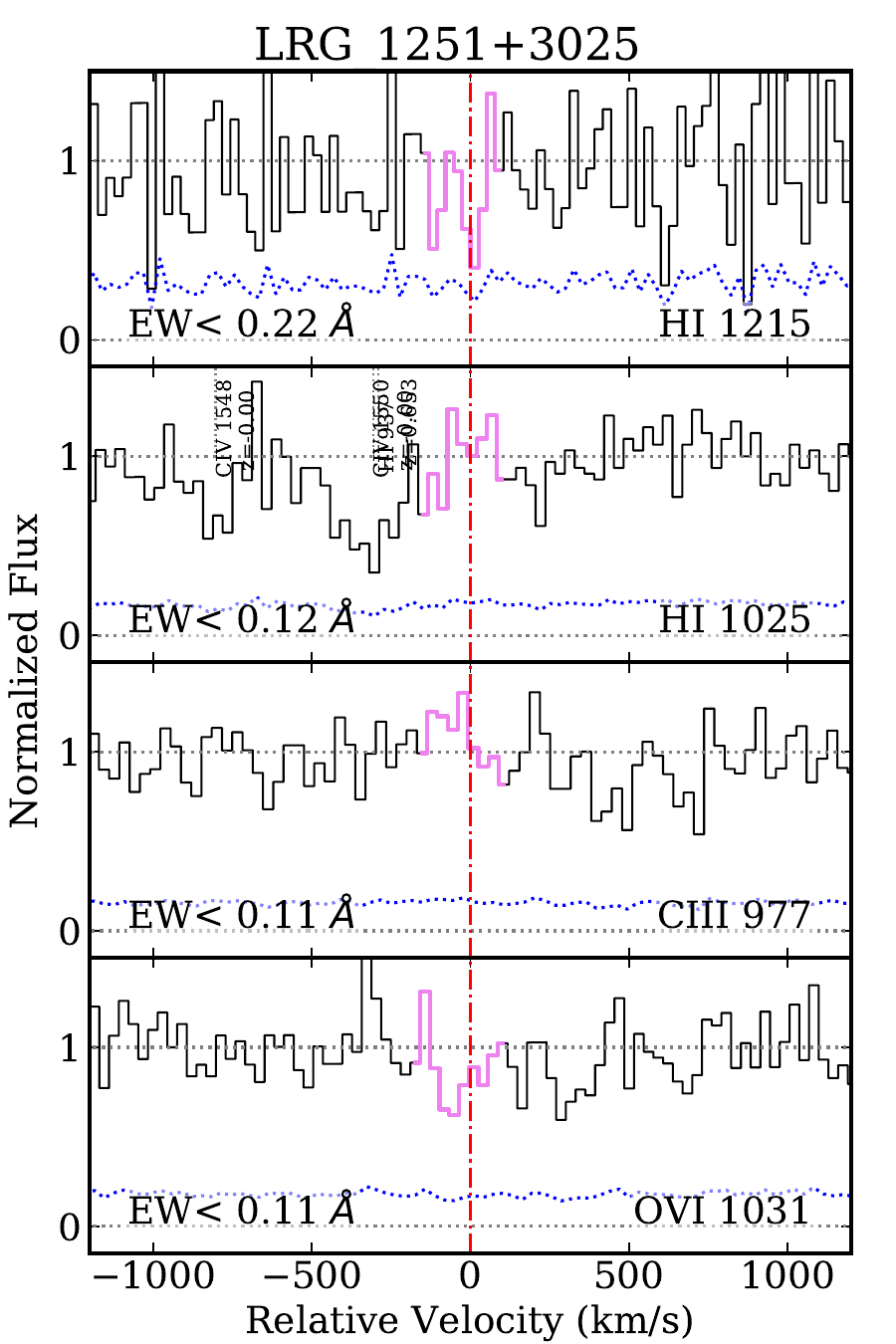}
\includegraphics[width=0.30\textwidth]{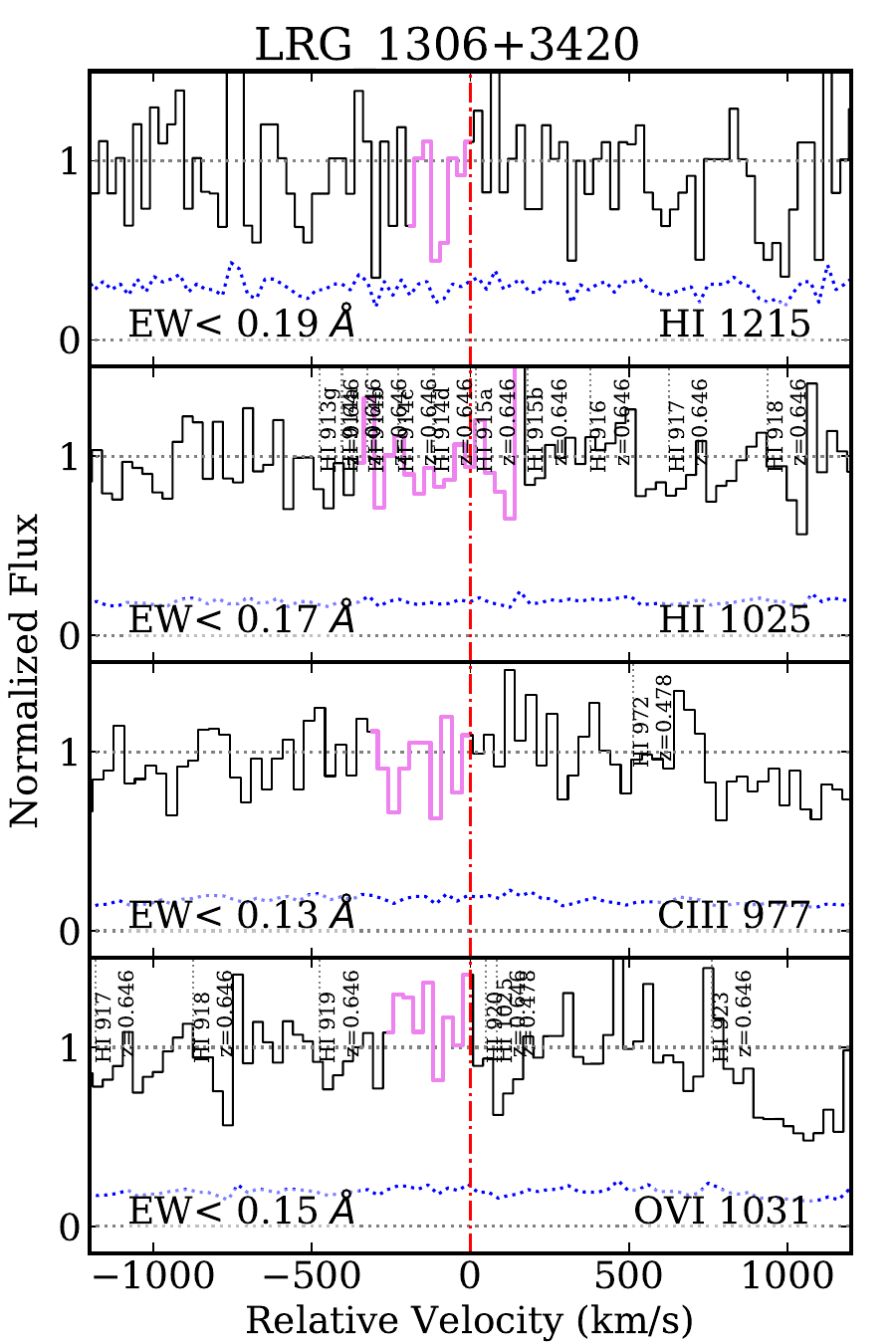}
\includegraphics[width=0.30\textwidth]{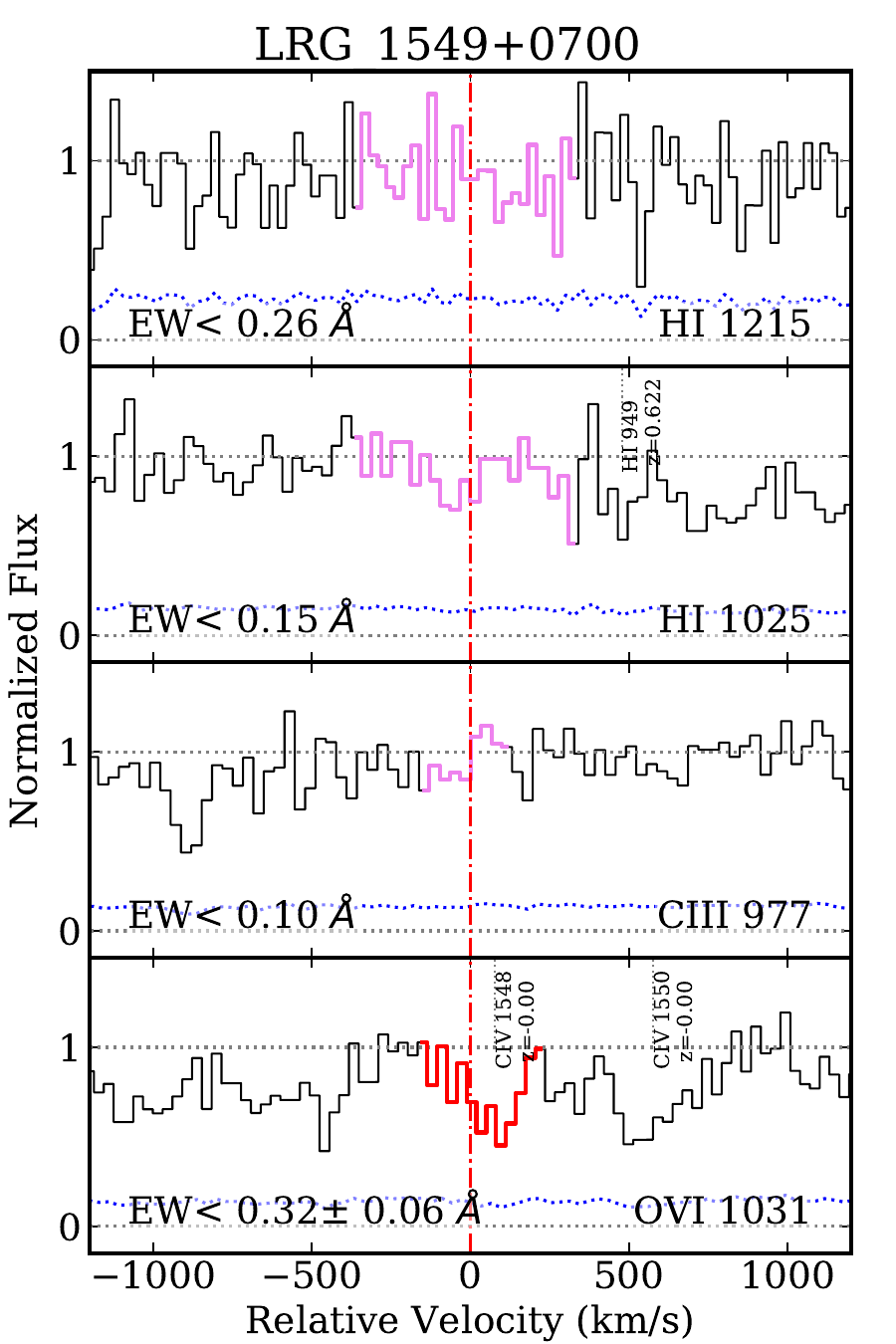}
\caption{\small 
The same as Figure \ref{fig:stacksmgii2}, but for Baseline LRGs without or with only one absorption line (which is \ion{H}{1} 1215).  
}
\label{fig:stacksno}
\end{figure*}

Figures \ref{fig:stacksmgii2} - \ref{fig:stacksmgii7} present absorption lines detected in the CGM around MgII-LRGs. These lines include \ion{H}{1} transitions and metal lines, such as \ion{C}{3}, \ion{Si}{3}, \ion{Si}{2}, \ion{C}{2}, \ion{O}{6} transitions. 
Figures \ref{fig:stacksnomg} and \ref{fig:stacksno} present absorption lines in the CGM around Baseline LRGs. 
Baseline LRGs also show \ion{H}{1} line transitions and some of the metal lines. In contrast to MgII-LRGs, around Baseline LRGs we do not detect low-ionization lines such as \ion{C}{2} and \ion{Si}{2}.

% figures with absorption lines in all LRG-CGM systems 

% MgII-LRGs

\clearpage

% Baseline LRGs (without MgII)

% non-detections and uncertain HI (only LyA)

\clearpage

\subsection{MgII-LRGs}

\textit{LRG\_1059+4039:} QSO sightline is located at 29 kpc from the LRG. 
Absorption lines from the LRG-CGM system include: \ion{H}{1} transitions (strong \ion{H}{1} with EW(\ion{H}{1} 1215) $\sim 5.63$ \AA , and $\log N_{HI}$ between 17.69 and 20.4); intermediate ions \ion{C}{3} and \ion{Si}{3}; \ion{O}{6}; low ions \ion{C}{2}, \ion{Si}{2}, \ion{N}{2}, \ion{Mg}{2}, and \ion{Ca}{2}; and \ion{Mg}{1}. 
We note that we measure EW(\ion{C}{3} 977) in two ways, by including and by not including the central part of the line, since that part might be blended. Then, we found an average value between these two measurements. Because we do not know if the line is blended, the error in EW includes both cases, and is larger than it would be for the individual measurements. 
One additional component was found at $\sim 2500$ km s$^{-1}$, which shows \ion{H}{1} and \ion{C}{3} lines. 
Around the QSO sightline and close to the LRG's redshift, a few other galaxies are located: one blue NG, one NG close to the QSO sightline, one another LRG, and two red NGs. 

\textit{LRG\_1144+0714:} QSO sightline is located at 98 kpc from the LRG.
Absorption lines from the LRG-CGM system include: \ion{H}{1} transitions (strong \ion{H}{1} with EW(\ion{H}{1} 1215) $\sim 2.06$ \AA , and $\log N_{HI}$ between 17.91 and 19.4); intermediate ions \ion{C}{3} and \ion{Si}{3}; \ion{O}{6}; and low ions \ion{C}{2}, \ion{Si}{2}, and \ion{Mg}{2}. 
No any NGs are found around the QSO sightline. 

\textit{LRG\_1351+2753:} QSO sightline is located at 99 kpc from the LRG.
Absorption lines from the LRG-CGM system include: \ion{H}{1} transitions (strong \ion{H}{1} with EW(\ion{H}{1} 1215) $\sim 1.45$ \AA , and $\log N_{HI} \sim 17.37$); intermediate ions \ion{C}{3} and \ion{Si}{3}; \ion{O}{6}; and low ions \ion{C}{2} and \ion{Mg}{2}. We also find possible \ion{Fe}{3}, but define it as an upper limit.  
Two blue NGs are found around the QSO sightline. 

\textit{LRG\_1520+2534:} QSO sightline is located at 225 kpc from the LRG.
Absorption lines from the LRG-CGM system include: \ion{H}{1} transitions (strong \ion{H}{1} with EW(\ion{H}{1} 1215) $\sim 7.41$ \AA , and $\log N_{HI}$ between 17.86 and 20.65); intermediate ions \ion{C}{3} and \ion{Si}{3}; and low ions \ion{Mg}{2} and \ion{Fe}{2} . 
One red NG is found around the QSO sightline. 

\textit{LRG\_1106-0115:} QSO sightline is located at 383 kpc from the LRG.
Absorption lines from the LRG-CGM system include: \ion{H}{1} transitions strong \ion{H}{1} with EW(\ion{H}{1} 1215) $\sim 0.55$ \AA , and $\log N_{HI}\sim 17.26$ ); intermediate ions \ion{C}{3} and \ion{Si}{3}; low ions \ion{C}{2}, \ion{Si}{2}, and \ion{Mg}{2}.
One additional component was found at $\sim 600$ km s$^{-1}$ (shows only \ion{H}{1}). 
One red NG is found around the QSO sightline.

\textit{LRG\_1440-0157:} QSO sightline is located at 343 kpc from the LRG.
Absorption lines from the LRG-CGM system include: \ion{H}{1} transitions (strong \ion{H}{1} with EW(\ion{H}{1} 1215) $\sim 1.7$ \AA , and $\log N_{HI} \sim 17.02$)); intermediate ions \ion{C}{3} and \ion{Si}{3}; \ion{O}{6}; and \ion{Mg}{2}. We also detect possible \ion{O}{4} and \ion{Ne}{8}. 
Two red and two blue NGs are found around the QSO sightline.

\textit{LRG\_1121+4345:} QSO sightline is located at 80 kpc from the LRG.
Absorption lines from the LRG-CGM system include: \ion{H}{1} transitions (strong \ion{H}{1} with EW(\ion{H}{1} 1215) $\sim 1.72$ \AA , and $\log N_{HI}$ between 17.89 and 19.2); intermediate ions \ion{C}{3} and \ion{Si}{3} (\ion{Si}{3} is blended with a strong \ion{H}{1} 1025 from a higher redshift); \ion{O}{6}; and low ions \ion{C}{2}, \ion{Si}{2}, \ion{N}{2}, and \ion{Mg}{2}. We also possibly detect \ion{Fe}{2} and \ion{Fe}{3}. 
One blue and one red NG are found around the QSO sightline.

\subsection{Baseline LRGs} 

\textit{LRG\_1237+1106:} QSO sightline is located at 23 kpc from the LRG. 
Only \ion{H}{1} 1215 line is detected at a significance level $> 3 \sigma$. Since we detected only one line at wavelength that corresponds to the \ion{H}{1} 1215 at the LRG redshift, we compared apparent column density profiles for \ion{H}{1} 1215 and \ion{H}{1} 1025, and concluded that the absorption line is likely \ion{H}{1} 1215 at the LRG redshift. 
One red NG is found around the QSO sightline. 

\textit{LRG\_1549+0701:} QSO sightline is located at 120 kpc from the LRG. 
No any CGM lines are detected. 
Three red NGs are found around the QSO sightline. 

\textit{LRG\_1306+3421:} QSO sightline is located at 184 kpc from the LRG. 
No any CGM lines are detected. 
Two red NGs are found around the QSO sightline.

\textit{LRG\_1217+4931:} QSO sightline is located at 208 kpc from the LRG. 
Absorption lines from the LRG-CGM system include: \ion{H}{1} transitions (relatively strong \ion{H}{1} 1215 with EW $\sim 1.45$ \AA , however for $N_{HI}$ we measure only an upper limit), and intermediate ion \ion{C}{3}. We found an absorption line close to the wavelenght that corresponds to the \ion{O}{6} 1031 at the LRG's redshift, but concluded that the line is likely an interloping line from another redshift or blended. 
Three red NGs are found around the QSO sightline. 

\textit{LRG\_1102+4543:} QSO sightline is located at 193 kpc from the LRG. 
Absorption lines from the LRG-CGM system include: \ion{H}{1} transitions (relatively strong \ion{H}{1} 1215 with EW $\sim 2.52$ \AA , however for $N_{HI}$ we measure only an upper limit), and intermediate ions \ion{C}{3} and \ion{Si}{3}.  
We found a line close to the wavelenght that corresponds to \ion{O}{6} 1031 at the LRG's redshift, but concluded that the line is likely an interloping line from another redshift or blended. 
We also found that there is a possible \ion{Ne}{8} 770, however we were not able to confirm this detection, so we define it as an upper limit. 
One blue and three red NGs are found around the QSO sightline. 

\textit{LRG\_1251+3025:} QSO sightline is located at 281 kpc from the LRG. 
No any CGM lines are detected. 
One blue and one red NG are found around the QSO sightline. 

\textit{LRG\_0226+0014:} QSO sightline is located at 357 kpc from the LRG. 
No any CGM lines are detected. One red NG is found around the QSO sightline. 

\textit{LRG\_0855+5615:} QSO sightline is located at 315 kpc from the LRG. 
Absorption lines from the LRG-CGM system include: \ion{H}{1} transitions (relatively strong \ion{H}{1} 1215 with EW $\sim 1.32$ \AA , however for $N_{HI}$ we measure only an upper limit), \ion{O}{6}, and possibly weak \ion{N}{2} and blended \ion{C}{3}. 
We find one additional component (absorption system) at $\sim 1000$ km s$^{-1}$, that shows \ion{H}{1} and \ion{O}{6} absorption. 
Four red and two blue NGs are found around the QSO sightline.

\vspace{22pt}

\section{Additional components associated with LRGs}\label{sec:appe}

% add cmps - probability random 
In section \ref{sec:kin}, we found additional absorption line components around two LRGs, with velocities similar or larger than the corresponding escape velocity of the LRGs' host halos. We discussed if the additional components originate from the same halos as the target LRGs. Here we provide more details.    

We calculate the probability to detect a random Ly $\alpha$ absorption with EW as in these two additional components or higher. \citet{Danforth16} analyzed 85 spectra of active galactic nuclei at $z < 0.85$, which included 2611 different absorption systems at redshifts up to 0.75. From these data, the authors derived \ion{H}{1} column density distribution per redshift path, and parametrized it as a power law. 
We access \citet{Danforth16} data at \url{http://vizier.cfa.harvard.edu/viz-bin/VizieR?-source=J/ApJ/817/111}, derive EW(\ion{H}{1} 1215) distribution, and fit this distribution with a power law. We obtain $\log (N(>EW(HI 1215))/dz) = a EW(HI 1215) + b$, where $a = -0.0029$ and $b = 2.129$. 
We found that the probabilities to detect a random Lyman-alpha absorption with the EW as in the two additional components or higher, in the minimum velocity range that encompasses the additional components and is symmetric around the LRG redshift, are 0.072 and 0.00047. 
When taking into account that the additional components were found in one of seven MgII-LRGs and in one of eight Baseline LRGs, these probabilities are 0.58 and 0.0033. 
% add cmps - probability random - vs z
We note that the redshift of our LRGs ($\sim 0.45$) is somewhat higher than in \citet{Danforth16} sample ($\sim 0.16$), but if we assume same redshift dependence as \citet{Shull17} found, our results do not change significantly. 
For LRG\_0855+5615, we could not conclude if the additional component is associated with LRG halo. For LRG\_1059+4039, we conclude that the additional component likely originates from the same halo as the LRG, with significance $\sim 3 \sigma$.   

\vspace{22pt}

% adding for arxiv 
\section{LRG LYMAN-LIMIT DECREMENTS IN OUR QSO SPECTRA}\label{sec:appf}

\begin{figure*}[t!]
\centering
\includegraphics[width=0.9\textwidth]{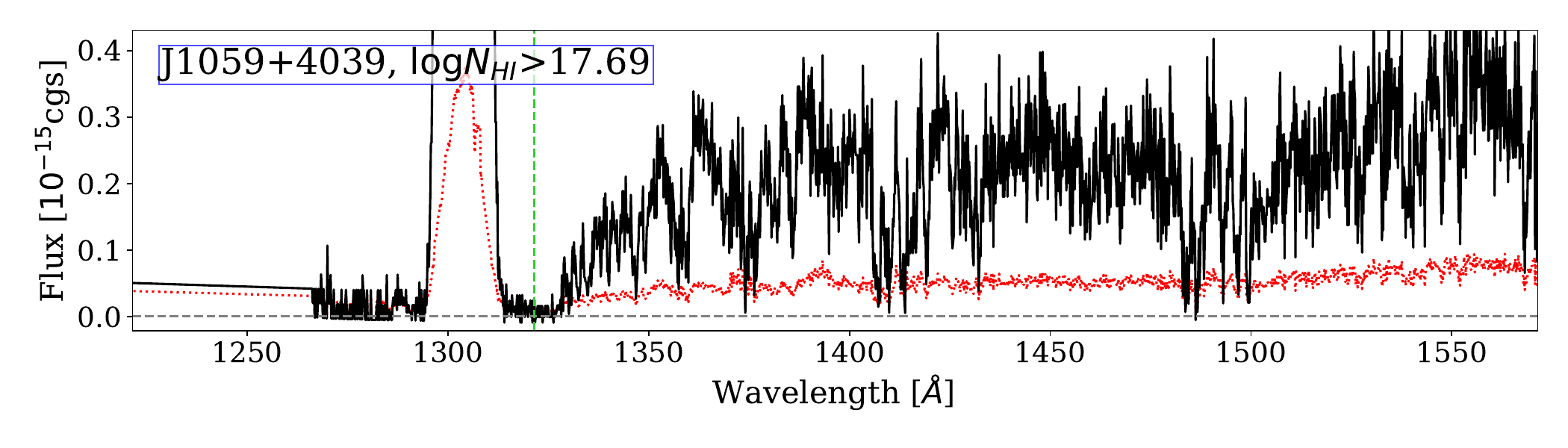}
\caption{\small 
The LL decrement in the J1059+4039 spectrum. The black line is flux, the red dotted line is the $1\sigma$ error in the flux, and the vertical green dashed line is the expected location of the LL. The measured \ion{H}{1} column density $N_{HI}$ is shown in the upper left corner. % The complete figure set (15 images with the LL decrement in all our LRG-CGM systems) is available in the online journal.
}
\label{fig:llsfig0a}
\end{figure*}

\begin{figure*}[t!]
\centering
\includegraphics[width=0.9\textwidth]{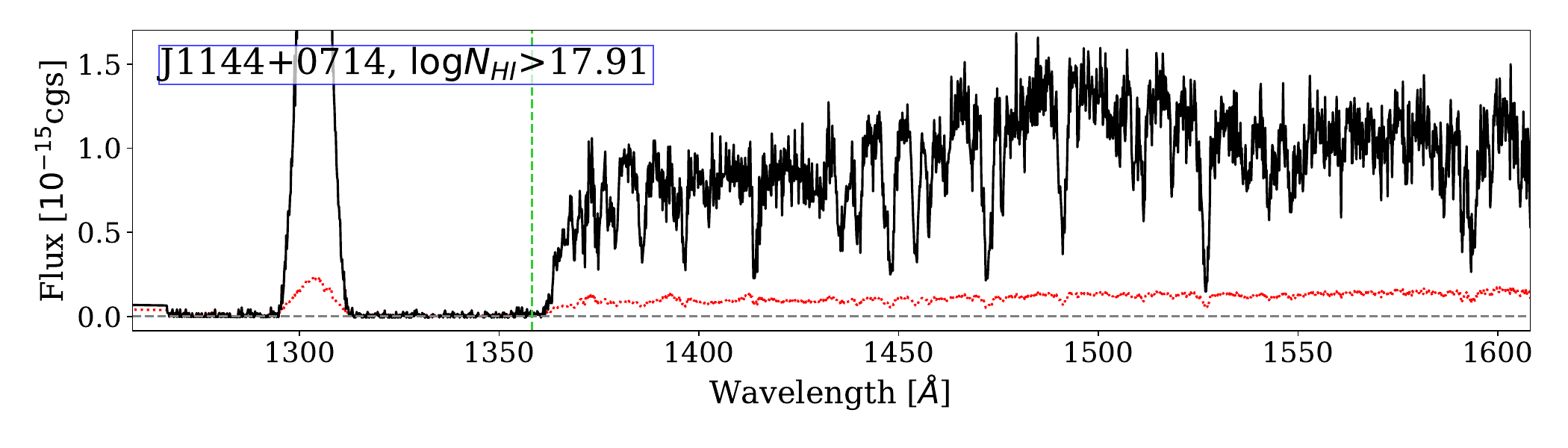}
\caption{\small 
The same as Figure \ref{fig:llsfig0}, but for J1144+0714.
}
\label{fig:llsfig1}
\end{figure*}

\begin{figure*}[t!]
\centering
\includegraphics[width=0.9\textwidth]{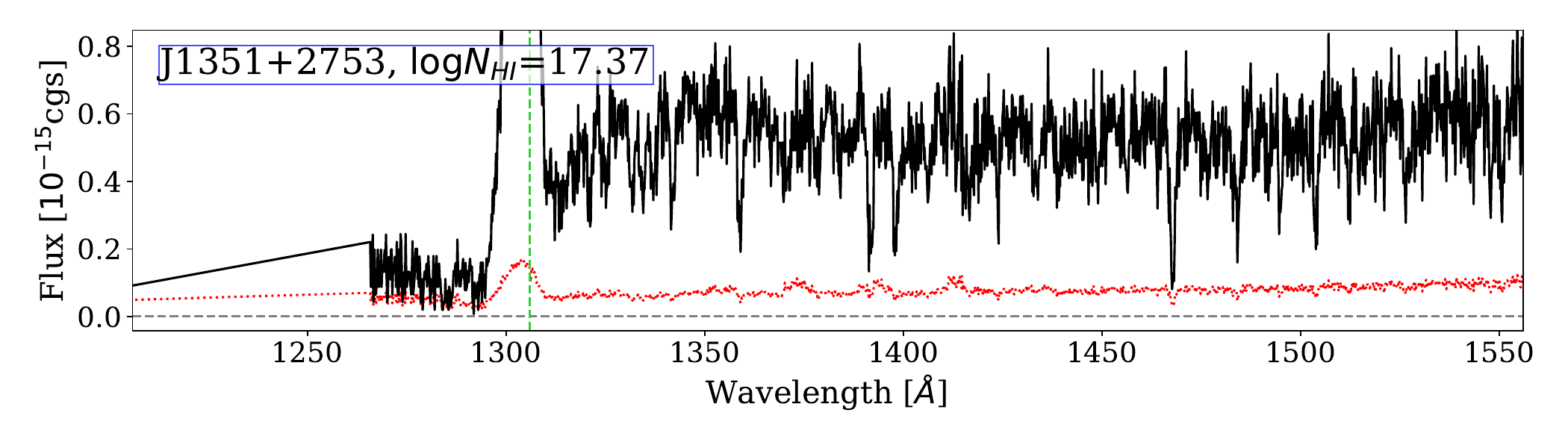}
\caption{\small 
The same as Figure \ref{fig:llsfig0}, but for J1351+2753.
}
\label{fig:llsfig2}
\end{figure*}

\begin{figure*}[t!]
\centering
\includegraphics[width=0.9\textwidth]{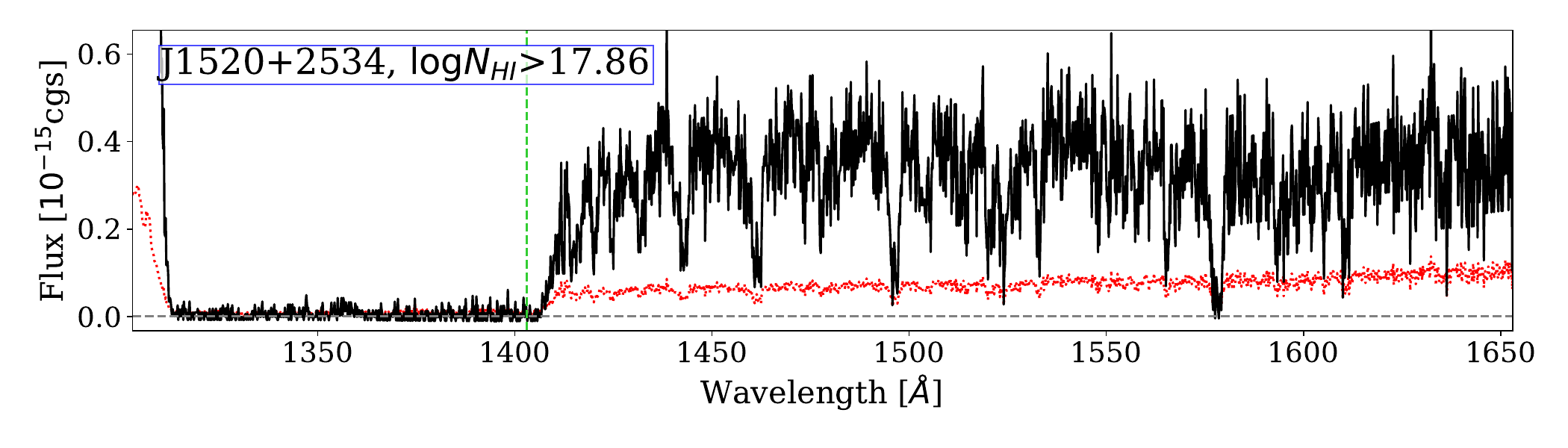}
\caption{\small 
The same as Figure \ref{fig:llsfig0}, but for J1520+2534.
}
\label{fig:llsfig3}
\end{figure*}

\begin{figure*}[t!]
\centering
\includegraphics[width=0.9\textwidth]{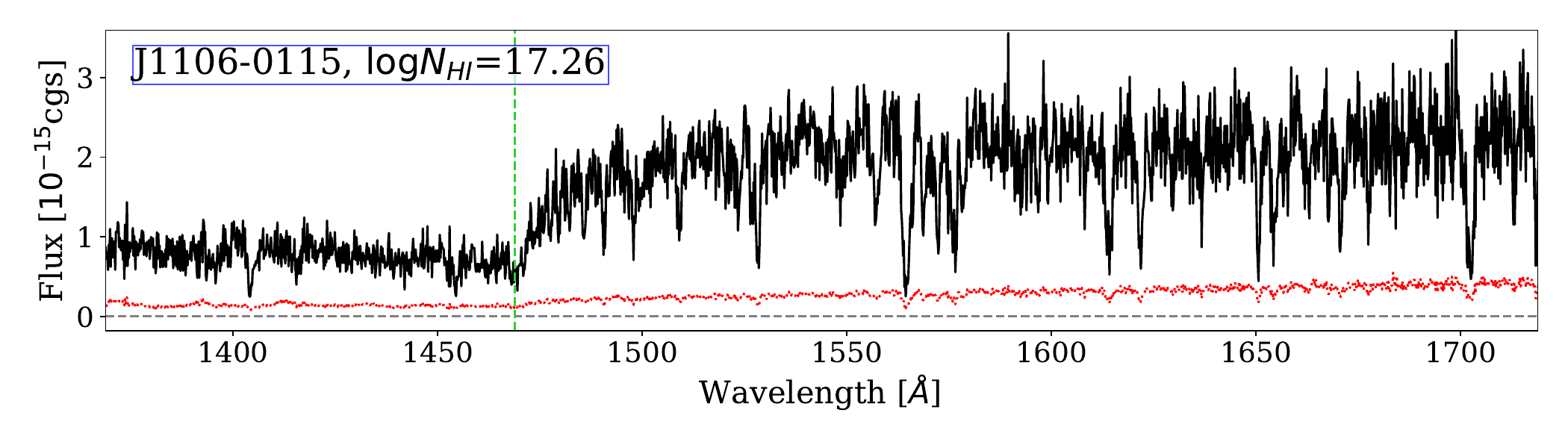}
\caption{\small 
The same as Figure \ref{fig:llsfig0}, but for J1106-0115.
}
\label{fig:llsfig4}
\end{figure*}

\begin{figure*}[t!]
\centering
\includegraphics[width=0.9\textwidth]{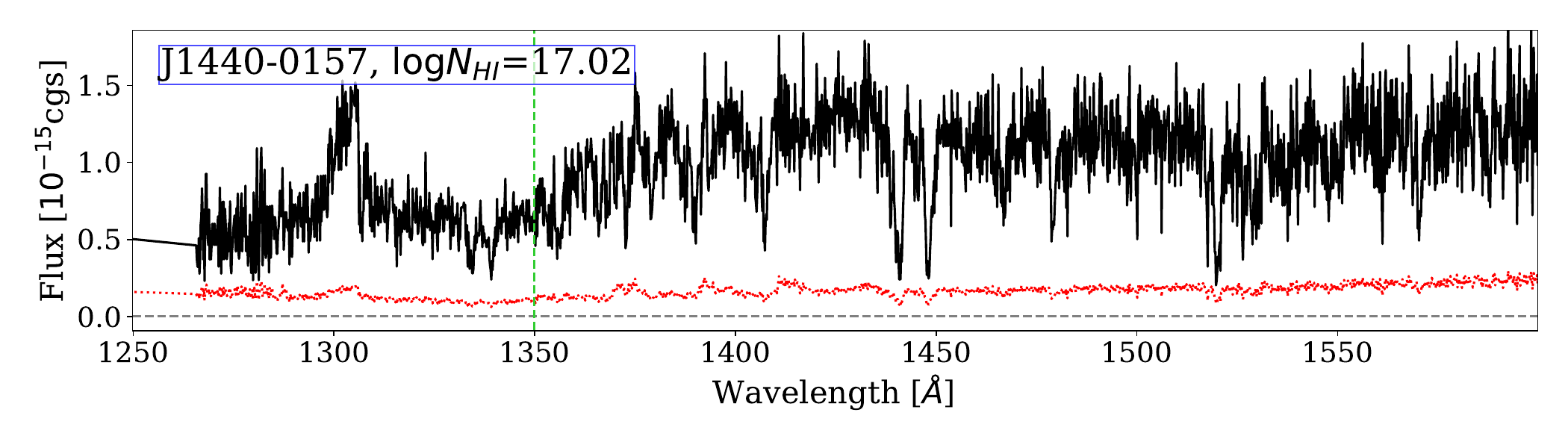}
\caption{\small 
The same as Figure \ref{fig:llsfig0}, but for J1440-0157.
}
\label{fig:llsfig5}
\end{figure*}

\begin{figure*}[t!]
\centering
\includegraphics[width=0.9\textwidth]{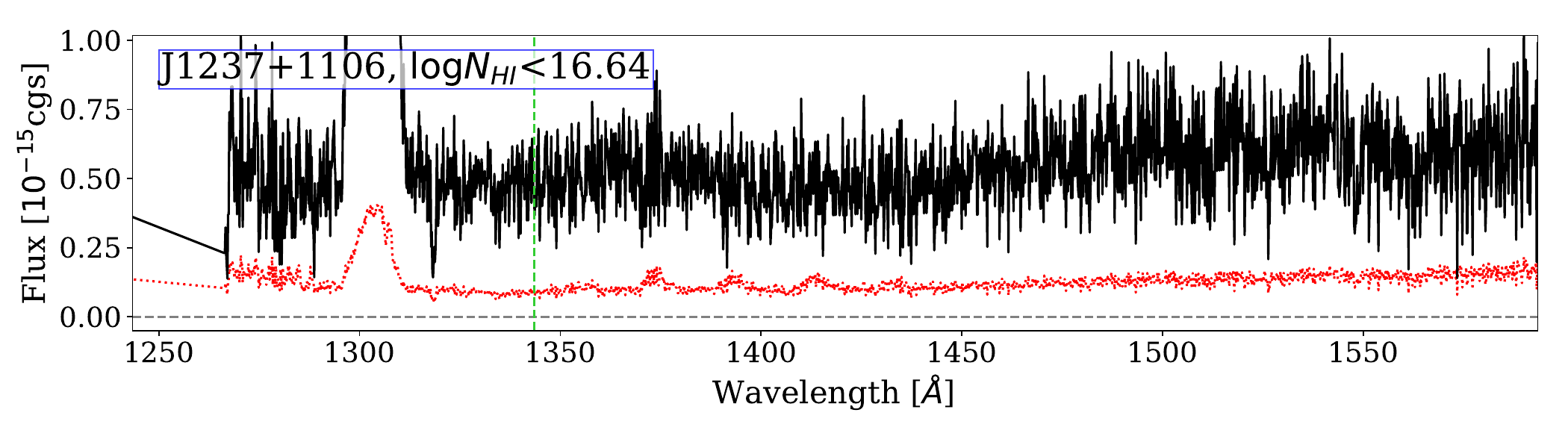}
\caption{\small 
The same as Figure \ref{fig:llsfig0}, but for J1237+1106.
}
\label{fig:llsfig6}
\end{figure*}

\begin{figure*}[t!]
\centering
\includegraphics[width=0.9\textwidth]{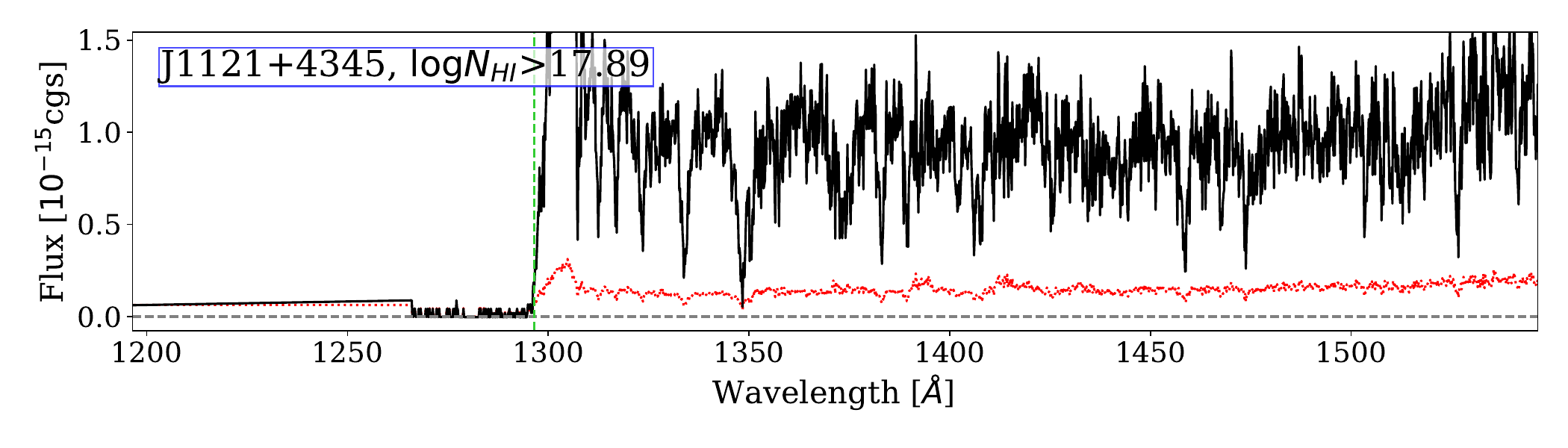}
\caption{\small 
The same as Figure \ref{fig:llsfig0}, but for J1121+4345.
}
\label{fig:llsfig7}
\end{figure*}

\begin{figure*}[t!]
\centering
\includegraphics[width=0.9\textwidth]{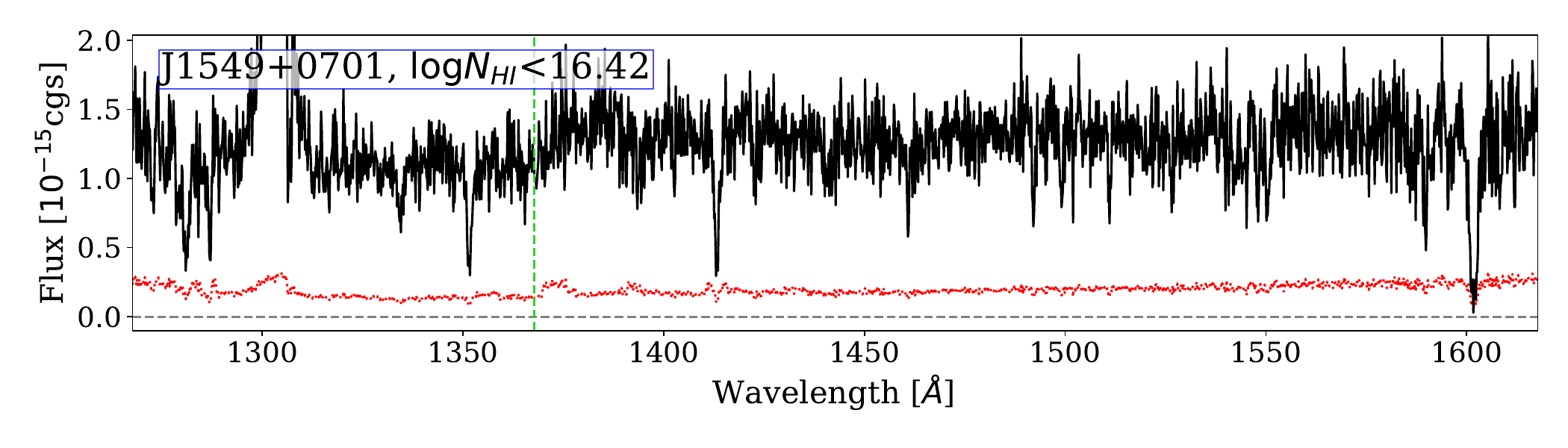}
\caption{\small 
The same as Figure \ref{fig:llsfig0}, but for J1549+0701.
}
\label{fig:llsfig8}
\end{figure*}

\begin{figure*}[t!]
\centering
\includegraphics[width=0.9\textwidth]{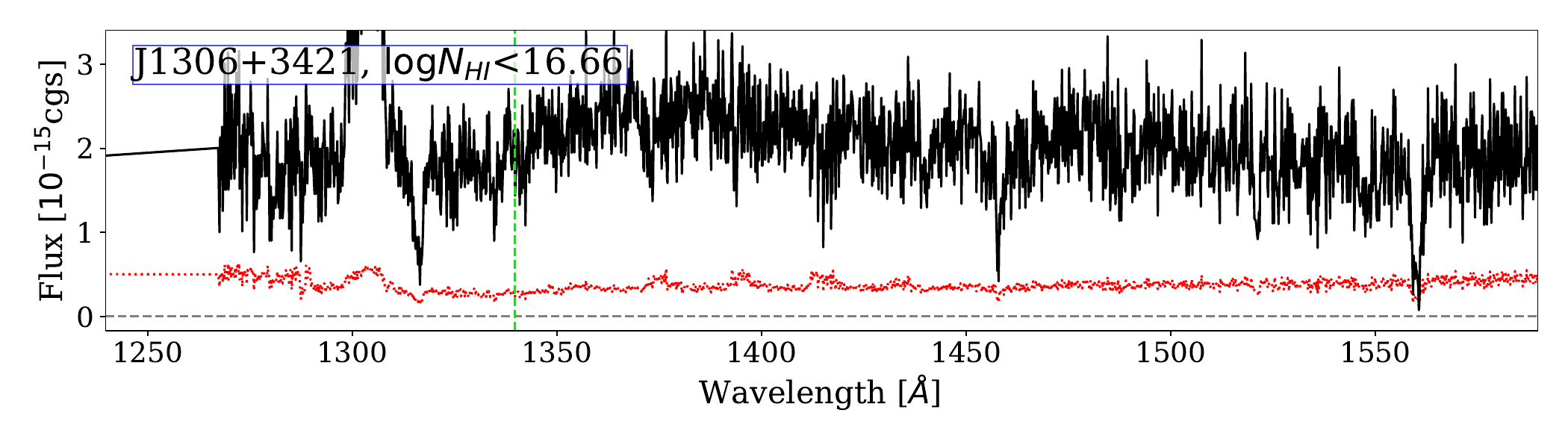}
\caption{\small 
The same as Figure \ref{fig:llsfig0}, but for J1306+3421.
}
\label{fig:llsfig9}
\end{figure*}

\begin{figure*}[t!]
\centering
\includegraphics[width=0.9\textwidth]{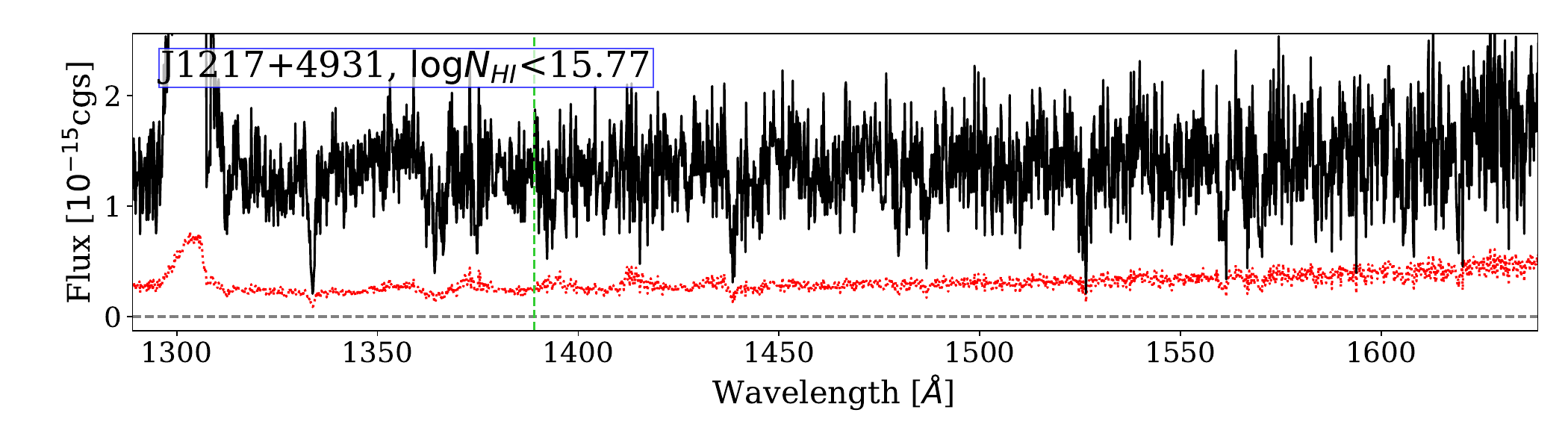}
\caption{\small 
The same as Figure \ref{fig:llsfig0}, but for J1217+4931.
}
\label{fig:llsfig10}
\end{figure*}

\begin{figure*}[t!]
\centering
\includegraphics[width=0.9\textwidth]{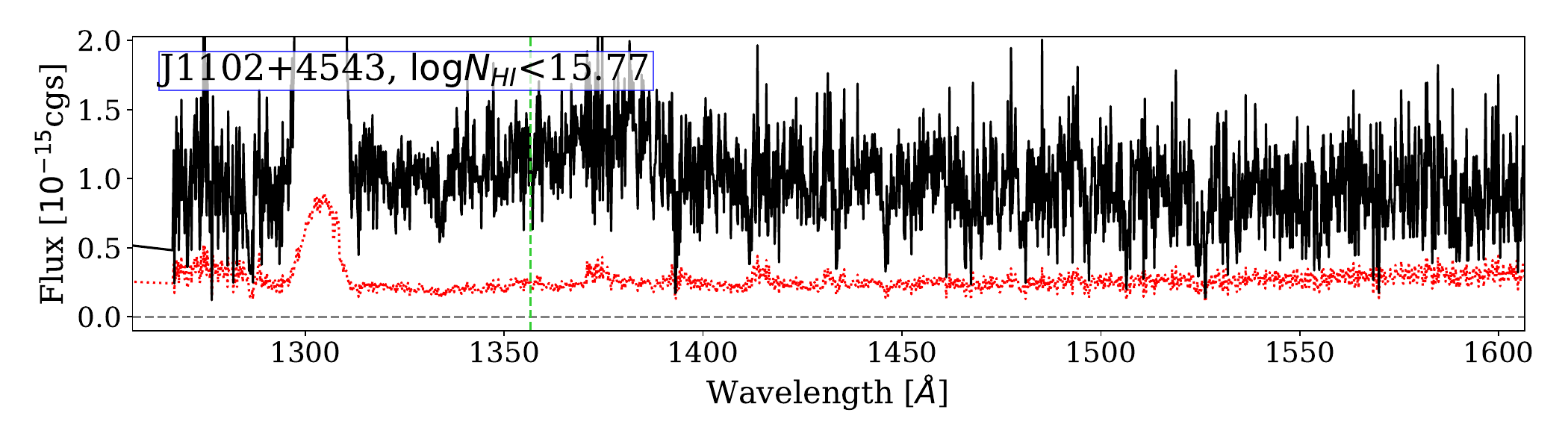}
\caption{\small 
The same as Figure \ref{fig:llsfig0}, but for J1102+4543.
}
\label{fig:llsfig11}
\end{figure*}

\begin{figure*}[t!]
\centering
\includegraphics[width=0.9\textwidth]{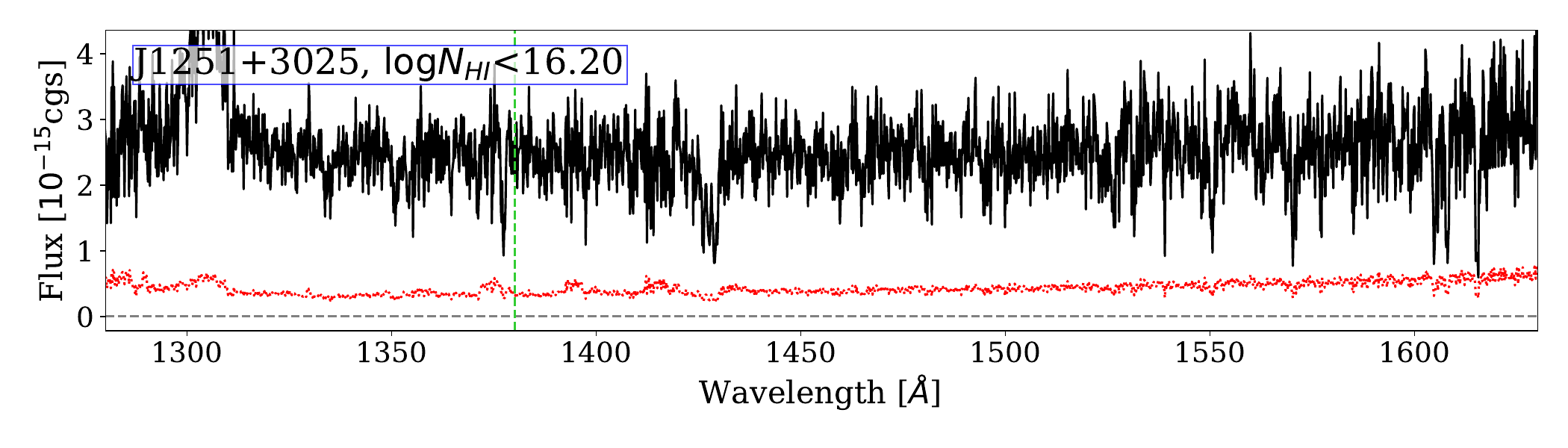}
\caption{\small 
The same as Figure \ref{fig:llsfig0}, but for J1251+3025.
}
\label{fig:llsfig12}
\end{figure*}

\begin{figure*}[t!]
\centering
\includegraphics[width=0.9\textwidth]{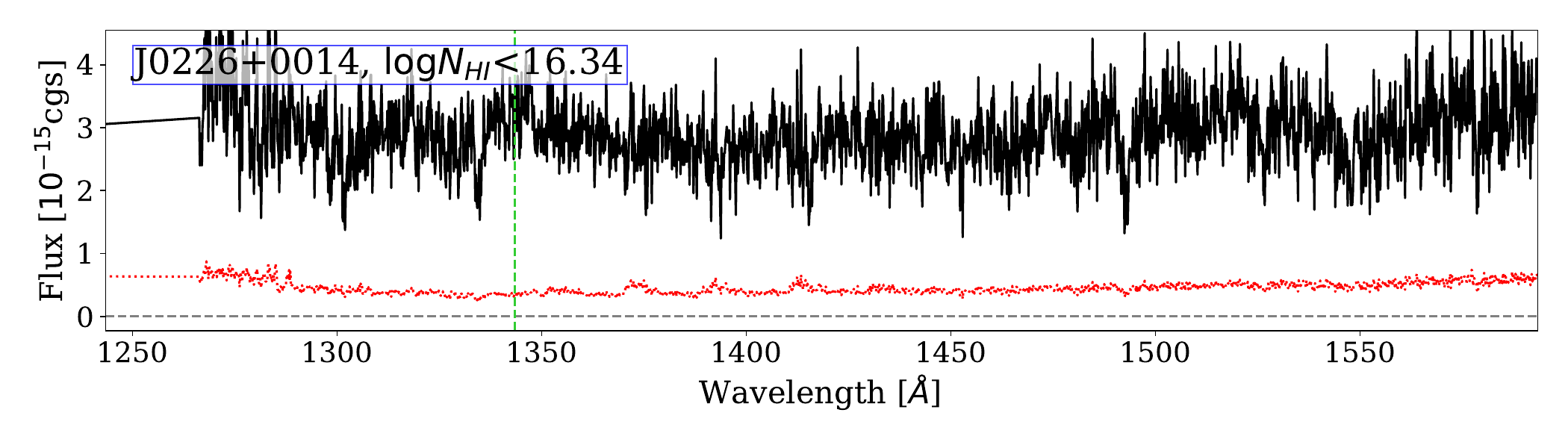}
\caption{\small 
The same as Figure \ref{fig:llsfig0}, but for J0226+0014.
}
\label{fig:llsfig13}
\end{figure*}

\begin{figure*}[t!]
\centering
\includegraphics[width=0.9\textwidth]{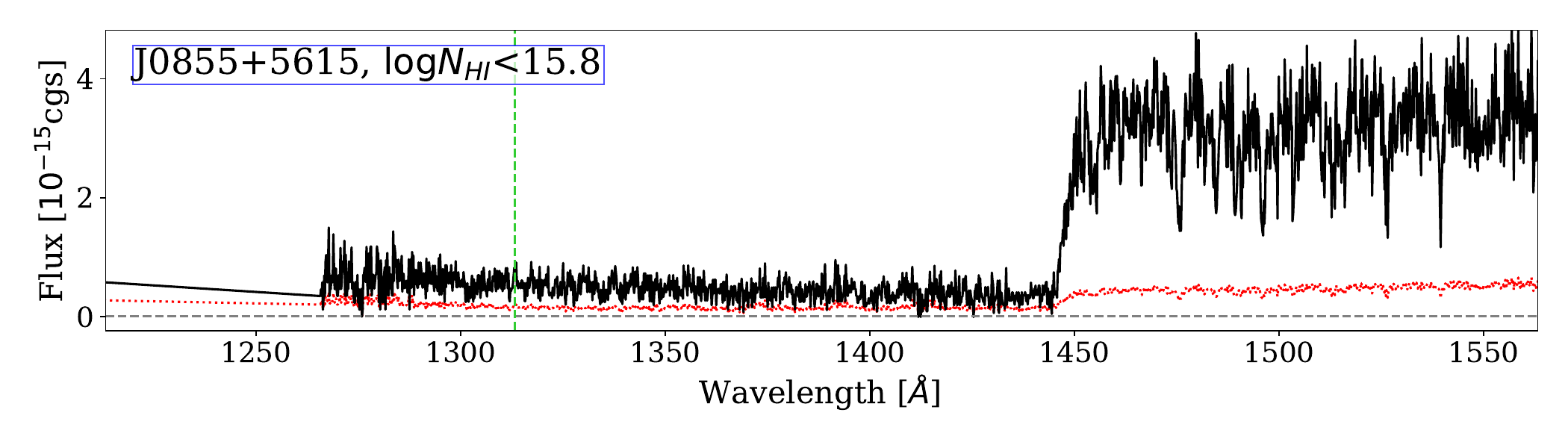}
\caption{\small 
The same as Figure \ref{fig:llsfig0}, but for J0855+5615.
}
\label{fig:llsfig14}
\end{figure*}

Figures \ref{fig:llsfig0a} - \ref{fig:llsfig14} show the LL decrement in our QSO spectra. 

\clearpage

%%% 

\clearpage

\end{document}